\documentclass[numberedappendix,apj]{emulateapj}
\usepackage{apjfonts}
\usepackage{amssymb}
\usepackage{amsmath}

\shorttitle{Velocity Centroids}
\shortauthors{Esquivel \& Lazarian}
\slugcomment{Draft version, \today}

\begin{document}

\title{Velocity centroids as tracers of the turbulent velocity statistics}

\author{Alejandro Esquivel and A. Lazarian}

\affil{Astronomy Department, University of Wisconsin-Madison, 475 N.Charter
St., Madison, WI 53706, USA}

\email{esquivel@astro.wisc.edu; lazarian@astro.wisc.edu}

\begin{abstract}
We use the results of magnetohydrodynamic (MHD) simulations to emulate
spectroscopic observations and use maps of centroids to study their
statistics.
In order to assess under which circumstances the scaling properties
of the velocity field can be retrieved from velocity centroids,
we compare two point statistics (structure functions and power-spectra)
of velocity centroids with those of the underlying velocity field and
analytic predictions presented in a previous paper \citep{LE03}.
We tested a criterion for recovering velocity spectral index from
velocity centroids derived in our previous work, and
propose an approximation of the early criterion using only the variances
of ``unnormalized'' velocity centroids and column density maps.
It was found that both criteria are necessary, however not sufficient to
determine if the centroids recover velocity statistics.
Both criteria are well fulfilled  for subsonic turbulence.
We find that for supersonic turbulence with sonic Mach numbers
$\mathcal{M}_s\gtrsim 2.5$ centroids fail to trace the spectral index
of velocity.
Asymptotically, however, we claim that recovery of velocity statistics
is always possible provided that the density spectrum is steep and the
observed inertial range is sufficiently extended.
In addition, we show that velocity centroids are useful for anisotropy
studies and determining the direction of magnetic field,
even if the turbulence is highly supersonic, but only if it
is sub-Alfv\'{e}nic. This provides a tool for mapping the magnetic
field direction,  and testing whether the turbulence
is sub-Alfv\'{e}nic or super-Alfv\'{e}nic.
\end{abstract}

\keywords{ISM: general --- ISM: structure --- MHD --- radio lines: ISM --- turbulence}

\section{Introduction}

It is well established that the interstellar medium (ISM) is turbulent.
From the theoretical point of view this arises from the very large
Reynolds numbers present in the ISM (the Reynolds number is defined
as the inverse ratio of the dynamical timescale -or {\it eddy turnover time}-
to the viscous damping timescale). From an observational standpoint
there is also plenty of evidence that supports a turbulent ISM, where
the turbulence expands over scales that range from Au to kpc
\citep*{Lar92,ARS95,DDG00,SL01}.
This turbulence is very important for many physical processes,
including star formation, cosmic ray propagation,
heat transport, and heating of the ISM (for a review see \citealt{VazOP00};
\citealt{CL05}; and references therein).

How to compare interstellar turbulence with the results of numerical
simulations and theoretical expectations is an important question that
must be addressed. After all, theoretical constructions involve necessary
simplifications, while numerical simulations of turbulence involve Reynolds,
and magnetic Reynolds numbers that are very different from those in the ISM.
Are numerical simulations of ISM of value? To what extent do they reproduce
interstellar turbulence? These sort of questions one attempts to answer
with observations.

Substantial advances in understanding of scaling of compressible
MHD turbulence\footnote{This scaling was used to solve important
astrophysical problems, for instance, finding the rates of
scattering of cosmic rays \citep{YL04} and acceleration of cosmic
dust \citep*{YLD04}.} (see reviews by \citealt{CL05,LC05}, and references
therein) allow to provide a direct comparison of the theoretical
expectations with
observations. How reliable are the turbulence spectra obtained via 
observations?

Studies of statistics of turbulence have been fruitful using interstellar
scintillations \citep{NG89,SG90}. However this technique is restricted
to the study of ionized media, and very importantly to density fluctuations
alone (see \citealt{C99}). Nowadays, radio spectroscopic observations
of neutral media provide us with an enormous amount of data containing
information about interstellar turbulence, including a more direct
physical quantity to study turbulence: velocity. But the emissivity
of a spectral line depends on both the velocity and density fields
simultaneously, and the separation of their individual contribution
is not trivial. Much effort has been put into this difficult task
and several statistical measures have been proposed to extract information
of velocity from spectroscopic data (see review by \citealt{L99}).
Among the simplest we can mention the use of line-widths
\citep{Lar81,Lar92,S84,S87}.
Velocity centroids have been around as measure of the velocity field
for a long time now \citep{H51,M58}. And they have been widely used
to study turbulence in molecular clouds \citep{KD85,DK85,MB94}.

Power-spectra, correlation, and structure functions have been
traditionally, and still are, the most widely used tools
to characterize the statistics of emissivity maps.
These statistical tools have been used to study the scaling
properties of turbulence, e.g. to determine its the spectral index. 
Recently, more elaborated
techniques have been proposed to analyze observational data, such as
``$\Delta$-variance'' wavelet transform \citep{SBHOZ98,MLO00,OML02}. 
Such techniques can be used  to obtain velocity information from spectral
data with some advantages, like being less sensitive to the effects of edges
or noise.
Regardless of the method used, it is usually assumed
that the map traces the velocity fluctuations, which
as we show below this is not always true.
To separate velocity from density contribution 
``Modified Velocity Centroids'' (MVCs) were derived in 
\citep[][hereafter LE03]{LE03}.

There has been an effort in parallel to develop new statistics
that trace velocity fluctuations. Here we can mention the ``Spectral
Correlation Function'' -SCF- \citep*{RGWW99,PRG01}, ``Velocity Channel
Analysis'' -VCA- \citep{LP00,LPVP01,ELPC03}, MVCs (LE03), and ``Velocity
Coordinate Spectrum'' -VCS- \citep{L04}.
Both VCA and VCS are good for studies of supersonic turbulence\footnote{
Using species heavier than hydrogen one can study subsonic turbulence as
well}. Although SCF
was introduced as an empirical tool, its properties, in what the
statistics of turbulence is concerned, can be derived
using a general theory in \citet{LP04}.

Synergy of different techniques is very advantageous for studies of
interstellar turbulence. \citet*{MLF03} attempted to test the
results obtained with VCA using velocity centroids.
However, as we show in the paper, 
without a reliable  criterion of whether velocity centroids
reflect the velocity statistics such studies deliver a rather limited
insight.

Numerical simulations provide us with an ideal 
testing ground for the statistical tools available for
application to observational data. However we must note that the situation
is rather complex. On one hand, real observations depend critically
on the physical properties of the object under study, such as variations
in the excitation state of the tracer and the radiation transfer within
it \citep[see][]{LP04}.
In addition observational limitations, like finite signal-to-noise
ratio and map size, griding effects, beam pattern, beam error, etc. 
are also present.
On the other hand, numerical simulations have their own limitations,
such as finite box size and resolution, numerical viscosity, and the
physics available to a particular code. This paper is mostly concerned
about the projection effects and the impact of density fluctuations
to centroid maps, which are shared by observations and simulations.

In LE03 we studied the maps of velocity centroids as tracers of the
turbulent velocity statistics. We derived analytical relations between
the two point statistics of velocity centroids and those of the 
underlying velocity field. We also identified an important term
in the structure function of centroids which includes
information of density, and that can be extracted from observables.
Subtraction of that term can isolate better the velocity contribution,
and this yielded to a new measure that we termed ``modified velocity
centroids'' (MVCs). In LE03 we proposed a criterion for determining
whether velocity centroids reflect the scaling properties of underlying
turbulent velocity (e.g. structure functions or spectra of velocity).
A major goal of this paper is to test the predictions in LE03 using synthetic
maps obtained via MHD simulations and to determine when velocity
centroids indeed reflect the velocity statistics.

Earlier on, in \citet*{LPE02} we showed how velocity
centroids can be used to reveal the anisotropy of MHD
turbulence and how this anisotropy can be used for studies of plane-of-sky magnetic
field. This technique was further discussed by \citet*{VOS03}.
In this paper we show how Mach number and Alfv\'{e}n Mach number affect
the anisotropy of velocity centroid statistics.

The results of LE03 obtained in terms of structure functions are trivially
recasted in terms of spectra and correlation functions. Therefore we use
structure, correlation functions and spectra interchangeably through our paper,
depending what measure is more convenient. While being interchangeable,
for practical statistical
data handling different measures have their own advantages
and disadvantages. We discuss those on the example of power-law scalar
field and thus benchmark our further velocity centroid study. We also
deal with a potentially  pernicious issue of non-uniformity of notations
and normalizations that plague the relevant literature by having details
of our derivations in the appendixes that  constitute an important part of
the paper.

In this work we perform a detailed numerical study of the
ability of velocity centroids to extract turbulent velocity statistics,
We study the issues of velocity-density correlations and
outline the relation of velocity centroids to other techniques.
In \S2 we review the basic problem of the density and velocity
contributions to spectroscopic observations.
We summarize LE03 in \S3, we include in this work appendices
with mathematical derivations omitted in our earlier short communication.
In \S4 we test the analytical predictions, and the spectral indices from
our numerical data. In \S5 we show how centroids can be used
for turbulence anisotropy studies and determination of the
plane-of-sky direction of magnetic field.
A discussion of the results can be found in \S6, and a summary in \S7.

\section{Turbulence statistics and spectral line data}

Due to the stochastic nature of turbulence it is best described by statistical
measures. Among these we have two point statistics such as structure functions,
correlation functions, and power spectra \citep[see for instance][]{MY75}.
Their definition and a more comprehensive discussion can be found in Appendix
A.
Structure and correlation functions depend in general on a ``lag''
$\mathbf{r}$, the separation between two points $\mathbf{x_1}$
and $\mathbf{x_2}$, such that $\mathbf{r}=\mathbf{x_2}-\mathbf{x_1}$. 
Power spectrum is defined as the Fourier Transform of the correlation function,
and is function of the wave-number vector $\mathbf{k}$. With amplitude 
$k=\vert\mathbf{k}\vert\sim 2\pi/r$, where $r=\vert\mathbf{r}\vert$.
Additional simplification is achieved if the turbulent field is  isotropic, 
in which case structure and correlation functions depend only on the
magnitude of the separation $r$ (and not on the direction), similarly
power spectrum is only function of $k$.
This is not strictly true for magnetized media, as the presence of a magnetic
field introduces a preferential direction for motion.
In fact, MHD turbulence becomes axisymmetric in a system of reference
defined by the direction of the {\it local} magnetic field (see reviews
by \citealt{CL05} and \citealt*{CLV03a}), thus breaking the isotropy.
However, since the local direction changes from one place to another,
the anisotropy is rather modest and it is still possible
to characterize the turbulence with isotropic statistics (see \citealt{ELPC03}).

\subsection{Three-dimensional power-law statistics}

In the simplest realization of turbulence we have injection of energy
at the largest scales. The energy cascades down without losses to the
small scales, at which viscous forces become important and turbulence
is dissipated.
At intermediate scales, between the injection and the dissipation
scales, the turbulent cascade is {\it self-similar}.
This range constitutes the so called {\it inertial-range}.
There the physical variables are proportional to simple powers of
eddy sizes, and the two point statistics can be described by power-laws. 
For power-law statistics \citet{LP00}
discussed two regimes, a {\it short-wave-dominated} regime, corresponding
to a {\it shallow} spectrum, and a {\it long-wave-dominated} spectrum,
with a {\it steep} spectrum.

While dealing with numerical data one encounters a few non-trivial
effects that we find advantageous to discuss below. The insight into
the limitations of numerical procedures that involve conversion from
mathematically equivalent statistics helps us for the rest of the paper.

\subsubsection{Steep (long-wave-dominated) spectrum}

Consider an isotropic power-law one-dimensional power-spectrum
of the form
\begin{equation}
P_{1D}(k) = C\  k^{\gamma_{1D}}.
\label{eq:Ekpowlaw}
\end{equation}
A {\it steep} spectrum corresponds to spectral indices $\gamma_{1D}<-1$.
The structure function $D(r)$ can be written in terms of the spectrum
as
\begin{equation}
D(r)=4\int_{0}^{\infty}\left[1-\cos(k\  r)\right]\  P_{1D}(k)\ {\rm d}k.
\label{eq:drEK}
\end{equation}
For a power-law steep power spectrum (substituting eq.[\ref{eq:Ekpowlaw}]
into eq.[\ref{eq:drEK}]), the structure function also follows
a power-law:
\begin{equation}
D(r)=A\  r^{\xi}=A\  r^{-1-\gamma_{1D}}\ \ \ \ \ \ \ \ \ \ 0<\xi<2,
\label{eq:Drpowlaw}
\end{equation}
where $A=C~2 \pi/[\Gamma(1+\gamma)\sin(\pi\gamma/2)]$, and $\Gamma(x)$
is the Euler Gamma function. The relation between the spectral index
of the structure function and power spectrum can be generalized for
isotropic fields to $P_{ND}\propto k^{\gamma_{ND}}$, with
\begin{equation}
\gamma_{ND}=-N-\xi.
\label{eq:gammaND}
\end{equation}
Where N is the number of dimensions (see Appendix A).
For instance, the velocity ($v$) in Kolmogorov turbulence scales
as $v\propto r^{1/3}$, which corresponds to a spectral index for
the structure function of $\xi=2/3$, to a three-dimensional power-spectrum
($P_{3D}$) index $\gamma_{3D}=-11/3$, a two-dimensional power-spectrum 
($P_{2D}$) index $\gamma_{2D}=-8/3$, and a one-dimensional power-spectrum 
$P_{1D}$ index $\gamma_{1D}=-5/3$.
Note that Kolmogorov spectrum falls into the steep spectra category.

Structure functions given by equation (\ref{eq:Drpowlaw}) are
well defined only for $\xi>0$, which allows to satisfy $D(0)=0,$ and $\xi<2$,
so that the representation in terms of Fourier integrals is possible
(see \citealt{MY75}).\footnote{Correlation functions in this regime are maximal, and finite
at $r=0$.  It is important to notice also that for a power-law steep spectrum
the structure function is a power-law, but the correlation
function is not (see eq. {[}\ref{eq:Dij_dec}{]}). The correlation
function is a constant minus a growing (positive index)
power-law, therefore in a log-log scale is flat at small scales,
and drops at large scales.}

\begin{figure}
\epsscale{1.0}
\plotone{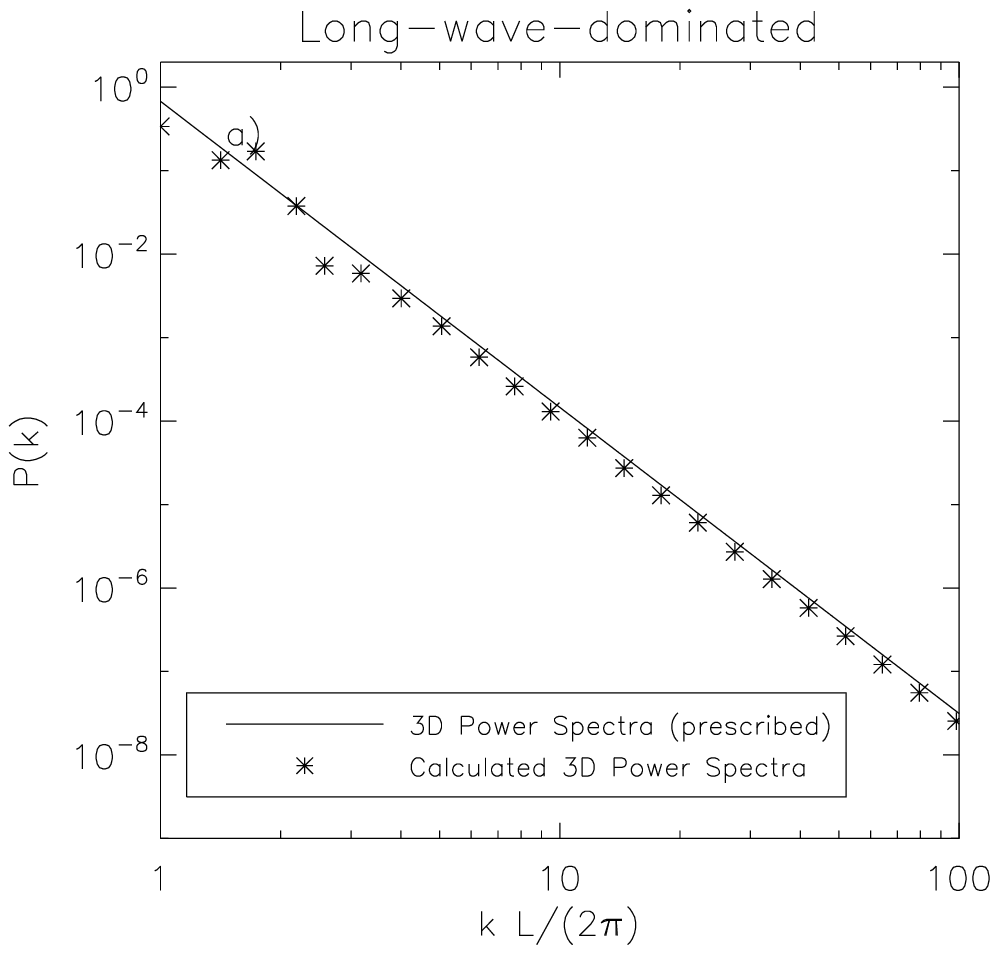}\plotone{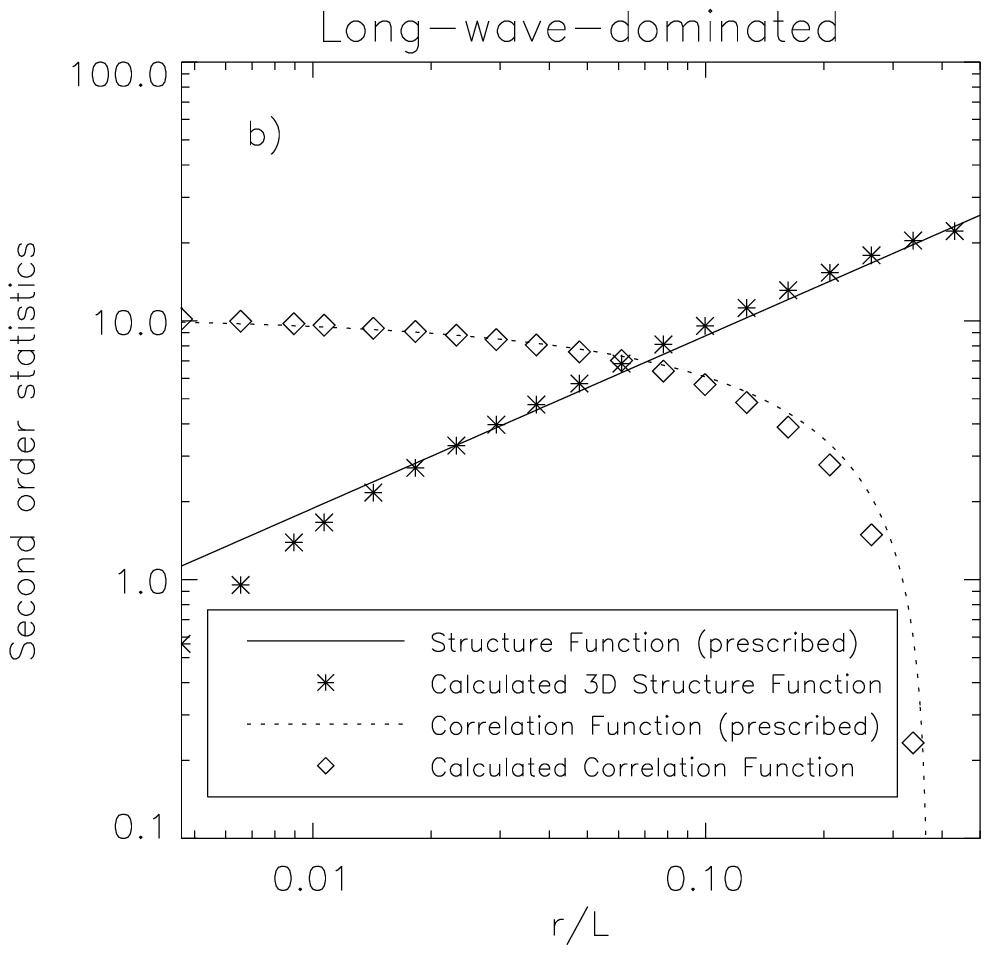}
\figcaption{Three-dimensional two point statistics of a long-wave dominated Gaussian field (steep spectrum). In panel ({\it a}) the three-dimensional power spectrum, the solid line is the prescribed spectrum, with a spectral index of $\gamma_{3D}=-11/3$, the stars correspond to the calculated. In panel ({\it b}) the expected structure function ({\it solid} line), the computed structure function ({\it stars}); the expected correlation function of fluctuations ({\it dotted} line) , and the calculated correlation function of fluctuations({\it diamonds}).
\label{fig:3Dsteep}}
\end{figure}

To illustrate the relation between the two point statistics and power-law
spectrum, as well as the difficulties associated with handling numerical data. We produced a three-dimensional Gaussian cube with a prescribed
(3D) spectral index of $\gamma_{3D}=-11/3$, as described in \citet{ELPC03}.
This type of data-cubes are somewhat similar to the fractional Brownian
motion (fBms) fields used by \citet*{BH02}, or the {\it de-phased} fields
used in \citet{BHVSP03}. However, as in real observations, they do not have
perfect power-law spectrum for a particular realization, but only in
a statistical sense \citep[see][]{ELPC03}. 
In Figure  \ref{fig:3Dsteep} we show the calculated 3D power spectrum,
structure and correlation functions of our steep Gaussian cube, and
compare them with the prescribed scaling properties.
The power spectrum is computed using Fast Fourier Transform (FFT) in 3D
and then averaged in wave-number $k$. Ideally one can compute directly in
real space the 3D structure or correlation functions, however for a 3D field
of the dimensions used here ($216^3$) is already quite expensive computationally.
It would require looping over all the points in the data-cube to do the average
and repeat for all the possible values for the lag (in three dimensions as well).
Fortunately, since the data-cubes were produced using FFT we can safely compute
the correlation function with spectral methods.
The correlation function can be expressed as a convolution integral
$B(r)\propto \int {\rm d}r^{'}f(r)f(r+r^{'})$, which can be calculated as a simple
product of the Fourier tranformed fields.
That is, $B(r)\propto \mathcal{F}\{\hat{f}(k)~\hat{f}^{*}(k)\}$, where $\hat{f}(k)=\mathcal{F}\{f(r)\}$ is
the Fourier transform of $f(r)$, and $\hat{f}^{*}(k)$ its complex conjugate.
Then, with the use of equation (\ref{eq:Dij_dec}) we can obtain the
structure function.
The resulting correlation and structure functions in 3D are then averaged
in $r$.
An important thing to notice is that the Gaussian cubes have wrap-around
periodicity, and the largest variation available correspond to
scales of $L/2$, where $L$ is the size of the computational box.
In fact, we only plot the structure and correlation functions
up to such separations.
We see a fair agreement with prescribed and the measured scaling properties.
The power spectrum in Figure \ref{fig:3Dsteep}({\it a}) shows departures
from strict power-law which are more evident for small wave-numbers.
This is natural for this type of data-cubes, where random deviations from
strict power-law are expected at all scales.
But at large scales (small $k$) we have fewer points for the
statistics and the departures do not average to zero,
while at small scales they almost do.

\subsubsection{Shallow (short-wave-dominated) spectrum}

\begin{figure}
\epsscale{1.0}
\plotone{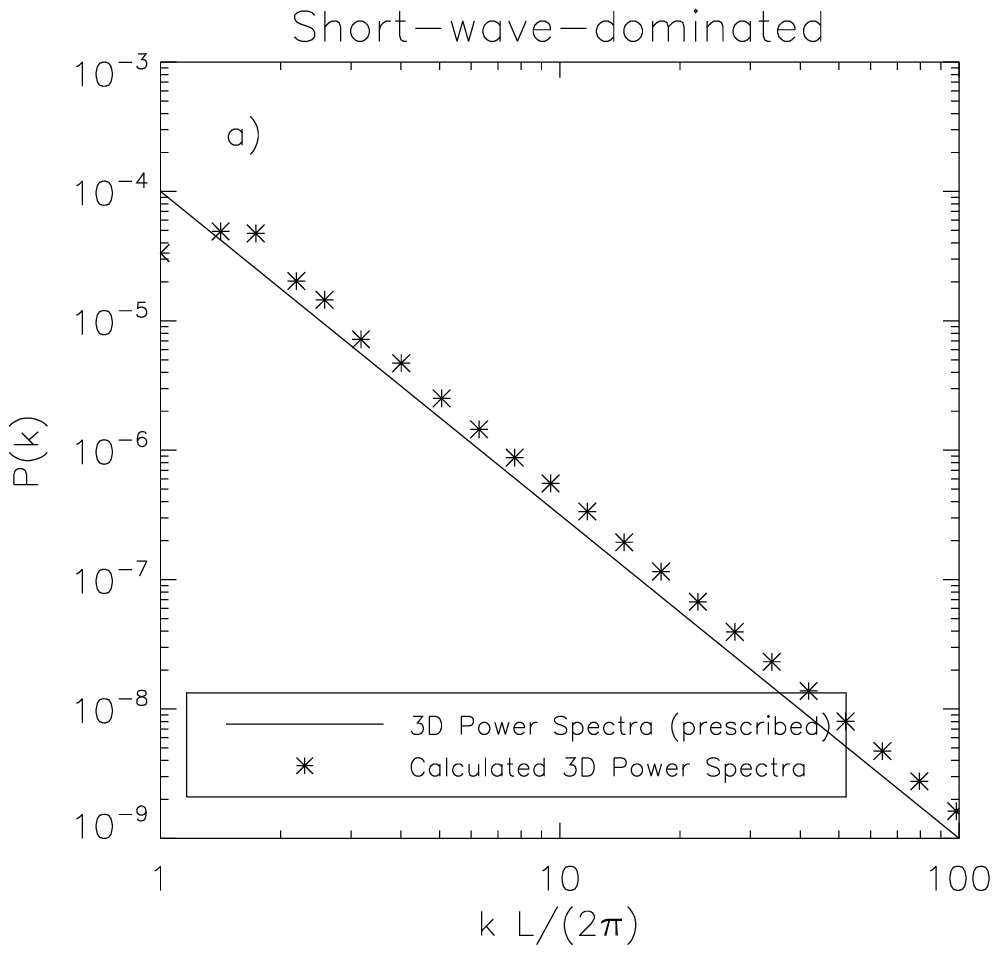}\plotone{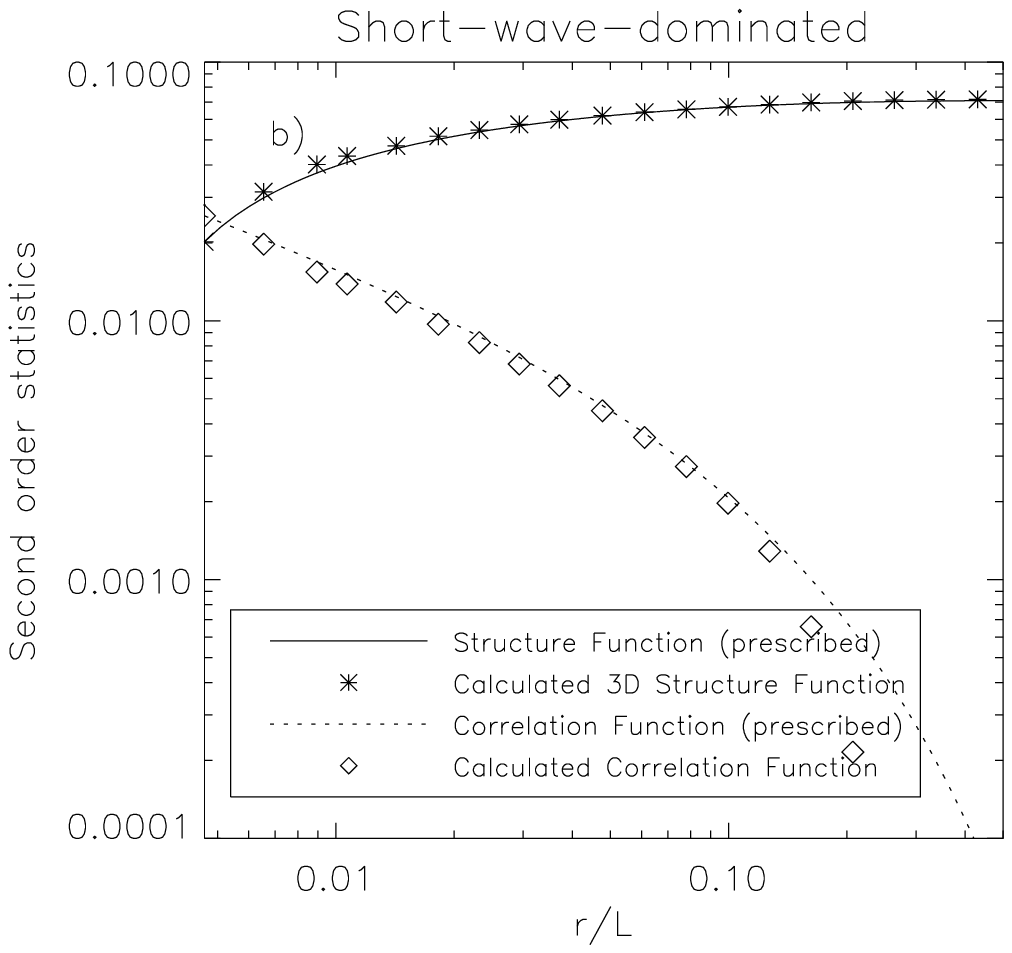}
\figcaption{Same as Fig. 1 but for a short-wave dominated spectrum (3D spectral index of $\gamma_{3D}=-2.5$). \label{fig:3Dshallow}}
\end{figure}

When the energy spectrum is shallow (i.e. $\gamma_{1D}>-1$), the
fluctuations of the field are dominated by small-scales, therefore
termed {\it short-wave-dominated} regime.
Density at high Mach number is an example of such shallow spectrum
\citep*[see][]{BLC05}.
In this case neither the
structure nor correlation functions can be strictly represented by
power-laws. In fact, in order for the Fourier Transforms to converge
in this case we need to introduce a cutoff for small wave-numbers, 
such that the power spectrum is only a power-law for $k > k_0$,
in other words
\begin{equation}
P_{1D}(k)=C^{'}\left(k_{0}^{2}+k^{2}\right)^{\gamma_{1D}/2}=C^{'}\ \left(k_{0}^{2}+k^{2}\right)^{(-\eta-1)/2}.
\label{eq:Ekshallow}
\end{equation}
To a power spectrum of this form corresponds a correlation function
of fluctuations:
\begin{equation}
\tilde{B}(r)=A^{'}\left(\frac{r}{r_{c}}\right)^{\eta/2}K_{\eta/2}\left(\frac{2\pi r}{r_{c}}\right),
\label{eq:Brshallow}
\end{equation}
 where, $r_c=2\pi/k_0$,  $K_{\eta}(x)$ is the $\eta$-order, modified Bessel function
of the second kind (also sometimes referred as hyperbolic Bessel function),
and $A^{'}=C^{'}\,2^{1-\eta}\,\pi^{-(\eta+1)/2}\,r_{c}^{\eta}/\Gamma[(\eta+1)/2]$.
The Nth-dimensional power spectrum index for a correlation function
of the form $B\propto(r/r_{c})^{\eta/2}K_{\eta/2}(2\pi r/r_{c})$ can also
be generalized to $P_{ND}\propto(k_{o}^{2}+k^{2})^{\gamma_{ND}/2}$,
with
\begin{equation}
\gamma_{ND}=-N-\eta.
\label{eq:gammaNDeta}
\end{equation}
This relation is very similar to that for the long wave dominated
case (eq{[}\ref{eq:gammaND}{]}). Actually for $r\ll r_{c}$,  $K_{\eta/2}$
can be expanded as $\sim(r/r_{c})^{\eta/2}$, thus the 3D correlation
function (as opposed to the structure function as in the long wave
dominated regime) goes as a power-law $\tilde{B}(r)\sim(r/r_{c})^{\eta}$
for small separations. Notice that for a shallow spectrum the structure
function grows rapidly at the smallest scales and then flattens. Similarly
to the steep spectrum, we produced a short-wave-dominated Gaussian
3D field, with a prescribed index of $\gamma_{3D}=-2.5$.
In Figure \ref{fig:3Dshallow} we present the expected, and calculated
two point statistics. Here the critical scale $r_{c}$ is determined
by the smallest wave-number ($k_0=2\pi /L$), in our case it corresponds
to the size of the computational box ($r_{c}=L$).
We see again a fair agreement with the prescribed and calculated spectra.
However, Figure  \ref{fig:3Dshallow} reveals a significant departure of
the calculated correlation function with the prediction from equation
(\ref{eq:Brshallow}) for large lags.
The explanation of such difference is that the analytical relations of
spectra (eqs. [\ref{eq:Ekpowlaw}, \ref{eq:Ekshallow}]) with structure
and correlation functions (eqs. [\ref{eq:Drpowlaw}, \ref{eq:Brshallow}])
is exact only in the limit of continuous integrals over infinite wave-numbers.
The data-sets presented in this section are constructed in Fourier space,
then translated to real space by means of discrete Fourier transforms of the
form:
\begin{equation}
\tilde{u}(x)=\sum_{k=1}^{L-1}\left|P_{1D}(k)\right|^{1/2}\exp\left[2\pi k x /L \right]
\label{eq:FT_sum}
\end{equation}
The sum runs from $k=1$ ensuring that $\langle \tilde{u}(x)\rangle=0$. 
In practice, we evaluate the Fourier Transforms via FFT and set explicitly the
$k=0$ component of the spectrum to zero to guarantee that the average of the
fluctuating part of the field is null. 
The resulting field has a limited range of harmonics available, determined
basically by the computational grid size ($L$).
We have constructed large 1D fields, in which we see that the gap
between the analytical and the computed correlation functions
gets smaller as we increase resolution.
Thus, it is not surprising that spectra shows a much  better correspondence
than correlation functions. At the same time, structure functions
do not deal with the lowest harmonics, which introduce largest errors
\citep[see][]{MY75}. 
And therefore, they are less affected by the lack of lower harmonics as can be
confirmed by the fact that they are closer to the analytical prediction than
correlation functions.
It is important to always keep in mind the issues that can arise from the
discrete nature of the data.
However, we must note that this particular problem lies within the generation
of the data-sets in frequency space and not in the computation of correlation or
structure functions via spectral methods. We obtain identical results
using FFT and looping in directly in real space to do the average required.
In real life, the limitation is likely to be in the opposite direction: the
finite wave-numbers available would show up as uncertainty in
determining the power spectra while structure and correlation functions
should be estimated with smaller errors (if measured directly in real space).

\subsection{Structure functions of quantities projected along the line of sight}

From spectroscopic observations we can not obtain
either the density or velocity fields in real space ($x$, $y$, $z$),
but we have to deal with projections along the line of sight (LOS).
Despite the fact that our main goal is to extract velocity statistics from the
centroids of velocity, we will discuss in this section the statistics of density
integrated along the LOS (column density). It will become clear later that the
same procedure can be applied to obtain velocity statistics.
The issue of projection has been previously discussed in \citet{L95}, here
we briefly state some results that are relevant to this work.
In Appendix B we exemplify the projection effects of
structure functions for the particular case of power-law statistics.
And in Appendix C the power-spectrum of a homogeneous
field that has been integrated along the LOS.

In what follows we will assume that the emissivity of our medium is
proportional to the first power of the density (this is true, for
instance, in the case of \ion{H}{1}).
We will consider an isothermal medium,
and neglect the effects of self-absorption. In this case the integrated
intensity of the emission (integrated along the velocity coordinate)
is proportional to the column density (see Appendix A for more details):
\begin{equation}
I(\mathbf{X})\equiv\int\alpha\ \rho_{s}(\mathbf{X},v_{z})\ {\rm d}v_{z}=\int\alpha\ \rho(\mathbf{x})\ {\rm d}z,
\label{eq:I}
\end{equation}
where $\alpha$ is a constant and $\rho(\mathbf{x})$ is the mass density.
The density of emitters $\rho_{s}(\mathbf{X},v_{z})$ can be identified
as the column density per velocity interval, commonly
referred as ${\rm d}N/{\rm d}v$. To distinguish between 2D and 3D vectors,
we will use capital letters to denote the former and lower case for the
latter (i.e. $\mathbf{X}=[x,y]$, $\mathbf{x}=[x,y,z]$).
Our assumption is satisfied for observational
data where the medium is optically thin, thermalized, and with constant
excitation conditions. However, for any observed map its applicability
has to be examined carefully. Even for \ion{H}{1} widespread
self-absorption has been detected (for example \citealt{J02,LG03}).

Consider the structure function of the integrated intensity described
in equation (\ref{eq:I})
\begin{equation}
\left\langle \left[I(\mathbf{X}_{1})-I(\mathbf{X}_{2})\right]^{2}\right\rangle =\left\langle \left(\int_{0}^{z_{tot}}\alpha\ \rho(\mathbf{x_{1}})\ {\rm d}z-\int_{0}^{z_{tot}}\alpha\ \rho(\mathbf{x_{2}})\ {\rm d}z\right)^{2}\right\rangle ,
\label{eq:I1I2def}
\end{equation}
where we have written explicitly the limits of integration, with $z_{tot}$
being the size of the object (in the LOS direction). 
Clearly $z_{tot}$ does not necessarily have to coincide with the transverse
size (in the plane of the sky) of the object under study, however that is
the case in our data-sets.
As described in \citet{L95}, we can expand the square in equation
(\ref{eq:I1I2def}) combining
\begin{equation}
\left(\int\chi(x){\rm d}x\right)^{2}=\iint\chi(x_{1})\chi(x_{2}){\rm d}x_{1}{\rm d}x_{2},
\label{eq:int2}
\end{equation}
and the elementary identity
\begin{equation}
\left(a-b\right)\left(c-d\right)=\frac{1}{2}\left[\left(a-d\right)^{2}+\left(b-c\right)^{2}-\left(a-c\right)^{2}-\left(b-d\right)^{2}\right],
\label{eq:elem}
\end{equation}
to obtain
\begin{equation}
\left\langle \left[I(\mathbf{X}_{1})-I(\mathbf{X}_{2})\right]^{2}\right\rangle =\alpha^{2}\int_{0}^{z_{tot}}\int_{0}^{z_{tot}}{\rm d}z_{1}{\rm d}z_{2}\left[d_{\rho}(\mathbf{r})-\left.d_{\rho}(\mathbf{r})\right|_{\mathbf{X}_{1}=\mathbf{X}_{2}}\right].
\label{eq:I1I2dr1dr2}
\end{equation}
Where $d_{\rho}(\mathbf{r})$ is the 3D structure function of
the density,
\begin{equation}
d_{\rho}(\mathbf{r})=\langle[\rho(\mathbf{x_{1}})-\rho(\mathbf{x_{2}})]^{2}\rangle.
\label{eq:drho}
\end{equation}
This definition is general and does not require any
particular functional form of the 3D structure functions (i.e. power-law
statistics). The problem of formally inverting equation
(\ref{eq:I1I2dr1dr2}) to obtain the underlying statistics,
allowing for anisotropic turbulence, with an arbitrary spectrum
(i.e. not a power-law), has been presented in \citet{L95}, but it is
somewhat mathematically involved.
For 3D fields with a power-law spectrum, homogeneous and isotropic,
the structure functions of the integrated fields (2D maps) can be simply
approximated by two power-laws, one at small the other at large separations
(see \citealt{M02}, and also Appendix B).
For instance, if the density has a power-spectrum
$P_{3D,\rho}\propto k^{\gamma_{3D}}$,
the structure function of column density will have the form
\begin{equation}
\left\langle \left[I(\mathbf{X}_{1})-I(\mathbf{X}_{2})\right]^{2}\right\rangle \propto R^{\mu},
\label{eq:I1I1pow-law}
\end{equation}
where $R$ is the separation in the plane of the sky
($R=\vert\mathbf{R}\vert=\vert\mathbf{X_{2}}-\mathbf{X_{1}}\vert$),
and
\begin{equation}
\mu\approx
\begin{cases}
-\gamma_{3D}-2 & \textrm{for }R\ll z_{tot}\textrm{ (either steep or shallow spectrum)},\\
-\gamma_{3D}-3 & \textrm{for }R\gg z_{tot},\textrm{ and }\gamma_{3D}<-3\textrm{ (steep spectrum)},\\
0 & \textrm{for }R\gg z_{tot},\textrm{ and }\gamma_{3D}>-3\textrm{ (shallow spectrum)}.
\end{cases}
\label{eq:mu}
\end{equation}

In contrast, the power spectrum of a field integrated along the LOS
corresponds to selecting only the $k_{LOS}=0$ components of the
underlying 3D spectra,
more precisely only the solenoidal part (see Appendix
C).
Thus, for isotropic and homogeneous power-law statistics the 2D power 
spectrum
will reflect the 3D spectral index.
If for instance the density has a power spectrum 
$P_{3D,\rho}\propto k^{\gamma_{3D}}$, the spectrum of column density will
scale as
\begin{equation}
P_{2D,I}\propto K^{\gamma_{3D}}.
\label{eq:P2D_gamma3D}
\end{equation}

\begin{figure}
\epsscale{1.0}
\plotone{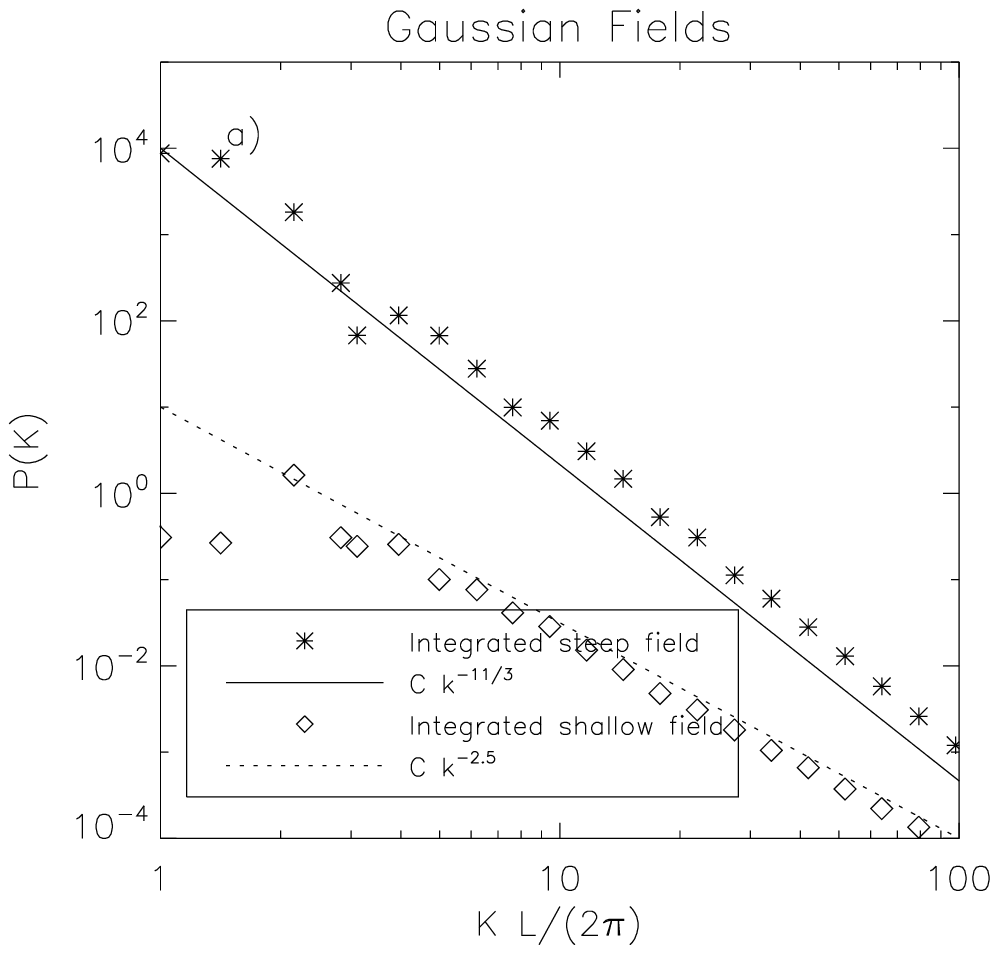}\plotone{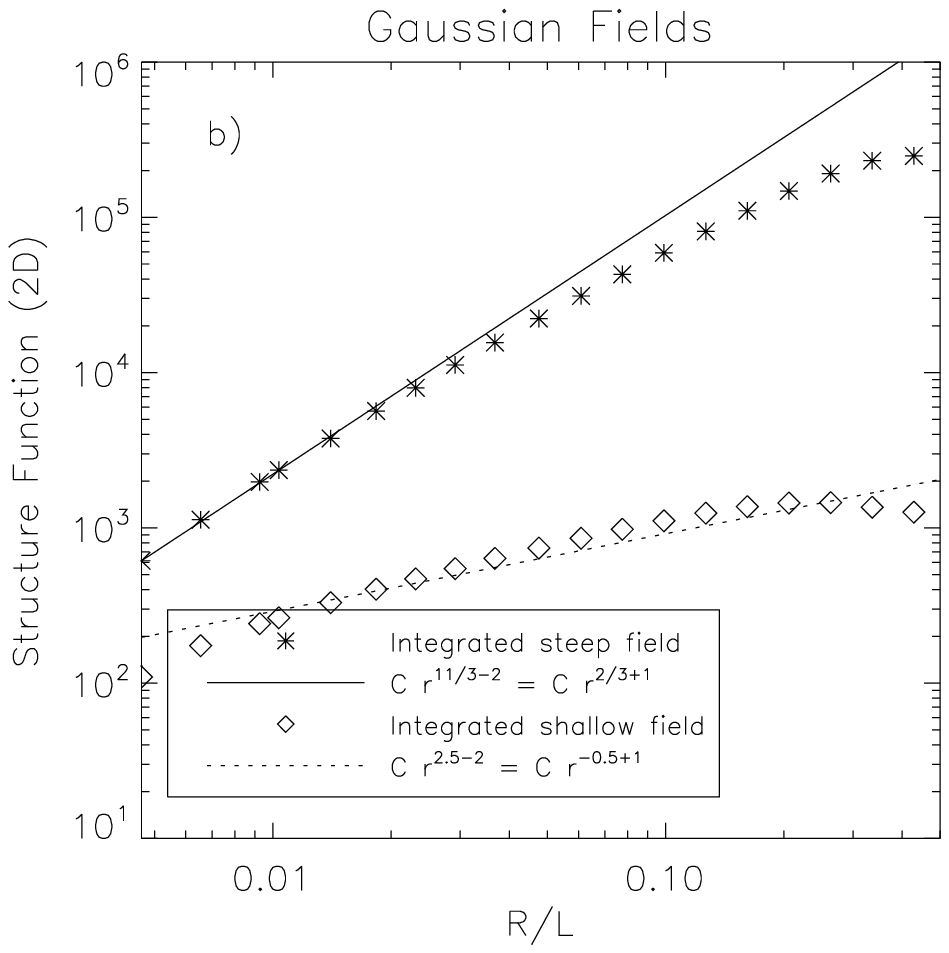}
\figcaption{Two point statistics for Gaussian fields integrated along the {\it z} direction. In panel ({\it a}) we show the 2D power spectra, panel ({\it b}) correspond to the second order structure functions. The {\it stars} correspond to a steep spectrum with a prescribed 3D power spectrum index of $\gamma_{3D}=-11/3$, the {\it diamonds} correspond to a shallow spectrum with a prescribed 3D power spectrum index of $\gamma_{3D}=-2.5$. In {\it solid} and {\it dotted} lines respectively, we plotted for reference the expectations for both long-wave or short-wave dominated cases. The vertical scales in panel ({\it b}) have been modified arbitrarily for visual purposes. \label{fig:2Dgauss}}
\end{figure}

We computed spectra and structure functions for the
Gaussian cubes used before, integrated along the $z$ direction. 
And presented them in Figure \ref{fig:2Dgauss} along with the the
expected behavior from eqs.(\ref{eq:mu}) and (\ref{eq:P2D_gamma3D}).
For the structure functions we show only the asymptotic behavior
for small separations ($R \ll z_{tot}$) as the $R \gg z_{tot}$ scales
are unavailable (the maximum scale not affected by wrap-around periodicity
is $L/2=z_{tot}/2$).

If the LOS size of the object under study is much smaller than its size
in the plane of the sky (so $R\gg z_{tot}$ is possible) we would see
the underlying 3D spectral index of the structure functions for large
separations (i.e. no projection effect). 
With enough resolution we could use the 2D spectral index of the column
density map for $R\ll z_{tot}$ to infer the  underlying 3D index of the
density.
However, for the  resolution used here ($216^3$ pixels), the projected
structure function for small lags is already in the transition between
the two asymptotics in eq.[\ref{eq:mu}].
Thus the measured index of the projected map is always shallower
(smaller) than the actual $\mu\approx -\gamma_{3D} -2$.
Another subtle, yet interesting point, is that
power spectrum applied to map of integrated velocity field, such as
\begin{equation}
V_z(\mathbf{X}) = \int v_z(\mathbf{x}){\rm d}z
\label{eq:V_z}
\end{equation}
will recover
only the incompressible (solenoidal) component of the field.
This could be potentially used to study the role of compressibility
in turbulence statistics, by combining velocity centroids and VCA (LE03).

\section{Modified velocity centroids revisited.}

Velocity centroids have been widely used to relate their statistics
with velocity, their conventional form is \citep{M58}
\begin{equation}
C(\mathbf{X})=\frac{\int v_{z}\ \rho_{s}(\mathbf{X},v_{z})\ {\rm d}v_{z}}{\int\ \rho_{s}(\mathbf{X},v_{z})\ {\rm d}v_{z}}.
\label{eq:C}
\end{equation}
We will refer to this definition as ``normalized'' centroids.
The denominator in equation (\ref{eq:C}) introduces an extra algebraic
complication for a direct analytical treatment of the two-point statistics.
For the sake of simplicity we start considering ``unnormalized''
velocity centroids:
\begin{equation}
S(\mathbf{X})=\int \alpha \ v_{z}\ \rho_{s}(\mathbf{X},v_{z})\ {\rm d}v_{z}
\label{eq:S}
\end{equation}
(notice that they have units of density times velocity as opposed to velocity alone).

Similarly to the expression in equation (\ref{eq:I}), with emissivity
proportional to the first power of the density, and no self absorption,
the structure function of the unnormalized velocity centroids is:
\begin{equation}
\left\langle \left[S(\mathbf{X_{1}})-S(\mathbf{X_{2}})\right]^{2}\right\rangle = \left\langle \left(\alpha\int v_{z}(\mathbf{x_{1}})\ \rho(\mathbf{x_{1}})\ {\rm d}z-\alpha\int v_{z}(\mathbf{x_{2}})\ \rho(\mathbf{x_{2}})\ {\rm d}z\right)^{2}\right\rangle .
\label{eq:S1S2}
\end{equation}
Presenting the density and velocity as a sum
of a mean value and a fluctuating part: $\rho=\rho_{0}+\tilde{\rho}$,
$v_{z}=v_{0}+\tilde{v}_{z}$. Where $\rho_{0}=\langle\rho\rangle$,
$v_{0}=\langle v_{z}\rangle$, and the fluctuations
satisfy $\langle\tilde{\rho}\rangle=0$,
$\langle\tilde{v}_{z}\rangle=0$. Analogous to
equation (\ref{eq:I1I2dr1dr2}), the structure function of the unnormalized
centroids can be written as
\begin{equation}
\left\langle \left[S(\mathbf{X_{1}})-S(\mathbf{X_{2}})\right]^{2}\right\rangle =\alpha^{2}\iint{\rm d}z_{1}{\rm d}z_{2}\left[D(\mathbf{r})-\left.D(\mathbf{r})\right|_{\mathbf{X_{1}}=\mathbf{X_{2}}}\right],
\label{eq:S1S22}
\end{equation}
with
\begin{equation}
D(\mathbf{r})=\left\langle \left[v_{z}(\mathbf{x_{1}})\rho(\mathbf{x_{1}})-v_{z}(\mathbf{x_{2}})\rho(\mathbf{x_{2}})\right]^{2}\right\rangle .
\label{eq:D}
\end{equation}
And $D(\mathbf{r})$ can be approximated as:
\begin{equation}
D(\mathbf{r})\approx\left\langle v_{z}^{2}\right\rangle d_{\rho}(\mathbf{r})+\left\langle \rho^{2}\right\rangle d_{v_{z}}(\mathbf{r})-\frac{1}{2}d_{\rho}(\mathbf{r})d_{v_{z}}(\mathbf{r})+c(\mathbf{r}),
\label{eq:D2}
\end{equation}
 which includes the underlying 3D structure function of the LOS velocity
\begin{equation}
d_{v_{z}}(\mathbf{r})=\left\langle \left[v_{z}(\mathbf{x_{1}})-v_{z}(\mathbf{x_{2}})\right]^{2}\right\rangle,
\label{eq:dv}
\end{equation}
and cross-correlations of velocity and density fluctuations
\footnote{Note that equation (\ref{eq:c}) is somewhat different from LE03,
where there was a misprint; which has no effect on the results presented.
}:
\begin{equation}
c(\mathbf{r})=2\left\langle \tilde{v}_{z}(\mathbf{x_{1}})\tilde{\rho}(\mathbf{x_{2}})\right\rangle ^{2}-4\rho_{0}\left\langle \tilde{\rho}(\mathbf{x_{1}})\tilde{v}_{z}(\mathbf{x_{1}})\tilde{v}_{z}(\mathbf{x_{2}})\right\rangle .
\label{eq:c}
\end{equation}
Because the derivation of equations (\ref{eq:S1S22}), (\ref{eq:D2}),
and (\ref{eq:c}) involves some tedious algebra, we place it in
Appendix D.
With all of this, the structure function of unnormalized velocity
centroids can be decomposed as
\begin{equation}
\left\langle \left[S(\mathbf{X_{1}})-S(\mathbf{X_{2}})\right]^{2}\right\rangle =I1(\mathbf{R})+I2(\mathbf{R})+I3(\mathbf{R})+I4(\mathbf{R}),
\label{eq:S1S2Dec}
\end{equation}
where
\begin{mathletters}
\begin{eqnarray}
I1(\mathbf{R}) & = & \alpha^{2}\left\langle v_{z}^{2}\right\rangle \iint{\rm d}z_{1}{\rm d}z_{2}\left[d_{\rho}(\mathbf{r})-\left.d_{\rho}(\mathbf{r})\right|_{\mathbf{X}_{1}=\mathbf{X}_{2}}\right],
\label{eq:I1}\\
I2(\mathbf{R}) & = & \alpha^{2}\left\langle \rho^{2}\right\rangle \iint{\rm d}z_{1}{\rm d}z_{2}\left[d_{v_{z}}(\mathbf{r})-\left.d_{v_{z}}(\mathbf{r})\right|_{\mathbf{X}_{1}=\mathbf{X}_{2}}\right],
\label{eq:I2}\\
I3(\mathbf{R}) & = &-\frac{1}{2} \alpha^{2}\iint{\rm d}z_{1}{\rm d}z_{2}\left[d_{\rho}(\mathbf{r})d_{v_{z}}(\mathbf{r})-\left.d_{\rho}(\mathbf{r})\right|_{\mathbf{X_{1}}=\mathbf{X_{2}}}\left.d_{v_{z}}(\mathbf{r})\right|_{\mathbf{X_{1}}=\mathbf{X_{2}}}\right],
\label{eq:I3}\\
I4(\mathbf{R}) & = & \alpha^{2}\iint{\rm d}z_{1}{\rm d}z_{2}\left[c(\mathbf{r})-\left.c(\mathbf{r})\right|_{\mathbf{X}_{1}=\mathbf{X}_{2}}\right].
\label{eq:I4}
\end{eqnarray}
\end{mathletters}
With this new decomposition is more evident the definition of the
structure function of ``modified'' velocity centroids (MVCs),
which is
\begin{eqnarray}
M(\mathbf{R})  & = & \left\langle \left[S(\mathbf{X_{1}})-S(\mathbf{X_{2}})\right]^{2}-\left\langle v_{z}^{2}\right\rangle \left[I(\mathbf{X_{1}})-I(\mathbf{X_{2}})\right]^{2}\right\rangle \nonumber \\
 & = & \left\langle \left[S(\mathbf{X_{1}})-S(\mathbf{X_{2}})\right]^{2}\right\rangle -I1(\mathbf{R})\nonumber \\
& = & I2(\mathbf{R})+I3(\mathbf{R})+I4(\mathbf{R}).
\label{eq:M}
\end{eqnarray}
The velocity dispersion $\langle v_{z}^{2}\rangle$ can be obtained directly from
observations using the second moment of the spectral lines
\begin{equation}
\left\langle v_{z}^{2}\right\rangle \equiv\frac{\int v_{z}^{2}\ \rho_{s}(\mathbf{X},v_{z})\ {\rm d}v_{z}}{\int\ \rho_{s}(\mathbf{X},v_{z})\ {\rm d}v_{z}}.
\label{eq:vsq}
\end{equation}
Thus also $I1(\mathbf{R})$, which can be related to the structure
function of column density as 
$I1(\mathbf{R})=\langle v_{z}^{2}\rangle\langle[I(\mathbf{X_{1}})-I(\mathbf{X_{2}})]^{2}\rangle$.

Similarly, the power spectrum of centroids can be decomposed as (details
in Appendix F):
\begin{eqnarray}
P_{2D,S}(\mathbf{K}) & =
& \langle \rho^2\rangle\langle v_z^2\rangle(\alpha z_{tot})^{2}\delta(\mathbf{K})+v_{0}^{2}P_{2D,I}(\mathbf{K})+\alpha^{2}\rho_{0}^{2}P_{2D,V_{z}}(\mathbf{K})\nonumber\\
& & +\mathcal{F}\left\{ B3(\mathbf{R})\right\} +\mathcal{F}\left\{ B4(\mathbf{R})\right\}.
\label{eq:PS_S_main}
\end{eqnarray}

The term $\langle \rho^2\rangle\langle v_z^2\rangle(\alpha z_{tot})^{2}\delta(\mathbf{R})$ has no effect
in the slope of the power spectrum because it only has power at
$\mathbf{K}=\mathbf{0}$. $P_{2D,I}(\mathbf{K})$, and
$P_{2D,V_z}(\mathbf{K})$ are the spectra
of column density, and integrated velocity respectively. They can be
used to obtain the 3D spectral index of density (or velocity) as shown
in Appendix C.
$\mathcal{F}\left\{ B3(\mathbf{R})\right\}$ is the Fourier transform of
$B3(\mathbf{R})$, a cross-term term analogous to $I3(\mathbf{R})$, but
in terms of correlation functions.
Similarly, $\mathcal{F}\left\{ B4(\mathbf{R})\right\}$ include the same
density-velocity cross-correlations as $I4(\mathbf{R})$.

The power spectrum of MVCs can be obtained by subtracting
$\langle v_z^2\rangle P_{2D,I}(\mathbf{K})$ from the spectrum of
centroids.
We derive in Appendix G a criterion for MVCs to trace
the statistics of velocity better than unnormalized centroids.
It was found that, with very little dependence on the spectral index,
MVCs are advantageous compared to unnormalized centroids at small lags.
This result is general and we tested it using analytical expressions
for the structure functions.
For simplicity we considered only the two physically motivated cases,
steep density with steep velocity, and shallow density with steep
velocity. The latter was not included explicitly in Appendix G because
we obtain almost identical results in both cases.

If $v_0=0$ and $\mathcal{F}\left\{ B3(\mathbf{R})\right\}+
\mathcal{F}\left\{ B4(\mathbf{R})\right\}$ can
be neglected the spectrum of unnormalized centroids will trace the
solenoidal component of the underlying velocity spectrum ($F_{NN}[K,0]$,
see Appendix C).
But if the turbulent velocity field is mostly solenoidal, as supported
by numerical simulations \citep*{MGOR96,PWP98,CL03}, the power-spectrum
is uniquely defined assuming isotropy
($E(k)=\int P_{ND}(\mathbf{k})d\mathbf{k}\approx4\pi k^{2}F_{NN}[k]$).
In the same way if $I1(\mathbf{R})$ can be eliminated (either for being
small compared to the structure function of centroids or by subtraction
-MVCs-) and if $I2(\mathbf{R})\gg I3(\mathbf{R})+I4(\mathbf{R})$,
the structure function of the remaining map will trace
the structure function of a map of integrated turbulent velocity.
And we can in principle recover the underlying 3D velocity statistics
(see Appendix B).
With this background (and disregarding the cross-terms $I3[\mathbf{R}]$,
$I4[\mathbf{R}]$)
we arrived in LE03 to a criterion for safe use of (unnormalized)
velocity centroids: {\it if} 
\begin{equation}
\left\langle\left[S(\mathbf{X_{1}})-S(\mathbf{X_{2}})\right]^{2}\right\rangle \gg
\left\langle v_{z}^{2}\right\rangle \left\langle\left[I(\mathbf{X_{1}})-I(\mathbf{X_{2}})\right]^{2}\right\rangle,
\label{eq:crit_full}
\end{equation}
{\it then the structure function of velocity centroids will mostly
trace the turbulent velocity statistics, otherwise the density
fluctuations are important and will be reflected in the centroid measures}.
When the structure function of velocity centroids is shallower or at least 
not much steeper than that of the column density, which can be verified by
the power spectrum, or computing the structure function directly in 3D for
a few of values for the lag.
Then the criterion proposed in LE03 can be simplified to use only the
variances of the two maps (and the velocity dispersion):
\begin{equation}
\left\langle \tilde{S}^{2}\right\rangle \gg \left\langle v_z^{2}\right\rangle \left\langle\tilde{I}^{2}\right\rangle.
\label{eq:crit_simp}
\end{equation}
If any of these two criteria is violated, one could in principle subtract
the contribution of density and the MVCs would trace velocity structure
function, provided that we could neglect the cross-terms.

The contribution of velocity-density cross-correlations
($c[\mathbf{r}]$) have been studied earlier. For
VCA it has been shown to be marginal \citep{LPVP01,ELPC03}.
However, a more detailed discussion of their effect in the context of
MVCs is necessary, and is provided below.

$I3(\mathbf{R})$ and $\mathcal{F}\left\{ B3(\mathbf{R})\right\}$ 
are in some sense ``cross-terms'', $I4(\mathbf{R})$ and
$\mathcal{F}\left\{ B4(\mathbf{R})\right\}$) are related
to correlations between density and velocity.
We expect both pairs to grow as we increase the ``interrelation'' between
density and velocity.
We will refer to $I3[\mathbf{R}]$ (or 
$\mathcal{F}\left\{ B3(\mathbf{R})\right\}$) simply as ``cross-term'',
and to  $I4[\mathbf{R}]$ (or $\mathcal{F}\left\{ B4(\mathbf{R})\right\}$)
as ``cross-correlations'' of density-velocity.
The latter should be zero for uncorrelated data.
At the same time the cross-term can be studied analytically for
power-law statistics as presented in Appendix E, and will not be zero,
even in the case of uncorrelated velocity and density fields.
Before computing them directly, in order to get a feeling of how
important the cross-term could become one can consider structure
functions.
First of all, note that
$\langle[S(\mathbf{X_{1}})-S(\mathbf{X_{2}})]^{2}\rangle$
is positive defined, and so are $I1(\mathbf{R})$ and $I2(\mathbf{R})$.
The remaining terms can be negative, in which case they must
be smaller than the sum of $I1(\mathbf{R})$ and $I2(\mathbf{R})$.
Let us focus for the moment on the contribution of $I3(\mathbf{R})$ and
disregard cross-correlations between density and velocity.
Its magnitude is maximal at large scales,
and so are $I1(\mathbf{R})$ and $I2(\mathbf{R})$.
At such scales $\vert I3(\mathbf{R})\vert$ is on the order
of $I1(\mathbf{R})$ and $I2(\mathbf{R})$.
However, in $I2(\mathbf{R})$ the structure function of velocity
is weighted by $\langle\rho^{2}\rangle=\rho_0^2+\langle\tilde{\rho}^2\rangle$
instead of only $\langle\tilde{\rho}^{2}\rangle$, enhancing the velocity
statistics compared to the cross-term.
The importance of the cross-term at the small scales (in which we are
most interested) will depend on details such as how steep the underlying
structure functions are (see Appendix E),
and the zero levels of
density and velocity \citep[see][]{OELS05}.
This is easy to understand for a particular case of power-law
statistics of the form $d_{\rho}(r)\propto r^{n}$, and 
$d_{v_{z}}(r)\propto r^{m}$, with ($m,~n>0$, i.e. both fields steep). 
Here the cross-term scales as $\propto r^{m+n}$,
steeper than both velocity and density.
If at large scales 
$I1(\mathbf{R})$ and $I2(\mathbf{R})$ are on the order
of  $\vert I3(\mathbf{R})\vert$, provided that the latter
falls more rapidly toward small scales, its contribution
will be smaller than both velocity and density structure functions,
at those scales.
But if the density or the LOS velocity (or both), have a shallow spectrum,
the cross-term can be larger than
$I1(\mathbf{R})$ or $I2(\mathbf{R})$, and can affect significantly the
statistics of centroids.
Measured spectral indices of density in
the literature, range from $\gamma_{3D}\sim-2.5$ to $\gamma_{3D}\sim-4.0$,
which include both shallow and steep. This is true for observations
in different environments in the ISM (for instance, \citealt{DDG00};
\citealt*{BSO01,SL01,OML02}),
as well as for numerical simulations (see \citealt{CL02,BML04,BLC05}).
The velocity spectral index is less known from observations, but has
been measured to be only in the steep regime (for example using VCA,
e.g. in \citealt{SL01}), also in agreement with simulations. From
the theoretical standpoint, at small scales when self-gravity is important
we might expect clumping that result on enhanced small scale structure
(yielding a shallow spectrum). On the other hand there are no clear
physical grounds to our knowledge that will produce a small scale
dominated (shallow) velocity field. However, even in the simple case
of steep density
and steep velocity spectra it is not clear beforehand how important
density-velocity cross-correlations ($I4[\mathbf{R}]$)
could be. Later, we will analyze the contribution of 
the cross-terms and density-velocity cross-correlations in more detail,
including spectra.

\section{Testing velocity centroids numerically}

In LE03 we performed some preliminary tests of the modified velocity
centroids using numerical simulations and compared the power-spectrum
with that of velocity field, normalized (equation [\ref{eq:C}]), and
unnormalized centroids (equation [\ref{eq:S}]). In this section we
provide a more detailed test to investigate under what conditions
velocity centroids can be used to recover the velocity statistics.

\subsection{The data}

We took compressible MHD data-cubes from the numerical simulations
of \citet{CL03}. This data-cubes correspond to fully-developed (driven)
turbulence. The turbulence is driven in Fourier Space (solenoidally)
at wave-numbers $2\leq (k_{driving} L/[2\pi])<3.4$. The data-cubes have
a resolution of $216^{3}$ pixels. We use four sets of simulations, the
parameters for each run are summarized in Table \ref{tb:param}.
\begin{deluxetable}{cccccccc}
\tabletypesize{\scriptsize} 
\tablecaption{Parameters of the four runs used. \label{tb:param}}
\tablewidth{0pt}
\tablehead{
\colhead{Model} & \colhead{$B_0$} 
& \colhead{$\langle P_{gas} \rangle$}& \colhead{$\beta$}
&\colhead{rms $v$} & \colhead{$\mathcal{M}_s$} & \colhead{$\mathcal{M}_A$} &
\colhead{$\langle \tilde{S}^2 \rangle /(\langle v_z^2 \rangle \langle \tilde{I}^2 \rangle)$}
          }
\startdata
A & $1.0$ & $2.0$  &$4$   & $\lesssim 0.7$ & $\sim 0.5$ & $\sim 0.7$ & $93.9$ \\
B & $1.0$ & $0.1$  &$0.2$ & $\lesssim 0.7$ & $\sim 2.5$ & $\sim 0.7$ & $4.6$  \\
C & $0.1$ & $0.1$  &$0.2$ & $\lesssim 0.7$ & $\sim 2.5$ & $\sim 8  $ & $4.2$  \\
D & $1.0$ & $0.01$ &$0.02$& $\lesssim 0.7$ & $\sim 7$   & $\sim 0.7$ & $2.2$ 
\enddata
\end{deluxetable}
The models include various values of the plasma $\beta$ (ratio of
gas to magnetic pressures), sonic Mach numbers $\mathcal{M}_s$,
and Alfv\'{e}n Mach number $\mathcal{M}_A$.
All of these parameters can be found in the ISM under different
situations.
For more details about the simulations we refer the reader to
\citet{CL03}.
The outcome of the simulations are density and velocity data-cubes that
we use to compute the centroids.
We will refer to this data-sets as ``original'', and the existing
correlations between density and velocity (consistent with MHD
evolution) are left intact.
The numerical simulations have a limited inertial range.
We do not have power-law statistics (i.e. inertial range) at the
largest scales (smallest wave-numbers) due to the driving of the
turbulence. And neither we have power-law statistics at the smallest
scales (largest wave-numbers), because of numerical dissipation.
Thus, it is very difficult to estimate spectral indices because the
measured log-log-slope is quite sensitive to the range in wave-numbers
(or lags) used. This poses a
problem of obtaining quantitative results.
For that reason, we created  another data-set by modifying the original
fields to have strict power-law spectra, following the procedure
in \citet{LPVP01}. The procedure consists in modifying the amplitude
of the Fourier components of the data so they follow a power-law,
while keeping the phases intact. This way we preserve most of the spatial
information. By keeping the phases we also minimize the effect of the
modification to the density-velocity correlations. In addition, these new
fields have the same mean value than the original data-sets, and the
magnitude of their power-spectra (vertical offset) was fixed to match the 
original variances as well. We will refer to this data-sets as ``reformed''.

\subsection{Results}

Column density and centroids of velocity are two-dimensional
maps, and therefore it is not computationally restrictive to obtain
their correlation or structure function directly in real space.
Power-spectrum is often computationally cheaper because FFT can
be used.
However, inherent difficulties of applying Fourier
analysis to real data, for instance the lack of periodic boundary
conditions and instrumentational response make power spectrum often
unreliable. To alleviate this problem more elaborate techniques
like wavelet transforms have been proposed \citep{ZS99}.
Moreover, if the observed maps are not naturally arranged in a
Cartesian grid one would need to smear the data onto that kind of
grid to use FFT. Which is not necessary for structure or correlation
functions if computed directly averaging in configuration space.
In despite of this, because the simulations we used are in a Cartesian
grid and indeed have periodic boundary conditions, we can compute spectra,
correlation, and structure functions using FFT (see \S2) with as good
accuracy than doing the average in real space.
The relation between the structure function, correlation function,
and power spectrum of unnormalized centroids can be found in Appendix
F.

\subsubsection{3D statistics}

\begin{figure*}
\epsscale{1.2}
\plotone{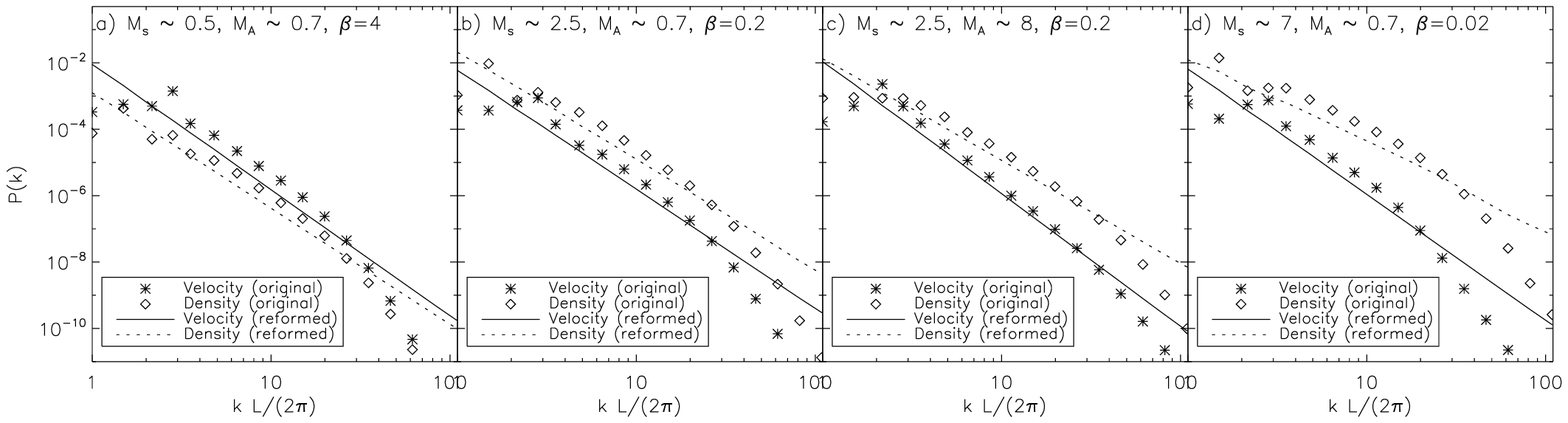}
\plotone{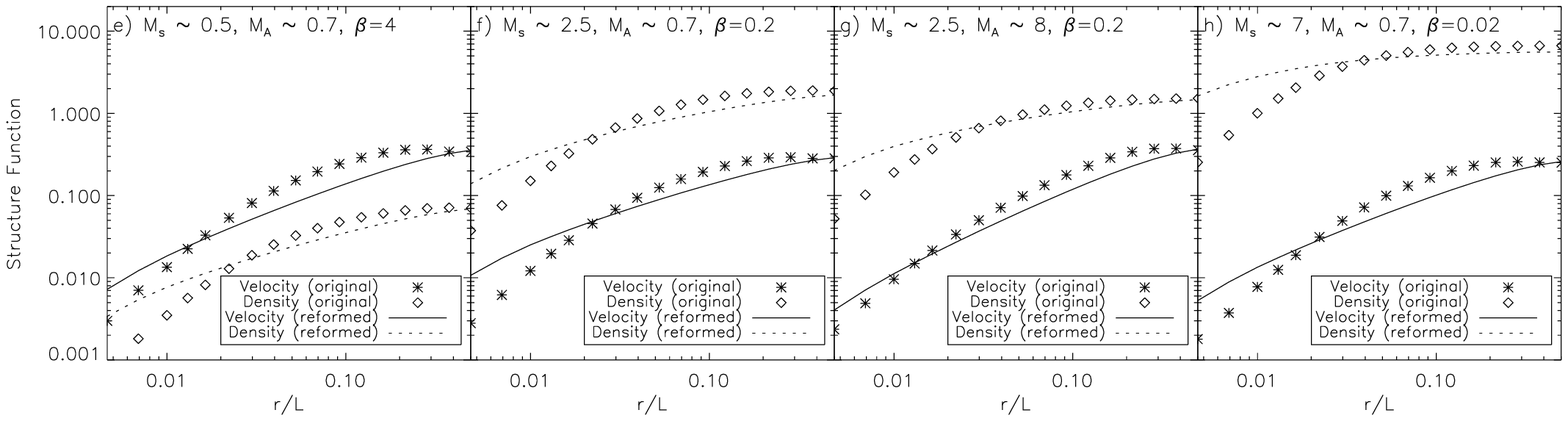}
\figcaption{Underlying 3D statistics of the MHD simulations. At the top the power spectra, and on the bottom the corresponding structure functions. The four runs are ordered in ascendant Mach number from left to right. \label{fig:3D_all}}
\end{figure*}

Before we study the 2D maps and try to extract from those the underlying
3D statistics we will start by computing the three-dimensional
statistics. This is shown in Figure \ref{fig:3D_all}.
It is noticeable that only for the case where the turbulence is subsonic
($\mathcal{M}_s\sim 0.5$) the level of the
velocity fluctuations is larger than that of the density.
In the figure is also evident the limited inertial range in the original
simulations (i.e. not perfect power-law statistics).
Spectral indices (log-log slopes), both for power-spectra and structure
functions from Figure \ref{fig:3D_all} are given in Table
\ref{tb:3Dindices}.
\begin{deluxetable}{cccccc}
\tabletypesize{\scriptsize}
\tablecaption{Three-dimensional spectral indices. The values in parentheses correspond to the {\it original} simulations, and in bold face to the {\it reformed} data-sets.
\label{tb:3Dindices}}
\tablecolumns{6}
\tablewidth{0pt}
\tablehead{
       \colhead{Model} & \multicolumn{2}{c}{Density} & \colhead{} &
\multicolumn{2}{c}{LOS Velocity} \\
\cline{2-3} \cline{5-6}
\colhead{} & \colhead{$- \gamma_{3D}$} & \colhead{$m$} & \colhead{} &
\colhead{$- \gamma_{3D}$} & \colhead{$m$} }
\startdata
A & $(3.5)~\mathbf{3.5}$ & $(0.4)~\mathbf{0.6}$ && $(3.8)~\mathbf{3.8}$ & $(0.5)~\mathbf{0.8}$ \\
B & $(3.3)~\mathbf{3.3}$ & $(0.3)~\mathbf{0.4}$ && $(3.6)~\mathbf{3.6}$ & $(0.5)~\mathbf{0.6}$ \\
C & $(3.1)~\mathbf{3.1}$ & $(0.1)~\mathbf{0.3}$ && $(4.0)~\mathbf{4.0}$ & $(0.8)~\mathbf{0.9}$ \\
D & $(2.6)~\mathbf{2.6}$ &      \nodata$^*$       && $(3.8)~\mathbf{3.8}$ & $(0.6)~\mathbf{0.8}$ 
\enddata
\tablenotetext{*}{The measured power-spectrum index in this case corresponds to a {\it shallow} spectrum. Thus, the correlation function is expected to follow a power-law, not the structure function.}
\end{deluxetable}
For the original data-sets the indices for power-spectra were obtained
in a range of wave-numbers $kL/(2\pi)\sim[5-15]$
(between the scale of injection and the scales at which dissipation
is dominant). The structure functions spectral indices were calculated with
the corresponding values for spatial separations, $r/L\sim [1/15-1/5]$.
The reformed data-sets were constructed using the power-spectra indices
estimated for the original MHD simulations.
We can see in Fig. \ref{fig:3D_all} the idealized power-law spectra of the
reformed data-sets. At the same time, although structure functions do not
have such perfect power-law behavior (see discussion at the end of \S\S2.1),
they show improvement compared to the original simulations. 
The range of scales used to measure the spectral indices for the reformed
data-sets is  $kL/(2\pi)\sim[3.5-100]$ or $r/L\sim [1/100-1/3.5]$, much wider
than for the original sets.
Notice that power-spectra for the density becomes shallower with the
increase of the Mach number.
At the same time the spectral index of velocity is always steep.

\subsubsection{Statistics of projected quantities (2D)}

\begin{figure}
\epsscale{1.3}
\plotone{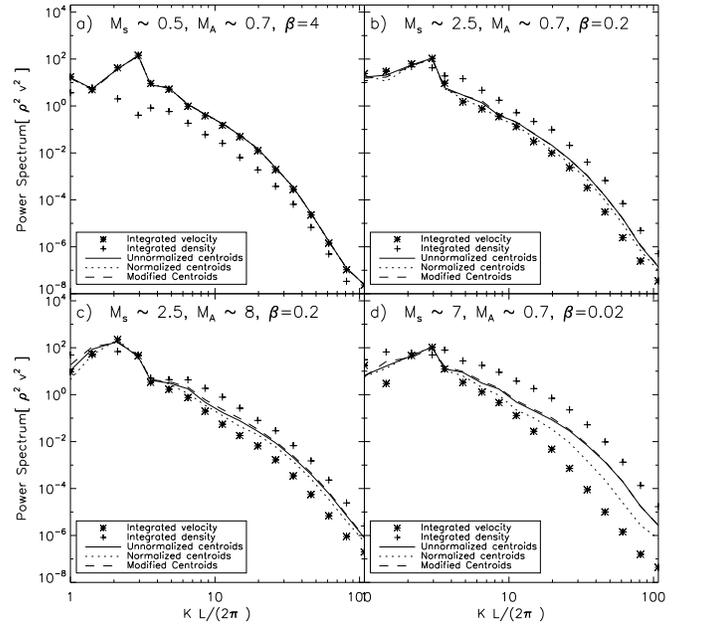}
\figcaption{Power spectra of the integrated density ({\it crosses}), integrated velocity ({\it stars}), unnormalized, normalized, and modified centroids ({\it solid}, {\it dotted}, and {\it dashed } lines respectively) for the {\it original} set of simulations.  We multiplied the spectrum of velocity fluctuations by $\rho_0^2$, and that of normalized centroids by $\langle I^2 \rangle$. We show the spectrum of integrated density for reference only, to be in the same units as the other quantities in the figure it should be multiplied by $v_0^2$, but since $v_0\approx 0$ the scaling is omitted here.
\label{fig:ps_all}}
\end{figure}

\begin{figure}
\epsscale{1.3}
\plotone{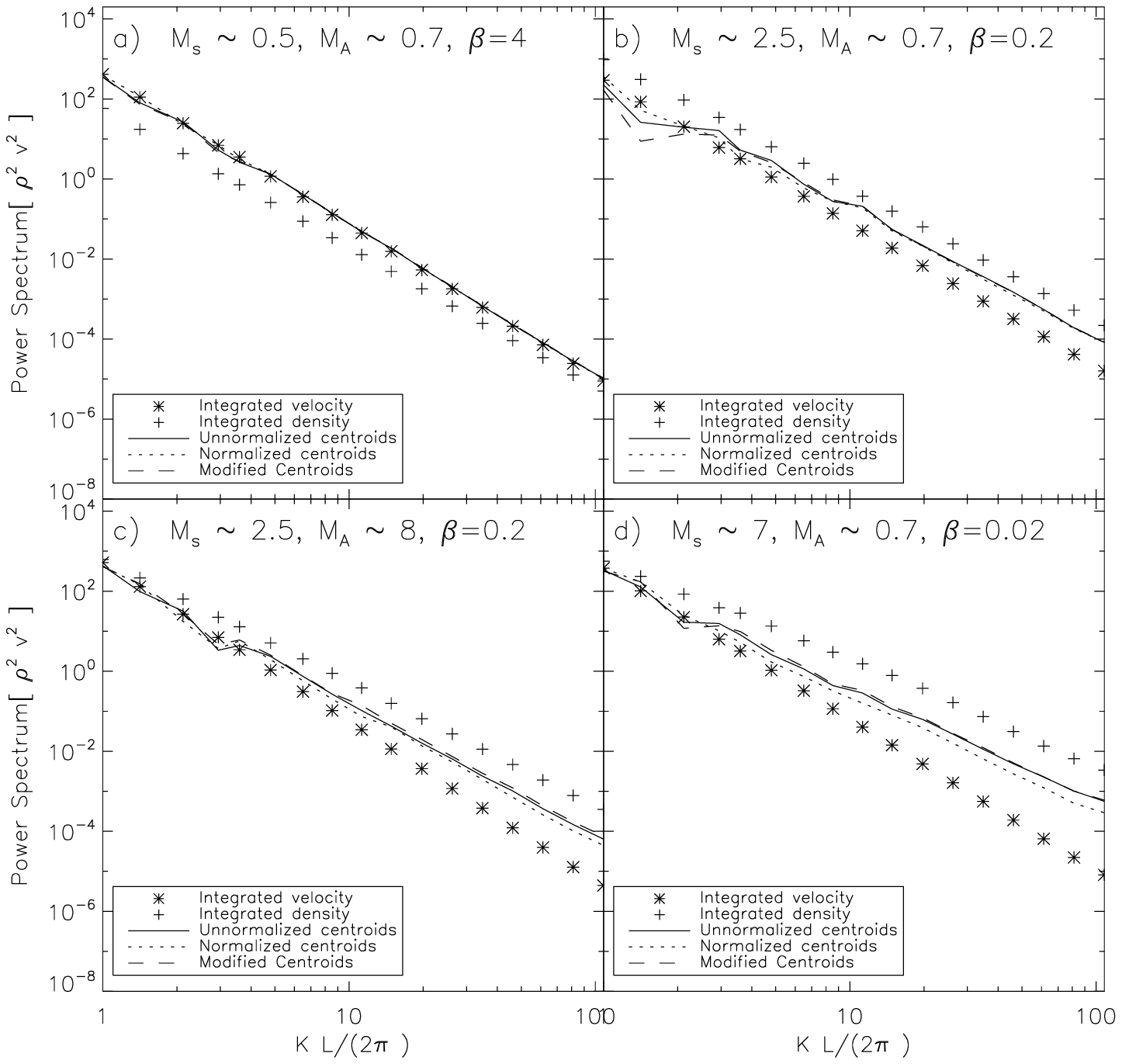}
\figcaption{Same as Fig. \ref{fig:ps_all}, for the {\it reformed} data-sets.
\label{fig:ps_all_mod}}
\end{figure}

\begin{figure}
\epsscale{1.3}
\plotone{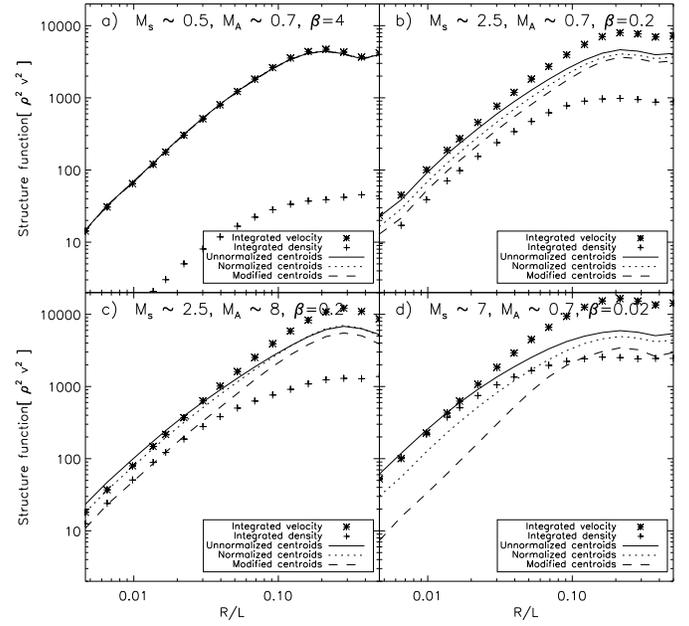}
\figcaption{Structure functions of integrated density ({\it crosses}), integrated velocity ({\it stars}), unnormalized, normalized, and modified centroids ({\it solid}, {\it dotted}, and {\it dashed } lines respectively) for the {\it original} set of simulations. . To allow for a direct comparison we scale the quantities to be all in units of $[ v^2 \rho^2 ]$ (i. e. we multiplied the structure function of the integrated velocity by $\langle \rho^2 \rangle$, the integrated density by $\langle v^2 \rangle$, and the normalized centroids by $\langle I^2 \rangle$). \label{fig:sf_all}}
\end{figure}

\begin{figure}
\epsscale{1.3}
\plotone{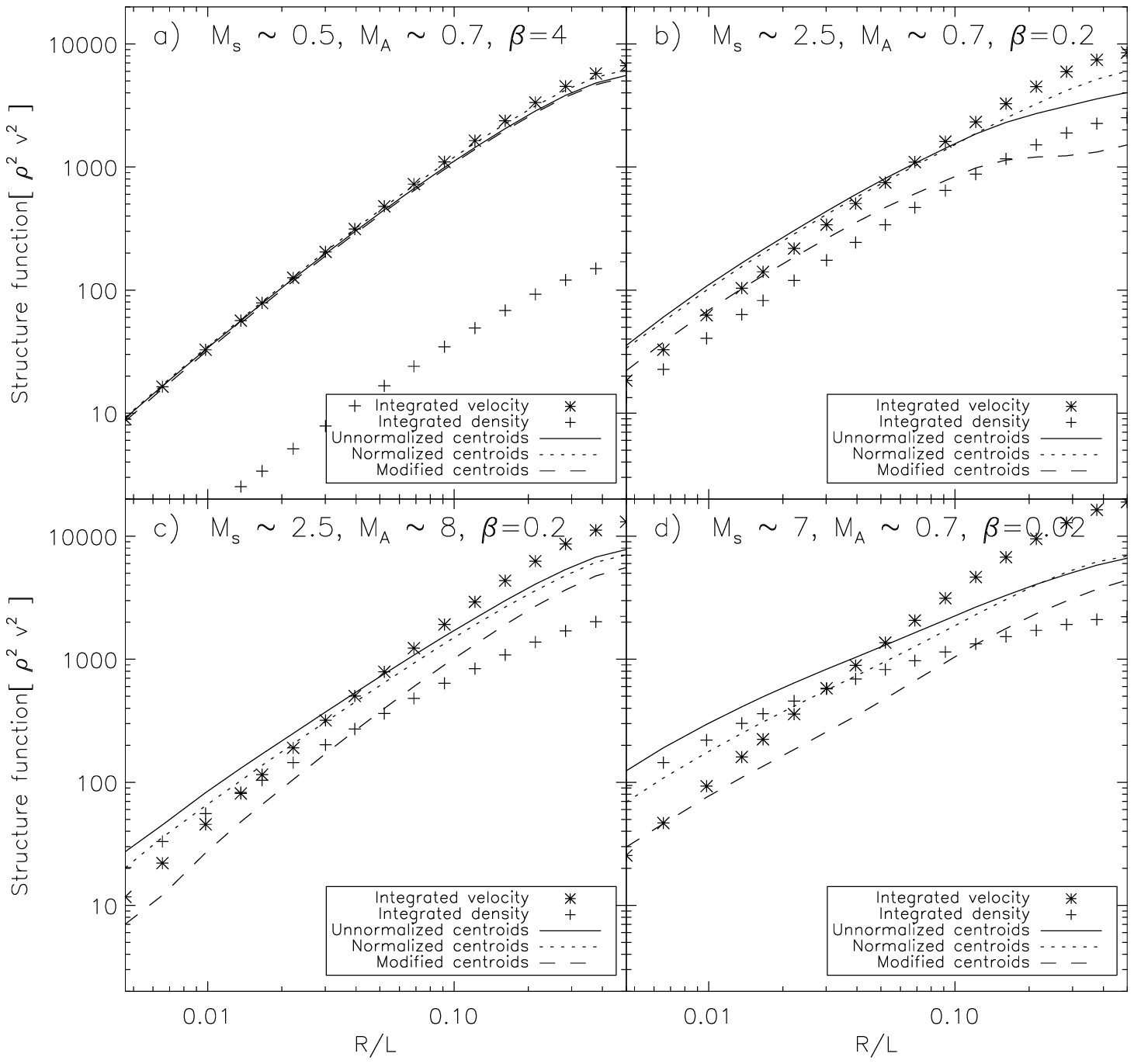}
\figcaption{Same as Fig \ref{fig:sf_all}, for the {\it reformed} data-sets.\label{fig:sf_all_mod}}
\end{figure}

A natural way to study how the velocity centroids trace the statistics
of velocity is to compare their two point statistics to those of an
integrated velocity map (equation \ref{eq:V_z}).
This map can be used to obtain the velocity spectral index in the
same way column density can be used to obtain that of density.
Therefore it is a direct measure
of the underlying velocity statistics (see Appendices
B and C). 
However, it is not observable,
while velocity centroids are.
We computed power-spectra and structure
functions of 2D maps of the various centroids (normalized, unnormalized,
and ``modified''), integrated velocity, and integrated density.
The results for power-spectra are shown in Figures \ref{fig:ps_all}, and
 \ref{fig:ps_all_mod}; for the original, and modified data-sets
respectively.
Similarly, Figures \ref{fig:sf_all}, and \ref{fig:sf_all_mod} show the
results for structure functions.
The most noticeable difference in the Figures is of course the larger
inertial range of the reformed data-set.
We also present in Table \ref{tb:2Dindices} a summary of spectral
indices (log-log slope)  measured over
$5 \lesssim K L/(2\pi) \lesssim 25$ (or $1/25 \lesssim R/L \lesssim 1/5$)
for the original data sets, and $kL/(2\pi)\sim [3.5-100]$ (or
$r/L\sim [1/100-1/3.5]$) for the reformed sets.
\begin{deluxetable*}{ccccccccccccccc}
\tabletypesize{\scriptsize}
\tablecaption{Spectral indices (2D) of quantities integrated along the LOS. The values in parentheses correspond to the {\it original} simulations, and in bold face to the {\it reformed} data-sets.
\label{tb:2Dindices}}
\tablecolumns{15}
\tablewidth{0pt}
\tablehead{
\colhead{} & \multicolumn{2}{c}{Integrated}  &
\colhead{} & \multicolumn{2}{c}{Integrated}  &
\colhead{} & \multicolumn{2}{c}{Unnormalized} & 
\colhead{} & \multicolumn{2}{c}{Normalized}  &
\colhead{} & \multicolumn{2}{c}{Modified}    \\
\colhead{Model} & \multicolumn{2}{c}{Density}     &
\colhead{} & \multicolumn{2}{c}{Velocity}    &
\colhead{} & \multicolumn{2}{c}{Centroids}   & 
\colhead{} & \multicolumn{2}{c}{Centroids}   &
\colhead{} & \multicolumn{2}{c}{Centroids}   \\
\cline{2-3} \cline{5-6} \cline{8-9} \cline{11-12} \cline{14-15}
\colhead{} & \colhead{$-\gamma_{3D}$} & \colhead{$\mu^*$} &
\colhead{} & \colhead{$-\gamma_{3D}$} & \colhead{$\mu^*$} &
\colhead{} & \colhead{$-\gamma_{3D}$} & \colhead{$\mu^*$} &
\colhead{} & \colhead{$-\gamma_{3D}$} & \colhead{$\mu^*$} &
\colhead{} & \colhead{$-\gamma_{3D}$} & \colhead{$\mu^*$} }
\startdata
A & $(4.0)~\mathbf{3.5}$ & $(0.5)~\mathbf{1.4}$ && $(3.6)~\mathbf{3.8}$ &$(0.9)~\mathbf{1.5}$ && $(3.6)~\mathbf{3.8}$ & $(0.8)~\mathbf{1.5}$ && $(3.5)~\mathbf{3.8}$ & $(0.8)~\mathbf{1.5}$ && $(3.6)~\mathbf{3.7}$ & $(0.8)~\mathbf{1.5}$ \\
B & $(3.8)~\mathbf{3.3}$ & $(0.4)~\mathbf{1.2}$ && $(3.9)~\mathbf{3.6}$ &$(1.0)~\mathbf{1.4}$ && $(3.5)~\mathbf{3.3}$ & $(0.8)~\mathbf{1.1}$ && $(3.5)~\mathbf{3.2}$ & $(0.9)~\mathbf{1.1}$ && $(3.8)~\mathbf{3.3}$ & $(1.0)~\mathbf{1.0}$ \\
C & $(3.3)~\mathbf{3.1}$ & $(0.6)~\mathbf{1.1}$ && $(4.5)~\mathbf{4.0}$ &$(1.3)~\mathbf{1.6}$ && $(3.8)~\mathbf{3.4}$ & $(1.1)~\mathbf{1.3}$ && $(3.9)~\mathbf{3.5}$ & $(1.2)~\mathbf{1.3}$ && $(3.7)~\mathbf{3.4}$ & $(1.2)~\mathbf{1.5}$ \\
D & $(2.8)~\mathbf{2.7}$ & $(0.2)~\mathbf{0.7}$ && $(4.7)~\mathbf{3.8}$ &$(0.8)~\mathbf{1.5}$ && $(3.4)~\mathbf{2.8}$ & $(0.5)~\mathbf{0.9}$ && $(3.8)~\mathbf{2.9}$ & $(0.7)~\mathbf{1.0}$ && $(3.5)~\mathbf{2.9}$ & $(0.8)~\mathbf{1.1}$ 
\enddata
\tablenotetext{*}{Since this index is not measured at scales corresponding to $R\ll z_{tot}$, but rather in the transition between the two asymptotic regimes in eq.(\ref{eq:mu}), this are only lower limits on the actual $\mu$ for small lags.}
\end{deluxetable*}

Comparing the spectral indices derived in 3D (Table \ref{tb:3Dindices})
with those of column density and integrated velocity in Table
\ref{tb:2Dindices}, one can notice a better correspondence for the
reformed data-sets. This is true for the
power-spectra index $\gamma_{3D}$; as well as for $m$, and $\mu$
for structure functions (related by equation[\ref{eq:mu}]).
Directly from Figures \ref{fig:ps_all}--\ref{fig:sf_all_mod} one can
see that only for the case of subsonic turbulence
(model A, $\mathcal{M}_{s}\sim 0.5$)
the spectrum of centroids clearly scales with that of integrated velocity.
In this case the power-spectra of all the variations of centroids recover
the spectral index of velocity, within $10\%$ error for the original
simulations, and $<3\%$ for the reformed data sets.
For the cases of mildly supersonic turbulence (models B, and C,
$\mathcal{M}_{s}\sim 2.5$) is not clear neither from the Figures,
nor from the measured slopes.
While for the strongly supersonic case (model D, $\mathcal{M}_{s}\sim 7$)
it is obvious that velocity centroids fail to recover the velocity scaling.

Due to the finite width effects discussed at the end of \S\S2.2, it is
more difficult to determine quantitatively the spectral index from centroid
maps using structure functions.
According to equation (\ref{eq:mu}), one should restrict to measure the
spectral index for lags either much smaller, or much larger than the
LOS extent of the object under study. The latter is not feasible with
our simulations because the maximum lag available, unaffected by 
wrap-around periodicity, is $L/2=z_{tot}/2$.
The other case ($R \ll z_{tot}$) is not strictly possible with the
resolution used here.
For the MHD data-cubes we have to avoid the smallest scales
because they are not within the inertial range (i.e. they are already
dominated by dissipation).
Actually, the lags used to measure $\mu$ in the original
data-sets are well in the transition between the two asymptotic
power-laws of equation (\ref{eq:mu}).
Thus, if we obtain the 3D spectral indices as $\gamma_{3D}\simeq -\mu -2$
(taking $\mu$ from Table \ref{tb:2Dindices}), we would only get upper
limits of the actual $\gamma_{3D}$.
Keeping in mind that our definition of the index includes the minus sign,
this means that the real index is going to be more negative than the
inferred $\gamma_{3D}$ from structure functions.
A lower limit on the spectral index can be
obtained as $\gamma_{3D}\simeq -\mu -3$, unfortunately this provides a
rather wide range of possible indices and it is not useful for practical
purposes (for instance distinguishing between two models of turbulence).
For the reformed data-sets the situation is better, because we can use the
smallest scales. This can be verified from the better agreement of
$\gamma_{3D}\simeq -\mu -2$ with $\gamma_{3D}$ measured directly from
power-spectra. However, $R \ll z_{tot}$ is still not strictly
fulfilled, and thus we get only lower limits of $\mu$.
We recognize this projection effect as an important drawback for
the structure functions compared  with spectra.
Power spectra do not suffer so strongly of finite width projection effects,
and could be used to obtain more accurately the spectral index of
the underlying 3D index field from integrated
quantities \citep[see also][]{OELS05}.
This is true for synthetic data,
but for real observations power spectrum is not as reliable
\citep[see][]{BSO01}.
At the same time, the problem of recovering quantitatively the
spectral index from structure functions might be overcome
for real observations with a larger inertial range.
With enough spatial resolution one can have easily over two
decades of inertial range in the plane of the
sky (below the size of a given object along the LOS direction).
In the spirit of keeping the description in terms of structure functions,
which might be better suited for some type of real observations.
And, using the limited range we have, it is still possible to do
a comparative analysis between the spectral index of structure functions
of centroids to that of integrated velocity or density.

From Figures  \ref{fig:ps_all}--\ref{fig:sf_all_mod} we see that
the relative importance of the density and velocity statistics
on the centroids changes with the Mach number of turbulence.
For the low $\mathcal{M}_s$ model (panels [{\it a}] in Figs.
\ref{fig:ps_all}--\ref{fig:sf_all_mod}) the level of
the density fluctuations is very small compared to the velocity
fluctuations (weighted by  $\langle v_{z}^{2}\rangle$ and
$\langle\rho^{2}\rangle$ respectively)
and all the centroids trace remarkably well the velocity statistics.
In this case the simplified criterion
$\langle \tilde{S}^{2}\rangle/(\langle v_{z}^{2}\rangle
\langle\tilde{I}^{2}\rangle)\gg1$
is well satisfied: $\langle \tilde{S}^{2}\rangle/(\langle v_{z}^{2}\rangle
\langle\tilde{I}^{2}\rangle)\gtrsim 90$ for the original data-set, and
$\gtrsim 30$ for the reformed one.
This result also agrees with the notion that velocity centroids trace
the statistics of velocity in the case of subsonic turbulence
(where density fluctuations are expected to be negligible, as the
turbulence is almost incompressible).
For the rest of the models density has an increasing impact which is
reflected in the centroids, and our criteria is either not entirely met,
or violated.
We can also see for supersonic turbulence the
slope of the structure function of unnormalized centroids is almost 
always steeper than that of column density.
This means that one need to use the full criterion as proposed
in LE03 to judge whether centroids will trace velocity or not.
We will see however that our simplification of the criterion seems to
discriminate when centroids trace the statistics of integrated
velocity, at least for the data sets employed here.
For all the supersonic cases the measured power-spectrum index from
the the different centroids fail to give the index of velocity.
And  in general, their centroids index is found to lie between those of
velocity and density, with more scatter for the original simulations
compared to the reformed fields.
For the original data-sets of models B, C, and D the ratio
$\langle \tilde{S}^{2}\rangle/(\langle v_{z}^{2}\rangle\langle\tilde{I}^{2}\rangle)$ is $\approx 5, 4, {\rm and}\,3$ respectively.
For the corresponding reformed data-cubes our simplified criterion gives
 $\langle \tilde{S}^{2}\rangle/(\langle v_{z}^{2}\rangle\langle\tilde{I}^{2}\rangle)\approx 2, 4, {\rm and}\,3$.
In all of these cases
$\langle \tilde{S}^{2}\rangle/(\langle v_{z}^{2}\rangle\langle\tilde{I}^{2}\rangle)\gg1$
is not strictly true, and indeed there is an important contribution of
density fluctuations. This is consistent with the results in LE03 where
we see evidence of a density dominated regime at small scales (rather
density contaminated regime as centroids do not show the same index as
density either).
Our results agree with those presented in \citet{BML04}, where they
found that centroids do not provide good velocity representation for
the supersonic turbulence they studied ($\mathcal{M}_s > 1.9$).

\subsubsection{Cross-terms and density-velocity cross-correlations}

\begin{figure}
\epsscale{1.3}
\plotone{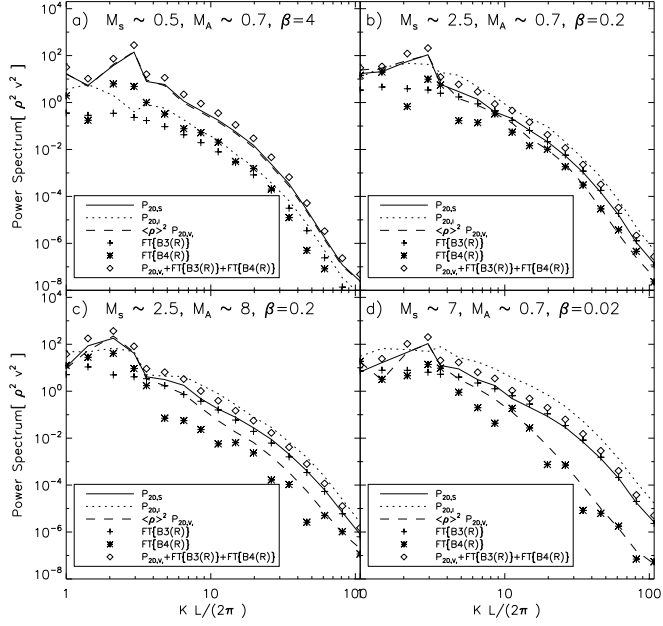}
\figcaption{Decomposition of the spectra of unnormalized velocity centroids as in equation \ref{eq:PS_S_main} (with $v_0\approx 0$), for the {\it original} data-sets.  The  {\it solid} line is the spectrum of the 2D map of velocity centroids. The {\it dotted} line  (for reference only) is the power spectrum of column density, the {\it dashed} is the spectrum of integrated velocity, both obtained from integrated (2D) maps. The {\it crosses} represent the  cross-term  $\mathcal{F}\left\{B3(\mathbf{R})\right\}$, and the {\it ``X''} correspond to density-velocity cross-correlations $\mathcal{F}\left\{B4(\mathbf{R})\right\}$. This last two were computed from 3D statistics, then integrated along the LOS. The {\it diamonds} show the sum of all the terms in the decomposition to compare with the {\it solid} line.
\label{fig:ps_I3_4}}
\end{figure}

\begin{figure}
\epsscale{1.3}
\plotone{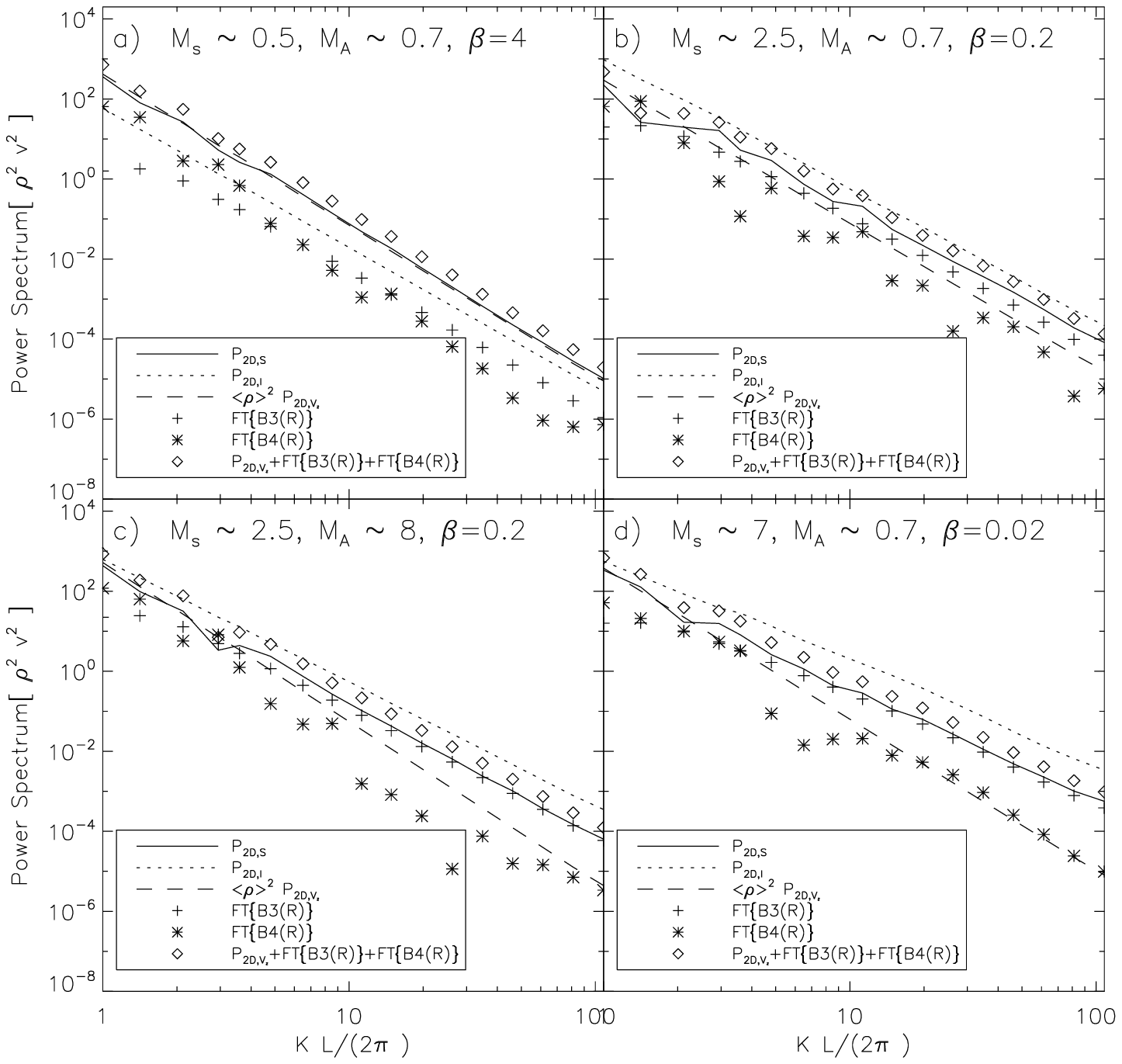}
\figcaption{Same as Fig. \ref{fig:ps_I3_4}, for the {\it reformed} data-sets.
\label{fig:ps_I3_4_mod}}
\end{figure}

\begin{figure}
\epsscale{1.3}
\plotone{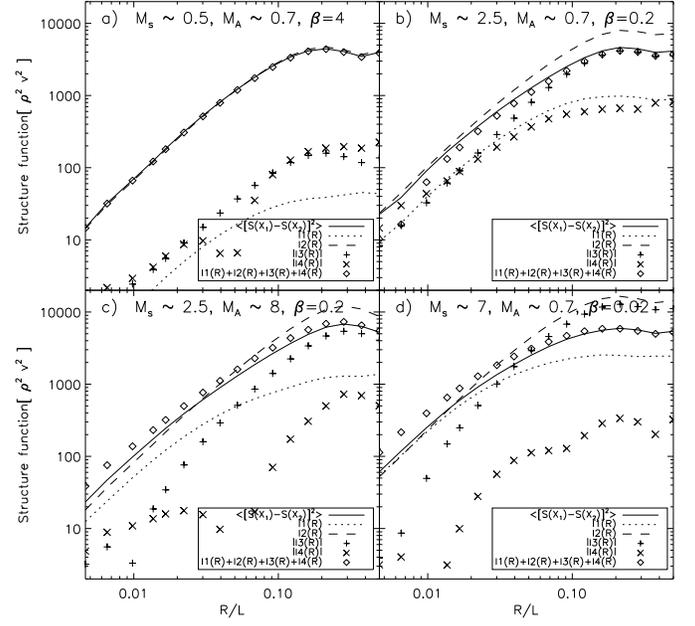}
\figcaption{Decomposition of the structure function of unnormalized velocity centroids (eq.[\ref{eq:S1S2Dec}]), for the {\it original} data-sets.  The {\it solid} line is the structure function of the 2D map of velocity centroids. The {\it dotted} line is $I1(\mathbf{R})$, the {\it dashed} $I2(\mathbf{R})$, both obtained from 2D maps. The cross-term  $\vert I3(\mathbf{R})\vert$ as {\it crosses}, while the  density-velocity cross-correlations ($\vert I4[\mathbf{R}]\vert$) are the {\it ``X''}. This last two were computed from 3D statistics, then integrated along the LOS. The {\it diamonds} show the sum of all the terms in the decomposition to compare with the {\it solid} line.
\label{fig:sf_I3_4}}
\end{figure}

\begin{figure}
\epsscale{1.3}
\plotone{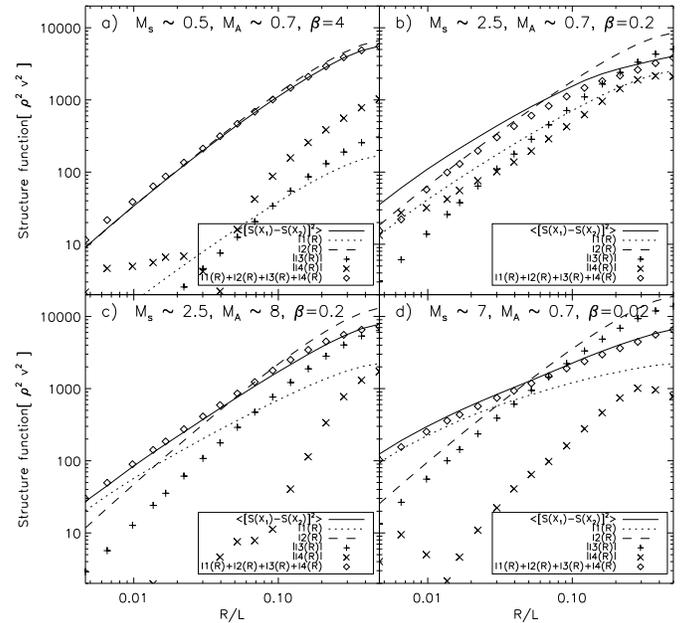}
\figcaption{Same as \ref{fig:sf_I3_4}, for the {\it reformed} data-sets.
\label{fig:sf_I3_4_mod}}
\end{figure}

The cross-terms, $I3(\mathbf{R})$,  $\mathcal{F}\left\{ B3(\mathbf{R})\right\}$, as well as the cross-correlations in 
$I4(\mathbf{R})$, and  $\mathcal{F}\left\{ B4(\mathbf{R})\right\}$
have contributions of velocity and density that we cannot disentangle
entirely from observables.
Furthermore, they cannot be expressed in terms of integrated quantities
and they have to be computed from 3D statistics.
We used Fourier transforms to obtain the structure, and correlation
functions in 3D needed to produce independently all the terms
in equations (\ref{eq:S1S2Dec}), and (\ref{eq:PS_S_main}).
These 3D statistics were integrated numerically to get
$\mathcal{F}\left\{ B3(\mathbf{R})\right\}$, $I3(\mathbf{R})$,
$\mathcal{F}\left\{ B4(\mathbf{R})\right\}$, and $I4(\mathbf{R})$.
To check the accuracy of the 3D quantities obtained and our numerical
integration to 2D we verified with the cases in which the statistics can
be also obtained directly in 2D (e.g. $I1[\mathbf{R}]$, $I2[\mathbf{R}]$),
finding always a good agreement.
The results of the decomposition in terms of power-spectra
(equation [\ref{eq:PS_S_main}], after $K$ average) are shown in
Figure \ref{fig:ps_I3_4} for the original simulations, and in Figure
\ref{fig:ps_I3_4_mod} for the reformed data-sets.
The decomposition for $R$ averaged structure-functions is presented in 
in Figures \ref{fig:sf_I3_4}, and \ref{fig:sf_I3_4_mod};
for the original, and modified data-sets respectively.
Something to notice is that, because structure functions span
over fewer decades in the vertical axis, the separation of all the terms
is generally more clear than for power-spectra.
It is also to be noticed, that for the  subsonic model (A),
the spectrum and structure function of centroids are almost unaffected
by density fluctuations, cross-terms, or density velocity
cross-correlations.
Cross-correlations are found to be larger for supersonic
turbulence compared to the subsonic case, but from this limited data
set we can not conclude that they scale in some particular way with
the sonic Mach number.
It is also significant that the cross-term increases relative to the
other terms with $\mathcal{M}_s$. For spectra, in all the supersonic
models (B, C , and D) the cross-term
$\mathcal{F}\left\{ B3(\mathbf{R})\right\}$ is dominant.
This would mean that the log-log slope measured from spectra in all
of this cases is not a direct reflection of the velocity spectral index.
At the same time we observe that the magnitude of density-velocity
cross-correlations can only be entirely ignored for model A (subsonic
turbulence).
For structure functions the cross-term is comparable or larger
in magnitude than the contribution of column density.
At the same time it gets closer, but does not become larger than
$I2(\mathbf{R})$.
In fact, it was found to be always steeper than the velocity term,
therefore its contribution at the small scales could be neglected.
Velocity-density cross-correlations in $I4(\mathbf{R})$ for all the cases
we considered here\footnote{The situations when velocity-density
  correlations are dominant are discussed in our next paper} never had
an important impact on the statistics of centroids.
Anyway, $I3(\mathbf{R})+I4(\mathbf{R})$ as a whole
(especially at the smallest separations, which we are most interested in)
has been found to be smaller than the integrated velocity term. 
Suggesting that there could be cases where $I3(\mathbf{R})+I4(\mathbf{R})$ 
--the cross-term and cross-correlations-- can be neglected.
However, since the indices from power-spectra of centroids
failed to give the velocity spectral index for models B, C, and D,
we conclude that the retrieval of spectral indices from velocity
centroids over the entire inertial range should be restricted to very
low sonic Mach numbers ($\mathcal{M}_s < 2.5$).
Furthermore, the relative importance of the cross-terms grow together with
the column density term as the strength of turbulence increases.
For $\mathcal{M}_s\sim7$ certainly the contribution of the column
density is considerable.
In this case MVCs could be used to remove the contribution from the column
density term in the structure function of centroids. However at such
high Mach number the cross-terms cannot be neglected either.

At a first glance it might seem surprising that the velocity centroids
could be dominated by velocity even when the 3D structure functions
of density had a zero point (offset) larger than that of velocity.
This effect arises primarily from the fact that the density is positive
defined, and necessarily has a non zero mean whereas the velocity
field can have a zero mean.
This results in the factors that multiply the density and
velocity structure functions in equation (\ref{eq:D2}). The factor
$\langle v_{z}^{2}\rangle=v_{0}^{2}+\langle\tilde{v}_{z}^{2}\rangle$
multiplies the density structure function, and we can minimize the
undesired density contribution by shifting our velocity axis in such
a way that $v_{0}=0$. For a more detailed discussion about the
velocity and density zero levels see \citet{OELS05}.

\section{Velocity centroids and anisotropy studies in supersonic turbulence}

\begin{figure}
\epsscale{0.55}
\plotone{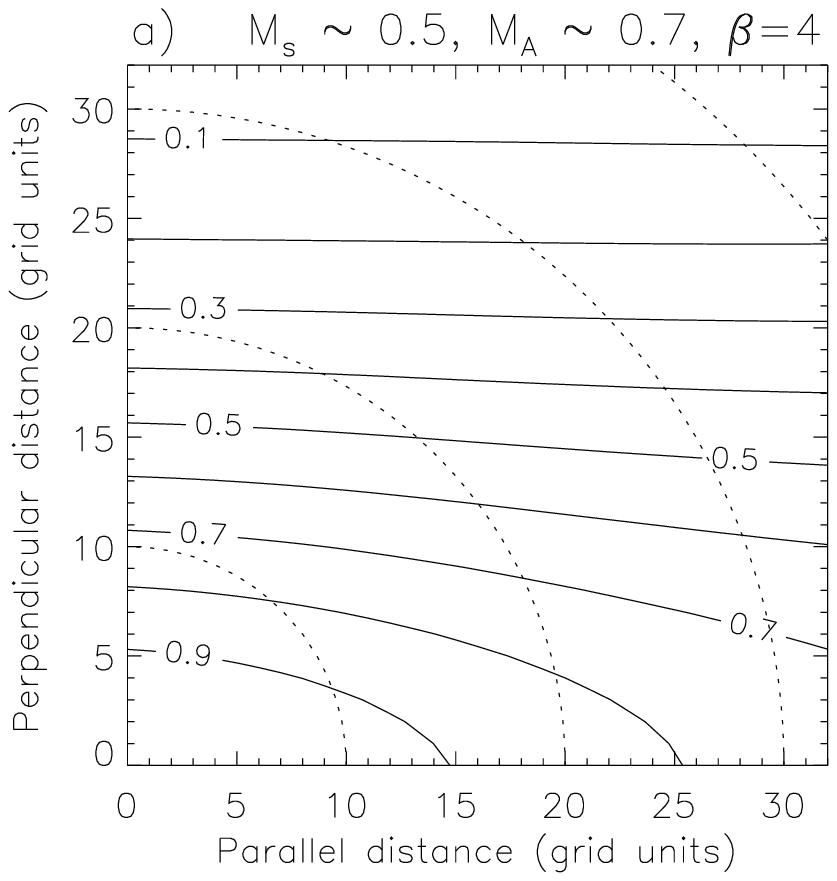}\plotone{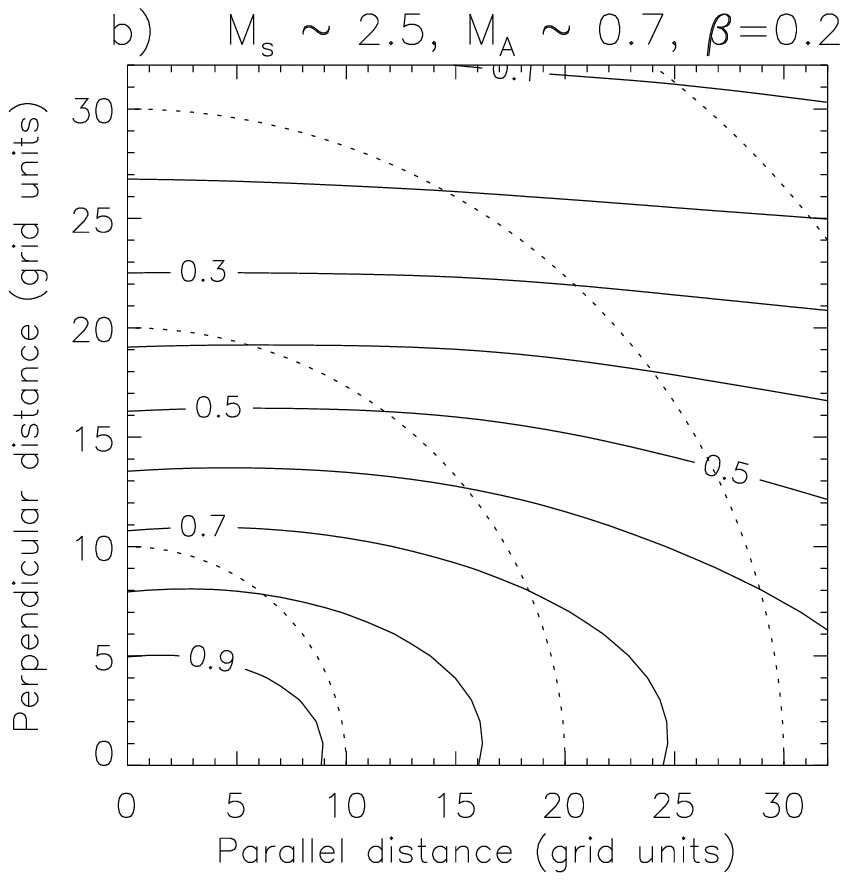}
\plotone{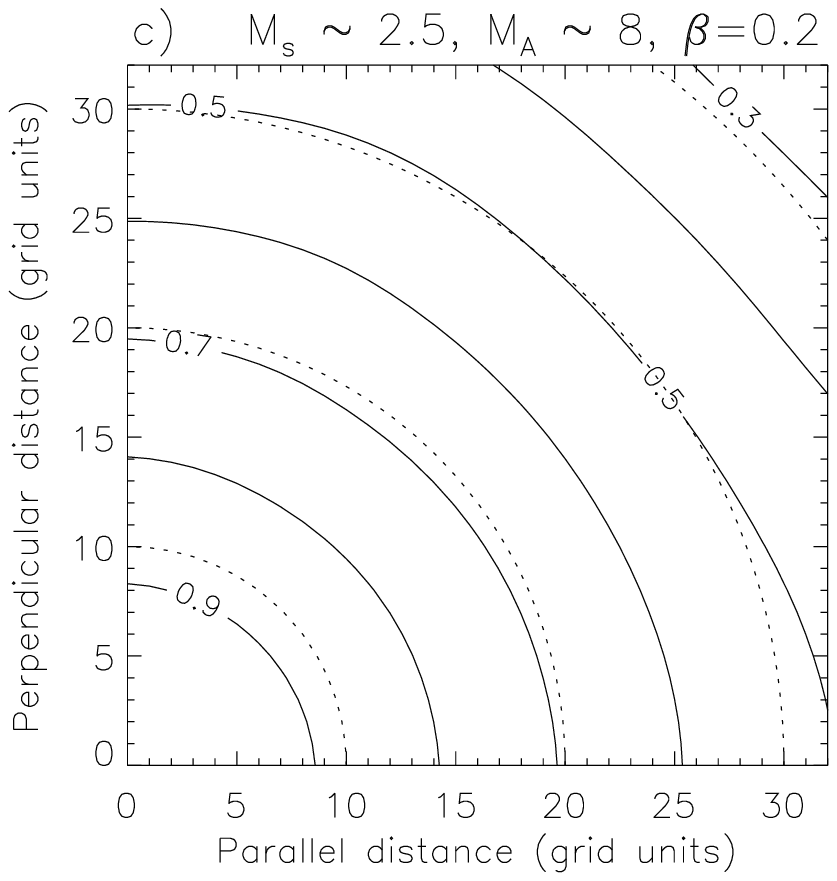}\plotone{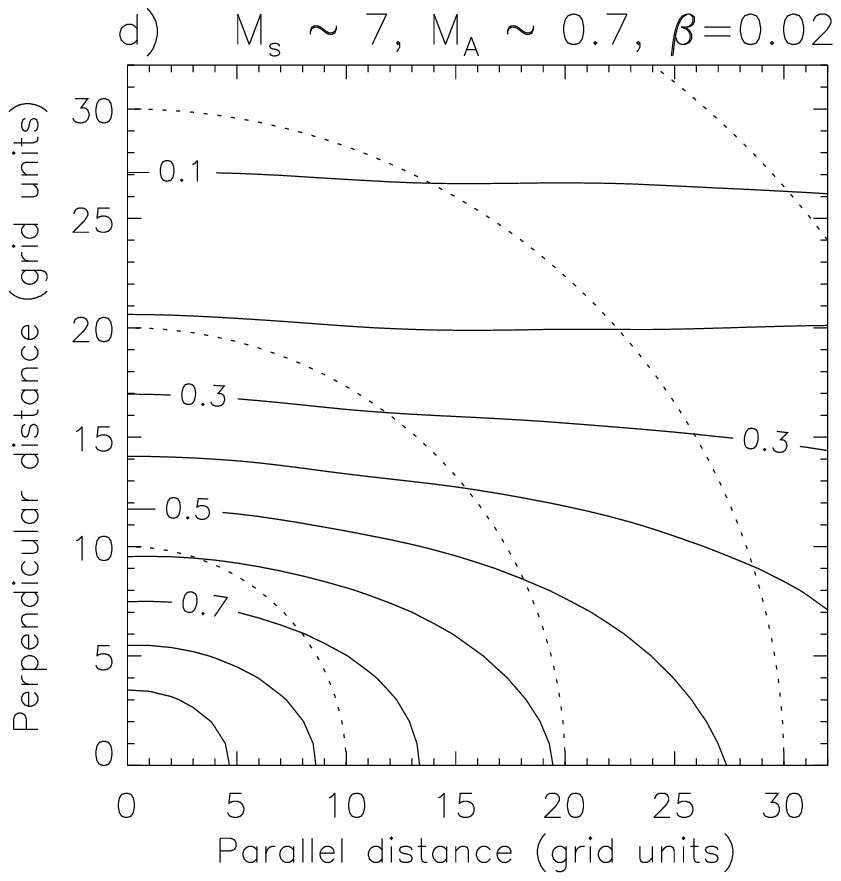}
\figcaption{Anisotropy in the correlation functions: contours of equal correlation for the simulations ({\it solid} lines). For reference we show isotropic contours as {\it dotted} lines. The sonic and Alfv\'{e}n Mach numbers ($\mathcal{M}_s$, $\mathcal{M}_A$ respectively) are indicated in the title of the plots. The anisotropy reveals the direction of the magnetic field for all the sub-Alfv\'{e}nic cases, regardless of the sonic Mach number.
\label{fig:anis}}
\end{figure}

We have seen how density fluctuations in supersonic turbulence affect
our ability to determine the velocity spectral index from observations.
But is there something else velocity centroids could be used for, even in
supersonic turbulence? The presence of a magnetic field introduces a
preferential direction of motion, thus breaking the isotropy in the
turbulent cascade. In a turbulent magnetized plasma, eddies become
elongated along the direction of the {\it local} magnetic field.
Velocity statistics have been suggested to study this anisotropy
\citep{LPE02,ELPC03,VOS03}\footnote{Both velocity centroids and
  spectral correlation functions were demonstrated to trace the
  magnetic field in \citet{LPE02}, channel maps were studied for the
  same purpose in \citet{ELPC03}, and velocity centroids in
  \citet{VOS03}}.
For instance, iso-contours of two point statistics of velocity centroids 
instead of being circular, as in the isotropic case, are ellipses 
with symmetry axis that reveals the direction of the
{\it mean} magnetic field.
We present in Figure \ref{fig:anis} contours of equal correlation
of velocity centroids (unnormalized) from our simulations.
The magnitude of the magnetic field determines how much anisotropy
will be present.
We can see from Figure \ref{fig:anis} that the anisotropy is very clear
for models A, B, and D, regardless of the large
differences in sonic Mach number and plasma $\beta$.
The only case in which the anisotropy is not evident (model C) is our
only super-Alfv\'{e}nic simulation. The concept of super-Alfv\'{e}nic
turbulence is advocated, for instance, by \citet{PJJN04} for molecular
clouds.

\begin{figure}
\epsscale{1.2}
\plotone{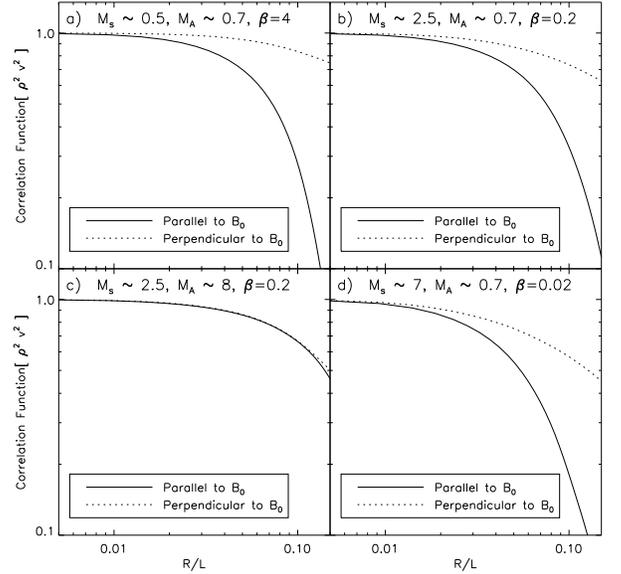}
\figcaption{Correlation functions taken in directions parallel and perpendicular to the mean magnetic field. The anisotropy shows in the different scale-lengths for the distinct directions. It is noticeable the little difference in panels ({\it a}), ({\it b}), and ({\it d}). Which correspond to the same ratio $\tilde{B}/B_0$, but very different sonic Mach numbers. Panel ({\it c}) corresponds to super-Alfv\'{e}nic ($\tilde{B}>B_0$) turbulence, and the anisotropy is not evident in the centroid maps.
\label{fig:anis_2}}
\end{figure}

Another way to visualize the anisotropy is to plot the correlation
functions of centroids in the parallel or perpendicular directions
relative to $B_0$, this is shown in Figure \ref{fig:anis_2}.
In our simulations this corresponds to plot the value at the
intercepts in Figure \ref{fig:anis}, in observations it would mean to plot
the correlation function only along the major or minor axis of the contours
of equal correlation. the anisotropy is evident as a different scale-length
(correlation-length) for correlations in the parallel or perpendicular to
the mean magnetic field.
For subsonic turbulence the degree of anisotropy reflects the ratio
$(\tilde{B}/B_0)^2$. However for supersonic turbulence as the
contributions of density get important, and as density at high Mach
number gets more isotropic \citep[see][]{BLC05}, one could expect the
anisotropy to decrease while the ratio $\tilde{B}/B_0$ stays the same.
We found very little evidence of this decrease, as can be verified 
from models A, B, and D in Figure \ref{fig:anis_2}, all the three runs
have the same  $\tilde{B}/B_0$, and a large contrast in sonic Mach number.
In model C even though the magnetic pressure
is larger than the gas pressure ($\beta=0.2$), it corresponds to a weak
mean field $\tilde{B} > B_0$. In this case the magnetic field
has very chaotic structure at large scales \citep[see][]{CL03}.
As the fluctuations at large scale determine the anisotropy of the
projected data, the MHD anisotropy is erased after LOS averaging.

Whether the ISM is sub-Alfv\'{e}nic is still up for debate. The
fact that centroids are only anisotropic for sub-Alfv\'{e}nic turbulence
gives us an opportunity to study the conditions under which that is the
case.
It is certainly encouraging, that for sub-Alfv\'{e}nic turbulence
the degree of anisotropy does not seem to be strongly affected by density
fluctuations.
However, we should add a word of warning in trying to determine the ratio
$\tilde{B}/B_0$ from velocity centroids in supersonic turbulence.
The fact that centroids do not represent velocity for
$\mathcal{M}_s > 2.5$ calls for some caution as to what extent the ratio
$\tilde{B}/B_0$ can be obtained from observations.
Nonetheless, the result  on the direction of the mean field is rather robust,
including for highly supersonic turbulence, provided a relatively large
ordered component $B_0$.
Anisotropies of turbulence measured by different techniques, including
centroids, provide a promising way of measuring the direction of the 
perpendicular to the LOS component of the magnetic field.
For practical purposes the technique can be tested using polarized radiation
arising from aligned dust grains (see \citealt{L03} for a discussion of
grain  alignment), or CO polarization arising from Goldreich-Kylatis effect
\citep*[see][]{LGC03}.

For any particular observational data one should bear in mind that only
the plane-of-sky components of mean and fluctuating magnetic field are
available. Therefore, for instance, a cloud with sub-Alfv\'{e}nic turbulence
but with a mean magnetic field directed along the line of sight
would look like a cloud with super-Alfv\'{e}nic turbulence. 
The distinction between the two cases may be done, however, using
an ensemble of clouds. 
It is unlikely to have mean magnetic field always directed towards the observer.
If, indeed, the turbulence is typically sub-Alfv\'{e}nic, the anisotropy should
show up in centroid maps.

\section{Discussion}

We used MHD data-cubes to produce centroid maps, and explored the
limitations of velocity centroids for studying interstellar turbulence.

We investigated numerically the analytical predictions in our previous
study of velocity centroids (LE03). In that work, we
decomposed the structure function of velocity centroids into three
main contributions, namely integrated (or column) density, integrated
velocity, and ``cross-terms''. 
By calculating separately all the terms in the
structure function of unnormalized centroids we showed that the
decomposition works well.

We found two important restrictions on the procedure accuracy, both related
to the finite resolution in the numerical simulations.
First, structure functions of integrated quantities can in principle
be used to retrieve the underlying 3D spectral index if calculated for
lags either much greater or much smaller than the LOS extent of the
object.
We showed however, that with the resolution used here ($216^3$) structure
function of integrated quantities can only give an upper limit on the
actual spectral index.
Second, the limited inertial range in our simulations present an  additional
constraint on the accuracy of estimating the spectral index, mainly because
the measured log-log slope is very sensitive the range of scales used.

In regards to the finite width projection effects one can potentially still
do a comparative study of the centroid statistics with the statistics of
velocity and density.
That is, we can compare the spectral index of the centroid maps to that of
column density or integrated density, and see to which one is closer.
However, for real observational data with the possibility
to sample lags $R \ll z_{tot}$, obtaining the 3D spectral index
directly from structure functions is not a problem.

Power-spectrum is an alternative to obtain the scaling properties of the
turbulent velocity field that does not suffer of such projection
effects (i.e. the spectral index can be measured at any wave-numbers
within the inertial range).
In order to use power-spectra too, we constructed a power-spectra
decomposition analogous to that in LE03.
And, by calculating each term separately, we tested successfully the
validity of this decomposition.

We found that the scaling properties of the underlying velocity field 
(i.e. spectral index) can be reliably retrieved for subsonic turbulence.
However, the contamination from cross-terms clearly showed up in all the
supersonic cases for spectra, while for structure functions it was
only evident for $\mathcal{M}_s\sim 7$.
To alleviate the other problem (not enough inertial range) we introduced
reformed versions of the original simulations. The power spectra of
new data-set are strict power-laws, with almost the same spatial 
structure, and density-velocity cross-correlations. 
The spectral indices obtained with these reformed data sets are
in better agreement with the analytical predictions.

We also tested successfully a criterion for which velocity
centroids give trustworthy information (equation [\ref{eq:crit_full}]),
that can be obtained entirely from observations.
For practical purposes we suggest as the first approach an approximation
of the criterion in LE03 in terms only of the variances of the maps of
column density and velocity centroids instead of computing all the
structure function (equation [\ref{eq:crit_simp}]).
If this ratio is less than unity velocity centroids are not trustworthy.
When the ratio in eq.(\ref{eq:crit_simp}) is small but larger than unity
it is recommendable to use the full criterion, as proposed in LE03.
In our only case of subsonic turbulence (model A) centroids traced
the integrated velocity structure function extremely well. 
And, our simplified criteria was fulfilled:
$\langle \tilde{S}^{2} \rangle / (\langle v_{z}^{2}\rangle
\langle \tilde{I}^{2} \rangle) \gtrsim 90,~30$, for the original, and
reformed data-set respectively.
For the rest of the models (B, C, D) the criteria was not
satisfied ($\langle \tilde{S}^{2}\rangle/(\langle v_{z}^{2}\rangle
\langle\tilde{I}^{2}\rangle)\gtrsim [5-2]$), and the centroids failed
to trace the statistics of the velocity field.
In this models one can see a better correspondence between centroids and
integrated velocity for the $\mathcal{M}_s\sim 2.5$ runs compared to the
highly supersonic case $\mathcal{M}_s\sim 7$. Thus, one might hope
centroids to be able to trace the spectral index of velocity for
supersonic turbulence, but only for  $\mathcal{M}_s < 2.5$.
We must recognize that this study was done using a limited data set.
A more complete exploration of parameter space is desirable to determine 
better the range of applicability of velocity centroids,
including MVCs.

We have seen that velocity centroids are only reliable at
low Mach numbers. At the same time there are other techniques
available that work for strongly supersonic turbulence,
such as VCA and VCS. The techniques are complementary and can be
used simultaneously. For subsonic turbulence, VCA and VCS can be
used with higher mass species. Note, that the different techniques
pick up different  components of velocity tensor. This can potentially
allow a separation of compressional and solenoidal parts of the velocity
field (LE03). Determining these components are crucial for understanding
properties of interstellar turbulence, its sources and its dissipation.

We stress that every technique has its own advantages.
Velocity centroids can reproduce
velocity better than VCA and VCS when the velocity statistics
 is not a straight power-law. Therefore velocity centroids can better
pick up dissipation and injection of energy scales.

Nonetheless, one have to bear in mind that spectral indices derived
from all of the techniques above do not provide a complete description
of turbulence.
Anisotropy is another parameter that can be studied.
For instance, we showed that velocity centroids are useful to study the
anisotropy of the turbulent cascade, even for highly supersonic
turbulence. We showed however, that this is restricted to a relatively
large mean field. For super-Alfv\'{e}nic turbulence, a model favored by
some researchers \citep{PJNB04}, the anisotropy of centroids is
marginal, which allows to test these theories.
Anisotropy is not only present in velocity centroid maps, but in other
statistics as well, such as {\it spectral correlation function}
\citep[see][]{LPE02}. Combining the two measure can improve the
confidence in the results.

Additionally, we need tools to study the turbulence variations in
space (i.e. intermittency).
Higher order velocity structure functions (those described in this paper
are second order) have been shown to be a promising for such studies
\citep[see][]{MB00,CLV03b}. 
Velocity centroids can be easily recasted in terms of higher
order statistics, opening a new window for intermittency studies.
Is worth noting that higher moments can provide the directions
of the intermittent structures if those get oriented in respect to magnetic
field. Our results indicate that such studies can be carried
out even for high Mach number turbulence. In addition, studies of intermittency
with centroids can be incorporated to other techniques. For instance, the
interpretation of results of the
Principal Component Analysis (PCA) technique suggested as a tool
for turbulence studies by \citet{HS97} depends on the degree
of intermittency of turbulence (\citealt{BHVSP03}; Heyer 2004 private
communication).

Velocity centroids have been used for studies of turbulence for many decades.
How can we comment on these studies from the point of view of our present
theoretical understanding?  Let us glance at the available studies.
Turbulence in \ion{H}{2} regions is usually subsonic.
Therefore velocity centroids could be probably trustworthy
there \citep[see][]{OC87}.
Supersonic turbulence in molecular clouds \citep[see][]{MB94} is
a field where velocity centroids might be in error. The same is probably
true for \ion{H}{1} studies \citep[see][]{MJFB03}\footnote{An interesting
feature found in  \citet[see][]{MJFB03} is that the statistics of velocity
centroids is almost identical to the statistics of
density field. This could be due to the dominance of the $I1$ term, which
can be tested.} where turbulence is highly
supersonic in cold \ion{H}{1}.  

We feel that a more careful analysis of particular conditions 
present in the regions under study is necessary,
however. 
One one hand, while the turbulence is supersonic for molecular
clouds, the cores of molecular clouds are mostly subsonic 
\citep[see][]{TMCW04,ML98}. For such cores velocity
centroids provide a reliable tool for obtaining velocity statistics, 
if the resolution of the cores is adequate. On the other hand, our 
results are based on the analysis of isothermal numerical simulations.
In the presence of substantial density contrasts caused by the co-existence
of different phases the velocity centroids may get unreliable even for
subsonic turbulence. Therefore testing of the necessary criteria discussed
in this paper may be advantageous not only with the molecular and \ion{H}{1} data,
but also with data obtained for \ion{H}{2} regions. 
We also note, that for the
present analysis we assumed that the emission is optically thin and the
emissivity is proportional to the first power of density. Discussion
of more complex, but still observationally valuable cases will be done
elsewhere. The work started in this direction \citep[see][]{LP04} is
encouraging.

Recent work on centroids includes papers by \citet{OML02}
where they noticed that centroids poorly reproduce velocity statistics
for supersonic turbulence. Our work above confirms their finding and
also establishes a regime ($\mathcal{M}_s<2.5$) when the results by
centroids can be reliable.

A quite different and optimistic conclusion about centroids was obtained
in \citet{MLF03}. They used Brownian noise artificial data and obtained an
excellent correspondence between the centroids and the underlying velocity.
From the point of view of our analysis the origin of this correspondence
is in the choice of the mean density level. In these simulation in order
to make the density positively defined the authors were adding a substantial
mean density to the fluctuating density.
It is clear from eq.(\ref{eq:I2})
that adding mean density increases of the contribution of
$I2(\mathbf{R})$ term that is the term responsible for the velocity
contribution.

This work is complementary to the work on delta variance of centroids that
we do in collaboration with Ossenkopf and Stutzki \citep{OELS05}. There Brownian
noise simulations are used, but extra care is taken to avoid being misled
by the effect of adding mean density. An iterative procedure is proposed there,
which allows to correct for the contribution of the cross-terms when
density-velocity correlations are negligible.

In this paper we are dealing mostly with unnormalized centroids.
The modified centroids (MVCs) allow a different outlook at the problem 
of obtaining velocity statistics from the small scale asymptotics.
Indeed, our analysis in the paper shows that the term $I4(\mathbf{R})$
is unimportant for the cases that we studied.
In addition, we believe  that velocity has steep spectra.
Therefore, if the density is also steep, asymptotically
the term $I3(\mathbf{R})$, which is then steeper than both  velocity
and density, should be negligible (see Appendix E.1).
As the result, {\it if we see from the integrated intensity maps that
  density indeed is steep, we can use MVCs as suggested in LE03}, 
that for sufficiently small lags are bound to represent
the velocity statistics.
The critical scale at which this is true can be obtained using the
analytical expressions found in Appendix B.1.
Formally to find such critical scale one should know the mean density.
However, if the inertial range is sufficiently long, one should not be
worried about the exact value for that critical scale.

We have not seen much advantages of such a use in our numerical runs
because of the limited inertial range available.
However, unlike numerical simulations, astrophysical conditions
provide us with a substantially larger inertial range. For instance,
\citet{SL01} showed that the turbulent spectrum for velocity spans
from the size of the SMC, which is $\sim 4$ kpc, to the minimal
scale resolved, i.e. $\sim 40$ pc. In partially ionized gas we 
expect this spectrum to proceed to sub-parsec scales. In fully
ionized gas (see also discussion in \citealt{LVC04}) the scale of
Alfv\'{e}n and slow mode damping may be hundreds of kilometers only.

The limitations of this asymptotic approach stem from the fact that
for supersonic turbulence the density field tends to become shallower.
But, as we can see from the MHD simulations used here, at
$\mathcal{M}_s\sim 2.5$ we find that it is still steep and
therefore the MVCs should be reliable asymptotically. Therefore,
for handling observational data we can provide another prescription:
``If the inertial range is sufficiently long and density is steep
then MVCs asymptotically represent the velocity field.''
For instance, steep density was reported in \citet{MLF03}. Thus
for such density fields asymptotic use of MVCs should be advantageous.

In the cases when the underlying statistics is not a power law the
interpretation of centroids is more involved. This we have seen with
our analysis of non-power-law data from MHD simulations. While centroids
do represent injection and damping scales, the integration along the
line of sight does interfere with a more fine detail study.
In this situations one should use the full technique of inversion
as proposed in \citet{L95}. However, we expect that in most
cases astrophysical turbulence is self-similar over large expanses
of scales and therefore the power-law representation is adequate.

We have made use of the analytical expressions 
for unnormalized velocity centroids (LE03). Our work as well as a
study in \citet{Lev04}, show that normalization of centroids improve
them rather marginally, while it makes the analytical insight 
essentially impossible. In this paper we tested that
major conclusions reached for unnormalized centroids are applicable 
to the normalized ones.

\section{Summary}

In this paper we have provided a systematic study of velocity centroids
as a technique of retrieving velocity statistics from observations.
We used both results of MHD simulations as well as ``reformed'' data-sets
which have larger inertial range. We found:

\begin{enumerate}
\item Centroids of velocity can be successfully used to retrieve the
scaling properties of the underlying 3D velocity field for subsonic
turbulence. For supersonic turbulence with sonic Mach number
$\gtrsim 2.5$ velocity centroids failed to trace the spectral index
of velocity.
\item Our numerics confirmed the expression for centroid
statistics obtained in LE03. And, in particular, we tested successfully
the criterion in LE03 for the reliable use of velocity centroids.
We showed that it reflects a necessary condition for centroids to
reproduce the velocity statistics.
\item We studied modified velocity centroids (MVCs) proposed in LE03.
It is shown shown that MVCs reflect the statistics of velocity better than
unnormalized centroids for small lags. This result is valid for both steep
and shallow density fields with steep velocity.
\item We showed that velocity-density cross-correlations are marginally
important for our data set, at least for small lags. Combined with the
fact that products of density and velocity structure functions get
subdominant for steep velocity and density at small lags, this
provides a way to reliably study turbulence using MVCs.
\item We demonstrated that velocity centroids can be used for both
subsonic and supersonic turbulence to study the anisotropy introduced
by the magnetic field.
Even when they fail to retrieve the velocity spectral index in supersonic
turbulence. For up to at least $\mathcal{M}_s\sim 7$ they for
sub-Alfv\'{e}nic turbulence provide reliably
the direction of the component of the magnetic field perpendicular to the
LOS.

\item If turbulence is super-Alfv\'{e}nic it results in a marginal
anisotropy of velocity centroids which provides a good way to
test whether the Alfv\'{e}n Max number of turbulence in molecular
clouds is less or larger than unity.

\item Within their domain of applicability, centroids of velocity
provide a good tool for turbulence studies that should be used in
conjunction with other tools, e.g. VCA and VCS.
\end{enumerate}

\acknowledgements{We thank Jungyeon Cho for supplying us with the data cubes
of MHD turbulence
simulations, Anthony Minter, the referee of LE03, as well as the
anonymous referee of this paper,
for suggestions that benefited this work.
We would like to thank especially
V. Ossenkopf and J. Stutzki for many insightful discussions, in fact
great part of the improvements of
the manuscript have been made in collaboration with them.
A. E. acknowledges financial support from CONACYT (Mexico).
A. L. research is supported by NSF grant AST 0307869 and the NSF
Center for Magnetic Self Organization in Laboratory and
Astrophysical Plasmas.}

\begin{appendix}

\section{Turbulence statistics in three dimensions \label{app:basic}}

\subsection{Turbulence statistics in real ($xyz$) space.}

The two-point correlation function of a vector field
$\mathbf{u}(\mathbf{x})=[u_{x}(\mathbf{x}),u_{y}(\mathbf{x}),u_{z}(\mathbf{x})]$
is defined as:
\begin{equation}
B_{ij}(\mathbf{r})\equiv\left\langle u_{i}(\mathbf{x_{1}})u_{j}(\mathbf{x_{2}})\right\rangle ,
\label{eq:Bij}
\end{equation}
where $\mathbf{r}=\mathbf{x_{2}}-\mathbf{x_{1}}$ is the separation
or ``lag'', in Cartesian coordinates $i,\  j=x,\  y,\  z$, and
$\langle...\rangle$ denote
ensemble average over all space. An additional definition can be made
in terms of the fluctuations of the field. This can be obtained formally
by replacing $\mathbf{u}(\mathbf{r})$ by
$\tilde{\mathbf{u}}(\mathbf{x})=\mathbf{u}(\mathbf{x})-\langle\mathbf{u}(\mathbf{x})\rangle$
in equation (\ref{eq:Bij}). Note that this necessarily implies
$\langle\tilde{\mathbf{u}}\rangle=0$.
We will refer to this variation simply as correlation function of
fluctuations, and denote it by
\begin{equation}
\tilde{B}_{ij}(\mathbf{r})\equiv\left\langle \tilde{u}_{i}(\mathbf{x_{1}})\tilde{u}_{j}(\mathbf{x_{2}})\right\rangle =B_{ij}(\mathbf{r})-\left\langle u_{i}\right\rangle \left\langle u_{j}\right\rangle.
\label{eq:bij}
\end{equation}
In the same manner, it is customary to define the structure function
of the same vector field $\mathbf{u}(\mathbf{x})$ as
\begin{eqnarray}
D_{ij}(\mathbf{r}) & \equiv & \left\langle \left[u_{i}(\mathbf{x_{1}})-u_{i}(\mathbf{x_{2}})\right]\left[u_{j}(\mathbf{x_{1}})-u_{j}(\mathbf{x_{2}})\right]\right\rangle \nonumber \\
& = & \left\langle \left[\tilde{u}_{i}(\mathbf{x_{1}})-\tilde{u}_{i}(\mathbf{x_{2}})\right]\left[\tilde{u}_{j}(\mathbf{x_{1}})-\tilde{u}_{j}(\mathbf{x_{2}})\right]\right\rangle .
\label{eq:Dij}
\end{eqnarray}
Notice that structure functions, by definition, depend only
on the fluctuating part of the field and are insensitive to the mean
value of the field. Combining equations (\ref{eq:Bij}) and (\ref{eq:Dij})
it is trivial that:
\begin{equation}
D_{ij}(\mathbf{r})=2\left[\tilde{B}_{ij}(0)-\tilde{B}_{ij}(\mathbf{r})\right]=2\left[B_{ij}(0)-B_{ij}(\mathbf{r})\right].
\label{eq:Dij_dec}
\end{equation}
Where $\tilde{B}_{ij}(0)$ is also known as the variance. The correlation
of fluctuations must vanish at infinity, that is
$\tilde{B}_{ij}(\mathbf{r})\rightarrow0$,
as $r\rightarrow\infty$. Then from equation (\ref{eq:Dij_dec}) is
clear that $D_{ij}(\infty)=2\tilde{B}_{ij}(0)$. Thus, if we know
the structure function of a field we can obtain the correlation function
of fluctuations and vice versa. However, to get the correlation function
in general (as in equation {[}\ref{eq:Bij}{]})
we also need to know the mean value of the field 
($\langle\mathbf{u}\rangle$).
Correlation and structure functions are equivalent in theory.
In practice however, where we have a restricted
averaging space, it is easier to determine
$D_{ij}(\mathbf{r})$ more accurately compared to $B_{ij}(\mathbf{r})$.
At the same time, $\tilde{B}_{ij}(0)$ is usually better determined than
$D_{ij}[\infty]$ \citep[see discussion in][]{MY75}.

Alternatively, we can use spectral representation to describe turbulence.
The {\it spectral density tensor}, or {\it N-dimensional power-spectrum},
$F_{ij}(\mathbf{k})=P_{ND}(\mathbf{k})$, is given through a Fourier
transform of the correlation function of fluctuations:
\begin{equation}
F_{ij}(\mathbf{k})=P_{ND}(\mathbf{k})\equiv\frac{1}{(2\pi)^{N}}\int e^{-i\mathbf{k\cdot r}}\tilde{B}_{ij}(\mathbf{r})\  d^{N}\mathbf{r},
\label{eq:Fij}
\end{equation}
where $\mathbf{k}$ is wave-number-vector. The definitions presented
above are general, and also apply to scalar fields.

The power-spectrum of velocity (in the incompressible regime) has
an important physical interpretation as the distribution of kinetic
energy (per unit mass) as a function of scale.
If $\mathbf{u}(\mathbf{r})$ is the velocity field, the power-spectrum
is the energy in scales between
$\mathbf{k}$ and $\mathbf{k}+\delta\mathbf{k}$, and thus the total energy is
proportional to
$\langle\mathbf{u}(\mathbf{r})^{2}\rangle=\int P_{ND}(\mathbf{k}){\rm d}\mathbf{k}$.
Note however, that the physical interpretation is quite different
if we talk about the spectra of other quantity (e.g. the power-spectra
of density fluctuations).
For isotropic fields the correlation,
structure or spectral tensors can be expressed via longitudinal (parallel
to $\mathbf{r}$, denoted by subscript ``$LL$'') or transverse
(normal to $\mathbf{r}$, denoted by subscript ``$NN$'') components
\citep{MY75}:
\begin{mathletters}
\begin{eqnarray}
B_{ij}(r) & = & \left[B_{LL}(r)-B_{NN}(r)\right]\frac{r_{i}r_{j}}{r^{2}}+B_{NN}(r)\delta_{ij},
\label{eq:BNNBLL}\\
D_{ij}(r) & = & \left[D_{LL}(r)-D_{NN}(r)\right]\frac{r_{i}r_{j}}{r^{2}}+D_{NN}(r)\delta_{ij},
\label{eq:DNNDLL}\\
F_{ij}(k) & = & \left[F_{LL}(k)-F_{NN}(k)\right]\frac{k_{i}k_{j}}{k^{2}}+F_{NN}(k)\delta_{ij},
\label{eq:FNNFLL}
\end{eqnarray}
\end{mathletters}
where $\delta_{ij}$ is a Kronecker delta ($\delta_{ij}=1$
for $i=j$, and $\delta_{ij}=0$ for $i\ne j$). Solenoidal motions
(divergence free, therefore incompressible) correspond to the transverse
components whereas potential motions (curl free, compressible) correspond
to the longitudinal components.

\subsection{Turbulence statistics as observed (position-position-velocity space)}

Spectroscopic observations do not provide the distribution
of gas in real space coordinates ($\mathbf{x}\equiv[x,y,z]$). Instead
we observe the intensity of emission of a given spectral line at a
position $\mathbf{X}$ in the sky, and at a given velocity $v$ along
the LOS (we use capital letters for 2D vectors and lower case for 
3D vectors). 
Observational data are usually arranged in matrices with
coordinates $(\mathbf{X},v)$, also called position-position-velocity
(or simply PPV) cubes. We will identify the LOS with the coordinate
$z$. Thus, in the plane parallel approximation the relation between
real space and PPV space is that of a map
$(\mathbf{X},z)\rightarrow(\mathbf{X},v_z)$, with $\mathbf{X}=(x,y)$.

At any point the LOS velocity can be decomposed in a regular flow,
a thermal, and a turbulent components 
$[v_{z}(\mathbf{x})=v_{z,reg}(\mathbf{x})+v_{thermal}+v_{z,turb}(\mathbf{x})]$.
This way, the distribution of the Doppler shifted atoms follows is
a Maxwellian of the form:
\begin{equation}
\phi(\mathbf{x})\ {\rm d}v_{z}=\frac{1}{(2\pi\beta)^{1/2}}\exp\left\{ \frac{-\left[v_{z}-v_{z,reg}(\mathbf{x})-v_{turb,z}(\mathbf{x})\right]^{2}}{2\beta}\right\} {\rm d}v_{z},
\label{eq:Phi}
\end{equation}
where $\beta=\kappa_{B}T/m$, $\kappa_{B}$ is the Maxwell-Boltzmann
constant, $T$ the temperature, and $m$ the atomic mass. In PPV space,
the density of emitters $\rho_{s}(\mathbf{X},v_{z})$ can be obtained
integrating along the LOS
\begin{equation}
\rho_{s}(\mathbf{X},v_{z})\ {\rm d}\mathbf{X}\ {\rm d}v_{z}=\left[\int{\rm d}z\ \rho(\mathbf{x})\ \phi(\mathbf{x})\right]\ {\rm d}\mathbf{X}\ {\rm d}v_{z},
\label{eq:rhos}
\end{equation}
where $\rho(\mathbf{x})$ is the mass density of the gas in spatial
coordinates. The density of emitters $\rho_{s}(\mathbf{X},v_{z})$
can be identified as the column density per velocity interval, commonly
referred as ${\rm d}N/{\rm d}v$. Equation (\ref{eq:rhos}) simply
counts the number of atoms at a position in the plane of the sky $\mathbf{X}$,
with a $z$ component of velocity in the range $[v_{z},v_{z}+{\rm d}v_{z}]$,
and the limits of integration are defined by the LOS extent.
The integrated intensity of the emission (integrated along the velocity coordinate)
corresponds to the column density under the assumptions of optically thin media
and emissivity linearly proportional to the density:
\begin{equation}
I(\mathbf{X})\equiv\int\alpha\ \rho_{s}(\mathbf{X},v_{z})\ {\rm d}v_{z}=\int\alpha\ \rho(\mathbf{x})\ {\rm d}z.
\label{eq:IApp}
\end{equation}

\section{Projection of structure functions of a power-law spectrum field
\label{app:proj_sf}}

In this section we exemplify the long-wave dominated (steep) case
as if coming from a velocity field, while the shallow case as if from
a density field. The results are interchangeable, depending on
the specific spectral index they have (although there is no physical
motivation to consider a shallow the velocity field).

\subsection{Projection of a field with a steep power spectrum}

\begin{figure}
\epsscale{0.45}
\plotone{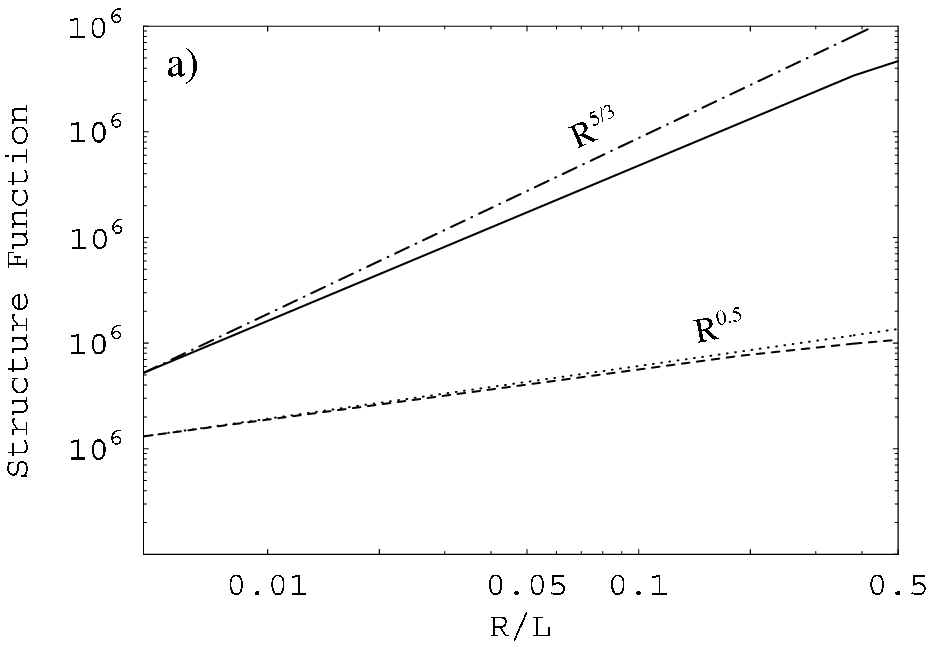}\plotone{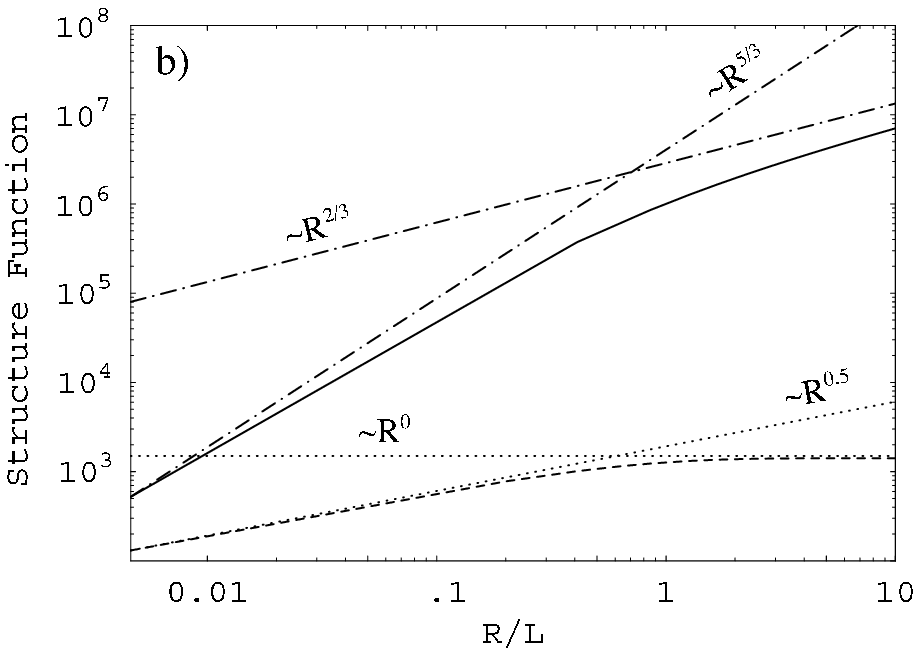}
\figcaption{Integrated structure functions from a field with a power-law spectrum. a)The {\it solid} line is the exact result (numerically integrating eq.[\ref{eq:V1V2}]) for a field with steep spectrum (3D power spectrum index of $\gamma_{3D}=-11/3$), the {\it dashed} is line for a shallow spectrum (3D power spectrum index $\gamma_{3D}=-2.5$, obtained numerically integrating eq.[\ref{eq:I1I2int}]). The {\it dash-dotted}, are the power-law asymptotics for the steep spectrum, and the {\it dotted} line for the shallow spectrum. Figure b) is the same as a) but we integrated for separations beyond the thickness of the object (chosen here to be $216$ to compare with the Gaussian Cubes in Figure \ref{fig:2Dgauss}), we can see clearly the effect of limited thickness discussed in the text.\label{fig:2Dmath}}
\end{figure}

Consider a homogeneous, and isotropic velocity field with a {\it steep}
power-law spectrum. The LOS (chosen to correspond with the $z$ direction)
velocity structure function will be of
the form:
\begin{equation}
\left\langle \left[v_{z}(\mathbf{x_{1}})-v_{z}(\mathbf{x_{2}})\right]^{2}\right\rangle =C_{1}\  r^{m}.
\label{eq:v1v2pl3d}
\end{equation}
Where $r=\sqrt{R^{2}+(z_{2}-z_{1})^{2}}$, $R$ is the separation
in the plane of the sky ($R^{2}=[x_{2}-x_{1}]^{2}+[y_{2}-y_{1}]^{2}$),
and $(z_{2}-z_{1})$, the separation along the LOS. This way we can
rewrite the 3D power-law structure function as
\begin{equation}
\left\langle \left[v_{z}(\mathbf{x_{1}})-v_{z}(\mathbf{x_{2}})\right]^{2}\right\rangle =C_{1}\left[R^{2}+\left(z_{2}-z_{1}\right)^{2}\right]^{m/2}.
\label{v1v2pl3d1}
\end{equation}
The projection of the velocity field, as described in \S\S2.2 results in
\begin{equation}
\left\langle \left[V_{z}(\mathbf{X_{1}})-V_{2}(\mathbf{X_{2}})\right]^{2}\right\rangle =C_{1}\ \int_{0}^{z_{tot}}\int_{0}^{z_{tot}}\left\{ \left[R^{2}+\left(z_{2}-z_{1}\right)^{2}\right]^{m/2}-\left(z_{2}-z_{1}\right)^{m}\right\} \ {\rm {\rm d}z_{2}{\rm d}z_{1}}.
\label{eq:V1V2}
\end{equation}
Where $z_{tot}$ is the largest scale along the LOS.
Since the field is isotropic, it is possible to evaluate this
integral changing variables from
$z_{1}$ and $z_{2}$, to $z_{+}=(z_{1}+z_{2})/2$
and $z_{-}=z_{2}-z_{1}$ . With this new variables, equation \ref{eq:V1V2}
becomes
\begin{eqnarray}
\left\langle \left[V_{z}(\mathbf{X_{1}})-V_{2}(\mathbf{X_{2}})\right]^{2}\right\rangle  & = & 2\  C_{1}\ \int_{0}^{z_{tot}}\int_{z_{-}/2}^{z_{tot}-(z_{-}/2)}\left[\left(R^{2}+z_{-}^{2}\right)^{m/2}-z_{-}^{m}\right]\ {\rm {\rm d}z_{+}{\rm d}z_{-}}\nonumber \\
 & = & 2\  C1\int_{0}^{z_{tot}}\left[\left(R^{2}+z_{-}^{2}\right)^{m/2}-z_{-}^{m}\right]\left(z_{tot}-z_{-}\right)\ {\rm {\rm d}z_{-}}.
\label{eq:V1V2zpzm}
\end{eqnarray}
The remaining integral can be solved analytically in terms of hypergeometric
functions, and converge only for $m>-1$ (which is automatically satisfied
since the spectrum is steep therefore with $m>0$), yielding (a similar
formula can be found in \citealt{SBHOZ98}):
\begin{eqnarray}
\left\langle \left[V_{z}(\mathbf{X_{1}})-V_{2}(\mathbf{X_{2}})\right]^{2}\right\rangle  & = & 2\  C_{1}\  z_{tot}^{2}\left\{ R^{m}\ _{2}F_{1}\left(\frac{1}{2},-\frac{m}{2},\frac{3}{2};-\frac{z_{tot}^{2}}{R^{2}}\right)+\left(\frac{R^{2}}{z_{tot}^{2}}\right)\frac{\left[R^{m}-\left(z_{tot}^{2}+R^{2}\right)^{m/2}\right]}{m+2}\right.\nonumber \\
 &  & \left.-\frac{z_{tot}^{m}\left(m+1\right)+\left(z_{tot}^{2}+R^{2}\right)^{m/2}}{m+2}\right\} .
\label{eq:V1V2after}
\end{eqnarray}
Where $_{2}F_{1}$ is the hypergeometric function, with a series
expansion (hypergeometric series) of the form:
\begin{eqnarray}
_{2}F_{1}(a,b,c;x) & = & y(x)=1+\frac{ab}{c}\frac{x}{1!}+\frac{a(a+1)b(b+1)}{c(c+1)}\frac{x^{2}}{2!}+...\label{eq:hypergeometric}\\
 &  & c\neq0,-1,-2,-3,...\nonumber 
\end{eqnarray}
However, for all practical purposes, we can calculate
it numerically, directly from the definition in equation (\ref{eq:V1V2zpzm}), yet
the zero point (determined by $C_{1}$) has to be estimated. At small
separations ($R\ll z_{tot}$), the projected structure functions is well
approximated by a power-law of the form
\begin{equation}
\left\langle \left[V_{z}(\mathbf{X_{1}})-V_{2}(\mathbf{X_{2}})\right]^{2}\right\rangle \approx C_1^{'}R^{m+1}~~~~~~R\ll z_{tot}.
\label{eq:V1V2approx}
\end{equation}
Where $C_1^{'}$ is a constant that can be related to $C_1$ by matching the
zero point of the 2D structure function from the data to a numerical computation
using equation (\ref{eq:V1V2}).
For large separations ($R\gg z_{tot}$) equation
(\ref{eq:V1V2after}), will follow a power law of the form
\begin{equation}
\left\langle \left[V_{z}(\mathbf{X_{1}})-V_{2}(\mathbf{X_{2}})\right]^{2}\right\rangle \approx C_1^{''}R^{m}~~~~~~R\gg z_{tot}.
\label{eq:V1V2approxlargeR}
\end{equation}
Where $C_1^{''}$ is another constant that will depend on the zero point of
velocity fluctuations and on $z_{tot}$ as well.
If we have enough inertial range below $z_{tot}$. The spectral index
$m$ can be obtained from the 2D structure function, and verified
with the power spectrum. An example of a numerically integrated
structure function using equation (\ref{eq:V1V2zpzm}) and the asymptotic
in equations (\ref{eq:V1V2approx}) and (\ref{eq:V1V2approxlargeR})
is shown in Figure \ref{fig:2Dmath} for a structure function index
of $m=2/3.$ This first panel is almost equivalent to Figure (\ref{fig:2Dgauss}),
where we calculate the structure function of integrated data-cubes
(Gaussian), but in Figure \ref{fig:2Dmath} only theoretical expressions
have been used, agreement of the two results provides us with a healthy
verification.

\subsection{Projection of a field with a shallow power spectrum}

Consider a shallow density field (also isotropic), with a 3D structure
function of the form (combining eqs. {[}\ref{eq:Dij_dec}{]} and
{[}\ref{eq:Brshallow}{]})
\begin{equation}
\left\langle \left(\rho(\mathbf{x_{1}})-\rho(\mathbf{x_{2}})\right)^{2}\right\rangle =2\left[B(0)-C_{2}\left(\frac{r}{r_{c}}\right)^{\eta/2}K_{\eta/2}\left(\frac{r}{r_{c}}\right)\right].
\label{eq:DrshallowA}
\end{equation}
The structure function of integrated density (column density) for
this case can be written as
\begin{equation}
\left\langle \left[I(\mathbf{X_{1}})-I(\mathbf{X_{2}})\right]^{2}\right\rangle =2C_{2}z_{tot}\int_{0}^{z_{tot}}\left[-\left(\frac{\sqrt{R^{2}+z^{2}}}{r_{c}}\right)^{\eta/2}K_{\eta/2}\left(\frac{\sqrt{R^{2}+z^{2}}}{r_{c}}\right)+\left(\frac{z}{r_{c}}\right)^{\eta/2}K_{\eta/2}\left(\frac{z}{r_{c}}\right)\right]\ {\rm {\rm d}z,}
\label{eq:I1I2int}
\end{equation}
where for a shallow spectrum $\eta<0.$ This integral is more difficult
to evaluate analytically than the long-wave dominated case, but for
practicality we can solve it numerically. We also find that for small
scales, the result can be well approximated by a simple power-law:
\begin{equation}
\left\langle \left[I(\mathbf{X_{1}})-I(\mathbf{X_{2}})\right]^{2}\right\rangle \approx C_{2}^{'}R^{m+1}~~~~R\ll z_{tot}.
\label{eq:I1I2approx}
\end{equation}
Similarly, the zero point can be found by matching the numerical result of eq.
(\ref{eq:I1I2approx}) to the calculated (2D) structure function from the
data. For large separations ($R\gg z_{tot}$) the resulting structure function
of the integrated field will follow the asymptotic scaling of the
3D structure function:
\begin{equation}
\left\langle \left[I(\mathbf{X_{1}})-I(\mathbf{X_{2}})\right]^{2}\right\rangle \approx constant=C_{2}^{''}R^{0}~~~~R\gg z_{tot}.
\label{eq:I1I2approxlargeR}
\end{equation}
In Figure \ref{fig:2Dmath} we also show an example of the numerical
calculation of an integrated structure function using (\ref{eq:I1I2int})
and the asymptotics in equations (\ref{eq:I1I2approx}) and (\ref{eq:I1I2approxlargeR}),
for an index $\eta=-0.5$ (corresponding to a 3D power spectrum with
an index $\gamma_{3D}=-2.5$).

\section{Power spectra of a field integrated along the line of sight
\label{app:proj_ps}}

The procedure presented here can be repeated analogously also for a
scalar field (e.g. density).
Consider the correlation function a velocity, projected along
the LOS direction (chosen here to be $z$): 
\begin{eqnarray}
\left\langle V_z(\mathbf{R})V_z(\mathbf{X}+\mathbf{R}) \right\rangle & = &
\left\langle \int dz v_z(\mathbf{x})\int dz v_z(\mathbf{x}+(\mathbf{r})\right\rangle \nonumber\\
& = & \iint dz_{1}dz_{2}\left\langle v_{z}(\mathbf{x})v_{z}(\mathbf{x}+\mathbf{r})\right\rangle = \iint dz_{1}dz_{2}\,B_{zz}(\mathbf{r}).
\label{eq:MA}
\end{eqnarray}

\begin{equation}
\left\langle V_z(\mathbf{R})V_z(\mathbf{X}+\mathbf{R}) \right\rangle
= \iint dz_{1}dz_{2}\left\langle v_{z}(\mathbf{x})v_{z}(\mathbf{x}+\mathbf{r})\right\rangle 
= \iint dz_{1}dz_{2}\,B_{zz}(\mathbf{r}).
\label{eq:CF_V}
\end{equation}
Or in terms of the underlying 3D spectrum (inverting equation \ref{eq:Fij})
\begin{equation}
\left\langle V_z(\mathbf{R})V_z(\mathbf{X}+\mathbf{R}) \right\rangle
= \iint dz_{1}dz_{2}\left[\int d^3\mathbf{k}~{\rm e}^{i\mathbf{k\cdot r}}F_{zz}(\mathbf{k}) \right].
\label{eq:CF_V2}
\end{equation}
The two-dimensional power spectrum can be obtained taking the Fourier
transform of the last expression
\begin{equation}
P_{2D,V_z}(\mathbf{K}) = \frac{1}{4\pi^2}\int d^2\mathbf{R}~{\rm e}^{-i\mathbf{K\cdot R}}
\left\{ \iint dz_{1}dz_{2}\left[\int d^3\mathbf{k}~{\rm e}^{i\mathbf{k\cdot r}}F_{zz}(\mathbf{k}) \right] \right\}.
\label{eq:PS_V}
\end{equation}
And using $\mathbf{k}=(\mathbf{K},k_{z})$,
$\mathbf{r}=(\mathbf{R},z_{2}-z_{1})$:
\begin{equation}
P_{2D,V_z}(\mathbf{K}) = \frac{1}{4\pi^2}\int d^2\mathbf{R}~{\rm e}^{-i\mathbf{K\cdot R}}
\left\{ \iint dz_{1}dz_{2}\left[\int d^3\mathbf{k}~{\rm e}^{i\mathbf{K\cdot R}+ik_z(z_2-z_1)}F_{zz}(\mathbf{K},k_z) \right] \right\}.
\label{eq:PS_V2}
\end{equation}
We can interchange the order of integration:
\begin{equation}
P_{2D,V_z}(\mathbf{K}) = \frac{1}{4\pi^{2}}
\iint dz_{1}dz_{2}
\left\{ \int d^{3}\mathbf{k^{'}}
\left[\int d^{2}\mathbf{R}\ {\rm e}^{i(\mathbf{K^{'}-K})\mathbf{\cdot R}+ik_{z}^{'}(z_{2}-z_{1})}F_{zz}(\mathbf{K^{'}},k_{z}^{'})\right]\right\}.
\label{eq:P2DA2}
\end{equation}
After performing the integral over $\mathbf{R}$ one has:
\begin{equation}
P_{2D,V_z}(\mathbf{K})=
\iint dz_{1}dz_{2}\int d^{3}\mathbf{k^{'}}\left[{\rm e}^{ik_{z}^{'}(z_{2}-z_{1})}\delta(\mathbf{K^{'}-K})\  F_{zz}(\mathbf{K^{'}},k_{z}^{'})\right].
\label{eq:P2DA3}
\end{equation}
Now integrate over $\mathbf{K}^{'}$:
\begin{equation}
P_{2D,V_z}(\mathbf{K})=
\iint dz_{1}dz_{2}\left[ \int dk_{z}\ {\rm e}^{ik_{z}^{'}(z_{2}-z_{1})}\  F_{zz}(-\mathbf{K},k_{z}^{'})\right].
\label{eq:P2DA4}
\end{equation}
Changing variables to $z_{+}=(z_{1}+z_{2})/2$, $z_{-}=z_{2}-z_{1}$,
using $F_{zz}(-\mathbf{k})=F_{zz}(\mathbf{k})$,  after of integration in $z_{-}$:
\begin{equation}
P_{2D,Vz}(\mathbf{K})= 4 \pi \int dz_{+}\left[\int dk_{z}\ \delta(k_{z})\  F_{zz}(\mathbf{K},k_{z}^{'})\right].
\label{eq:P2DA5}
\end{equation}
Now we do the integral over $k_{z}^{'}$, then finally over $z_{+}$:
\begin{eqnarray}
P_{2D,V_z}(\mathbf{K}) & =
& 4 \pi \int dz_{+}\left[F_{zz}(\mathbf{K},0)\right]\nonumber \\
& = & 4 \pi z_{tot}\  F_{zz}(\mathbf{K},0),
\label{eq:P2DA6}
\end{eqnarray}
lastly, replacing $k_{z}=0,$ in equation (\ref{eq:FNNFLL})
$F_{zz}(\mathbf{K},0)=F_{NN}(K)$,
we recover from the integrated structure function the spectrum of
the solenoidal component of the velocity:
\begin{equation}
P_{2D,V_Z}(\mathbf{K})=4\pi z_{tot}\  F_{NN}(K).
\label{eq:P2DAfinal}
\end{equation}

\section{Second order structure function of unnormalized velocity centroids
\label{app:algebra}}

Consider the structure function of the unnormalized velocity centroids
(equation \ref{eq:S1S2}):
\begin{equation}
\left\langle \left[S(\mathbf{X_{1}})-S(\mathbf{X_{2}})\right]^{2}\right\rangle =\left\langle \left(\alpha\int v_{z}(\mathbf{x_{1}})\ \rho(\mathbf{x_{1}})\ {\rm d}z-\alpha\int v_{z}(\mathbf{x_{2}})\ \rho(\mathbf{x_{2}})\ {\rm d}z\right)^{2}\right\rangle .
\label{eq:S1S2A}
\end{equation}
As described in \S\S2.2 (and in \citealt {L95}), we can rewrite last
line as in equations (\ref{eq:S1S22}) and (\ref{eq:D}), that we
rewrite here:
\begin{equation}
\left\langle \left[S(\mathbf{X_{1}})-S(\mathbf{X_{2}})\right]^{2}\right\rangle =\alpha^{2}\iint{\rm d}z_{1}{\rm d}z_{2}\left[D(\mathbf{r})-\left.D(\mathbf{r})\right|_{\mathbf{X_{1}}=\mathbf{X_{2}}}\right],
\label{eq:S1S22A}
\end{equation}
where
\begin{equation}
D(\mathbf{r})=\left\langle \left(v_{z}(\mathbf{x_{1}})\rho(\mathbf{x_{1}})-v_{z}(\mathbf{x_{2}})\rho(\mathbf{x_{2}})\right)^{2}\right\rangle .
\label{eq:DA}
\end{equation}
Using the definitions
$v_{z}=v_{0}+\tilde{v}_{z}$, $\rho=\rho_{0}+\tilde{\rho}$
in (\ref{eq:DA}) we have
\begin{eqnarray}
D(\mathbf{r}) & = & \left\langle \left[v_{0}\rho_{0}+\tilde{v}_{z}(\mathbf{x_{1}})\rho_{0}+v_{0}\tilde{\rho}(\mathbf{x_{1}})+\tilde{v}_{z}(\mathbf{x_{1}})\tilde{\rho}(\mathbf{x_{1}})-v_{0}\rho_{0}+\tilde{v}_{z}(\mathbf{x_{2}})\rho_{0}+v_{0}\tilde{\rho}(\mathbf{x_{2}})+\tilde{v}_{z}(\mathbf{x_{2}})\tilde{\rho}(\mathbf{x_{2}})\right]^{2}\right\rangle \nonumber \\
 & = & \left\langle \left\{ v_{0}\left[\tilde{\rho}(\mathbf{x_{1}})-\tilde{\rho}(\mathbf{x_{2}})\right]+\rho_{0}\left[\tilde{v}_{z}(\mathbf{x_{1}})-\tilde{v}_{z}(\mathbf{x_{2}})\right]+\left[\tilde{v}_{z}(\mathbf{x_{1}})\tilde{\rho}(\mathbf{x_{1}})-\tilde{v}_{z}(\mathbf{x_{1}})\tilde{\rho}(\mathbf{x_{2}})\right]\right\} ^{2}\right\rangle .
\label{eq:DA1}
\end{eqnarray}
Expansion of the last expression yields
\begin{eqnarray}
D(\mathbf{r}) & = & \left\langle v_{0}^{2}\left[\rho(\mathbf{x_{1}})-\rho(\mathbf{x_{2}})\right]^{2}+\rho_{0}^{2}\left[v_{z}(\mathbf{x_{1}})-v_{z}(\mathbf{x_{2}})\right]^{2}+\left[\tilde{v}_{z}(\mathbf{x_{1}})\tilde{\rho}(\mathbf{x_{1}})-\tilde{v}_{z}(\mathbf{x_{2}})\tilde{\rho}(\mathbf{x_{2}})\right]\right.\nonumber \\
 &  & +2v_{0}\rho_{0}\left[\rho(\mathbf{x_{1}})-\rho(\mathbf{x_{2}})\right]\left[v_{z}(\mathbf{x_{1}})-v_{z}(\mathbf{x_{2}})\right]+2\rho_{0}\left[v_{z}(\mathbf{x_{1}})-v_{z}(\mathbf{x_{2}})\right]\left[\tilde{v}_{z}(\mathbf{x_{1}})\tilde{\rho}(\mathbf{x_{1}})-\tilde{v}_{z}(\mathbf{x_{2}})\tilde{\rho}(\mathbf{x_{2}})\right]\nonumber \\
 &  & \left.+2v_{0}\left[v_{z}(\mathbf{x_{1}})-v_{z}(\mathbf{x_{2}})\right]\left[\tilde{v}_{z}(\mathbf{x_{1}})\tilde{\rho}(\mathbf{x_{1}})-\tilde{v}_{z}(\mathbf{x_{2}})\tilde{\rho}(\mathbf{x_{2}})\right]\right\rangle \nonumber \\
 & = & v_{0}^{2}\left\langle \left[\rho(\mathbf{x_{1}})-\rho(\mathbf{x_{2}})\right]^{2}\right\rangle +\rho_{0}^{2}\left\langle \left[v_{z}(\mathbf{x_{1}})-v_{z}(\mathbf{x_{2}})\right]^{2}\right\rangle +\left\langle \left[\tilde{v}_{z}(\mathbf{x_{1}})\tilde{\rho}(\mathbf{x_{1}})-\tilde{v}_{z}(\mathbf{x_{2}})\tilde{\rho}(\mathbf{x_{2}})\right]^{2}\right\rangle +\nonumber \\
 &  & 2v_{0}\rho_{0}\left\langle \left[\rho(\mathbf{x_{1}})-\rho(\mathbf{x_{2}})\right]\left[v_{z}(\mathbf{x_{1}})-v_{z}(\mathbf{x_{2}})\right]\right\rangle +2\rho_{0}\left\langle \left[v_{z}(\mathbf{x_{1}})-v_{z}(\mathbf{x_{2}})\right]\left[\tilde{v}_{z}(\mathbf{x_{1}})\tilde{\rho}(\mathbf{x_{1}})-\tilde{v}_{z}(\mathbf{x_{2}})\tilde{\rho}(\mathbf{x_{2}})\right]\right\rangle \nonumber \\
 &  & +2v_{0}\left\langle \left[\rho(\mathbf{x_{1}})-\rho(\mathbf{x_{2}})\right]\left[\tilde{v}_{z}(\mathbf{x_{1}})\tilde{\rho}(\mathbf{x_{1}})-\tilde{v}_{z}(\mathbf{x_{2}})\tilde{\rho}(\mathbf{x_{2}})\right]\right\rangle .
\label{eq:dexp}
\end{eqnarray}
Now consider the cross-terms one by one, the third term in equation
(\ref{eq:dexp}) is
\begin{equation}
\left\langle \left[\tilde{v}_{z}(\mathbf{x_{1}})\tilde{\rho}(\mathbf{x_{1}})-\tilde{v}_{z}(\mathbf{x_{2}})\tilde{\rho}(\mathbf{x_{2}})\right]^{2}\right\rangle =\left\langle \tilde{v}_{z}^{2}(\mathbf{x_{1}})\tilde{\rho}^{2}(\mathbf{x_{1}})\right\rangle +\left\langle \tilde{v}_{z}^{2}(\mathbf{x_{2}})\tilde{\rho}^{2}(\mathbf{x_{2}})\right\rangle -2\left\langle \tilde{v}_{z}(\mathbf{x_{1}})\tilde{\rho}(\mathbf{x_{1}})\tilde{v}_{z}(\mathbf{x_{2}})\tilde{\rho}(\mathbf{x_{2}})\right\rangle .
\label{eq:third}
\end{equation}
Relating the fourth order moments with the second order using the
Millionshikov hypothesis
\footnote{Evidently, this hypothesis is identically true for Gaussian fields,
strongly non Gaussian fields may show deviations.
} (see \citealt{MY75}) 
$\langle h_{1}h_{2}h_{3}h_{4}\rangle\approx\langle h_{1}h_{2}\rangle\langle h_{3}h_{4}\rangle+\langle h_{1}h_{3}\rangle\langle h_{2}h_{4}\rangle+\langle h_{1}h_{4}\rangle\langle h_{2}h_{3}\rangle$
we have
\begin{mathletters}
\begin{eqnarray}
\left\langle \tilde{v}_{z}^{2}(\mathbf{x_{1}})\tilde{\rho}^{2}(\mathbf{x_{1}})\right\rangle =\left\langle \tilde{v}_{z}(\mathbf{x_{1}})\tilde{\rho}(\mathbf{x_{1}})\tilde{v}_{z}(\mathbf{x_{1}})\tilde{\rho}(\mathbf{x_{1}})\right\rangle  & \approx & \left\langle \tilde{v}_{z}^{2}(\mathbf{x_{1}})\right\rangle \left\langle \tilde{\rho}^{2}(\mathbf{x_{1}})\right\rangle +2\left\langle \tilde{v}_{z}(\mathbf{x_{1}})\tilde{\rho}(\mathbf{x_{1}})\right\rangle ^{2},\label{eq:third1a}\\
\left\langle \tilde{v}_{z}^{2}(\mathbf{x_{2}})\tilde{\rho}^{2}(\mathbf{x_{2}})\right\rangle =\left\langle \tilde{v}_{z}(\mathbf{x_{2}})\tilde{\rho}(\mathbf{x_{2}})\tilde{v}_{z}(\mathbf{x_{2}})\tilde{\rho}(\mathbf{x_{2}})\right\rangle  & \approx & \left\langle \tilde{v}_{z}^{2}(\mathbf{x_{1}})\right\rangle \left\langle \tilde{\rho}^{2}(\mathbf{x_{2}})\right\rangle +2\left\langle \tilde{v}_{z}(\mathbf{x_{2}})\tilde{\rho}(\mathbf{x_{2}})\right\rangle ^{2},\label{eq:third1b}\\
2\left\langle \tilde{v}_{z}(\mathbf{x_{1}})\tilde{\rho}(\mathbf{x_{1}})\tilde{v}_{z}(\mathbf{x_{2}})\tilde{\rho}(\mathbf{x_{2}})\right\rangle  & \approx & 2\langle\tilde{v}_{z}(\mathbf{x_{1}})\tilde{\rho}(\mathbf{x_{1}})\rangle\langle\tilde{v}_{z}(\mathbf{x_{2}})\tilde{\rho}_{2}(\mathbf{x_{2}})\rangle\nonumber \\
 &  & +2\langle\tilde{v}_{z}(\mathbf{x_{1}})\tilde{v}_{z}(\mathbf{x_{2}})\rangle\langle\tilde{\rho}(\mathbf{x_{1}})\tilde{\rho}(\mathbf{x_{2}})\rangle\nonumber \\
 &  & +2\langle\tilde{v}_{z}(\mathbf{x_{1}})\tilde{\rho}(\mathbf{x_{2}})\rangle\langle\tilde{v}_{z}(\mathbf{x_{2}})\tilde{\rho}(\mathbf{x_{1}})\rangle.\label{eq:third1c}
\end{eqnarray}
\end{mathletters}
If we combine the last three lines using
\begin{mathletters}
\begin{eqnarray}
\left\langle \left[v_{z}(\mathbf{x_{1}})-v_{z}(\mathbf{x_{2}})\right]^{2}\right\rangle  & = & \left\langle \tilde{v}_{z}^{2}(\mathbf{x_{1}})\right\rangle -2\left\langle \tilde{v}_{z}(\mathbf{x_{1}})\tilde{v}_{z}(\mathbf{x_{2}})\right\rangle +\left\langle \tilde{v}_{z}^{2}(\mathbf{x_{2}})\right\rangle \nonumber \\
 & = & 2\left\langle \tilde{v_{z}}^{2}\right\rangle -2\left\langle \tilde{v}_{z}(\mathbf{x_{1}})\tilde{v}_{z}(\mathbf{x_{2}})\right\rangle,
\label{eq:third2a}\\
\left\langle \tilde{v}_{z}(\mathbf{x_{1}})\tilde{v}_{z}(\mathbf{x_{2}})\right\rangle  & = & \left\langle \tilde{v}_{z}^{2}\right\rangle -\frac{1}{2}\left\langle \left[v_{z}(\mathbf{x_{1}})-v_{z}(\mathbf{x_{2}})\right]^{2}\right\rangle,
\label{eq:third2b}\\
\left\langle \tilde{\rho}(\mathbf{x_{1}})\tilde{\rho}(\mathbf{x_{2}})\right\rangle  & = & \left\langle \tilde{\rho}^{2}\right\rangle -\frac{1}{2}\left\langle \left[\rho(\mathbf{x_{1}})-\rho(\mathbf{x_{2}})\right]^{2}\right\rangle.
\label{eq:third2c}
\end{eqnarray}
\end{mathletters}
Together with $\langle\tilde{v}_{z}(\mathbf{x_{1}})\tilde{\rho}(\mathbf{x_{1}})\rangle=\langle\tilde{v}_{z}(\mathbf{x_{2}})\tilde{\rho}(\mathbf{x_{2}})\rangle=0$,
$\langle\tilde{v}_{z}^{2}(\mathbf{x_{1}})\rangle=\langle\tilde{v}_{z}^{2}(\mathbf{x_{1}})\rangle=\langle\tilde{v}_{z}^{2}\rangle$,
and $\langle\tilde{\rho}^{2}(\mathbf{x_{1}})\rangle=\langle\tilde{\rho}^{2}(\mathbf{x_{2}})\rangle=\langle\tilde{\rho}^{2}\rangle$
equation (\ref{eq:third}) reduces to
\begin{eqnarray}
\left\langle \left[\tilde{v}_{z}(\mathbf{x_{1}})\tilde{\rho}(\mathbf{x_{1}})-\tilde{v}_{z}(\mathbf{x_{2}})\tilde{\rho}(\mathbf{x_{2}})\right]^{2}\right\rangle  & \approx & 2\left\langle \tilde{v}_{z}^{2}\right\rangle \left\langle \tilde{\rho}^{2}\right\rangle -2\langle\tilde{v}_{z}(\mathbf{x_{1}})\tilde{\rho}(\mathbf{x_{2}})\rangle\langle\tilde{v}_{z}(\mathbf{x_{2}})\tilde{\rho}(\mathbf{x_{1}})\rangle\nonumber \\
 &  & -2\left\{ \left[\left\langle \tilde{v_{z}}^{2}\right\rangle -\frac{1}{2}\left\langle \left[v_{z}(\mathbf{x_{1}})-v_{z}(\mathbf{x_{2}})\right]^{2}\right\rangle \right]\right.\nonumber \\
 &  & \left.\left[\left\langle \tilde{\rho}^{2}\right\rangle -\frac{1}{2}\left\langle \left[\rho(\mathbf{x_{1}})-\rho(\mathbf{x_{2}})\right]^{2}\right\rangle \right]\right\} .
\label{eq:third3}
\end{eqnarray}
To treat terms of the form $\langle\tilde{\rho}\tilde{v}_{z}\rangle$
we can generalize the correlation function defined in equation (\ref{eq:Bij}),
considering a four-dimensional vector of the form
$[v_{x}(\mathbf{x}),v_{y}(\mathbf{x}),v_{z}(\mathbf{x}),\rho(\mathbf{x})]$.
The resulting cross-correlation between the $z$ component of
the velocity and the density is 
\begin{equation}
B_{z\rho}(\mathbf{r})=\langle v_{z}(\mathbf{x_{1}})\rho(\mathbf{x_{2}})\rangle.
\label{eq:Bvrho}
\end{equation}
Similarly to the derivation of equation (\ref{eq:BNNBLL}) it can
be shown that for an isotropic (four-dimensional)
field $B_{j\rho}(r)=B_{L\rho}(r)r_{j}/r$, and
$B_{\rho j}(r)=B_{\rho L}(r)r_{j}/r=-B_{L\rho}(r)r_{j}/r$.
Thus
$\langle\tilde{v}_{z}(\mathbf{x_{1}})\tilde{\rho}(\mathbf{x_{2}})\rangle=-\langle\tilde{v}_{z}(\mathbf{x_{2}})\tilde{\rho}(\mathbf{x_{1}})\rangle$,
and equation (\ref{eq:third3}) simplifies to
\begin{eqnarray}
\left\langle \left[\tilde{v}_{z}(\mathbf{x_{1}})\tilde{\rho}(\mathbf{x_{1}})-\tilde{v}_{z}(\mathbf{x_{2}})\tilde{\rho}(\mathbf{x_{2}})\right]^{2}\right\rangle  & \approx & 2\langle\tilde{v}_{z}(\mathbf{x_{1}})\tilde{\rho}(\mathbf{x_{2}})\rangle^{2}+\left\langle \tilde{v_{z}}^{2}\right\rangle \left\langle \left[\rho(\mathbf{x_{1}})-\rho(\mathbf{x_{2}})\right]^{2}\right\rangle \nonumber \\
 &  & +\left\langle \tilde{\rho}^{2}\right\rangle \left\langle \left[v_{z}(\mathbf{x_{1}})-v_{z}(\mathbf{x_{2}})\right]^{2}\right\rangle \nonumber \\
 &  & -\frac{1}{2}\left\langle \left[v_{z}(\mathbf{x_{1}})-v_{z}(\mathbf{x_{2}})\right]^{2}\right\rangle \left\langle \left[\rho(\mathbf{x_{1}})-\rho(\mathbf{x_{2}})\right]^{2}\right\rangle .
\label{eq:third_final}
\end{eqnarray}
The next term of equation  (\ref{eq:dexp})
\begin{equation}
2v_{0}\rho_{0}\left\langle \left[\rho(\mathbf{x_{1}})-\rho(\mathbf{x_{2}})\right]\left[v_{z}(\mathbf{x_{1}})-v_{z}(\mathbf{x_{2}})\right]\right\rangle =0,
\label{eq:fourth}
\end{equation}
as shown explicitly in \citet{MY75}.
The fifth term in equation (\ref{eq:dexp}) is:
\begin{eqnarray}
2\rho_{0}\left\langle \left[\tilde{v}_{z}(\mathbf{x_{1}})-\tilde{v}_{z}(\mathbf{x_{2}})\right]\left[\tilde{v}_{z}(\mathbf{x_{1}})\tilde{\rho}(\mathbf{x_{1}})-\tilde{v}_{z}(\mathbf{x_{2}})\tilde{\rho}(\mathbf{x_{2}})\right]\right\rangle  & = & 2\rho_{0}\left\langle \tilde{v}_{z}(\mathbf{x_{1}})\tilde{\rho}(\mathbf{x_{1}})\right\rangle +2\rho_{0}\left\langle \tilde{v}_{z}(\mathbf{x_{2}})\tilde{\rho}(\mathbf{x_{2}})\right\rangle \nonumber \\
 &  & -2\rho_{0}\left\langle \tilde{v}_{z}(\mathbf{x_{1}})\tilde{v}_{z}(\mathbf{x_{2}})\tilde{\rho}(\mathbf{x_{1}})\right\rangle \nonumber \\
 &  & -2\rho_{0}\left\langle \tilde{v}_{z}(\mathbf{x_{1}})\tilde{v}_{z}(\mathbf{x_{2}})\tilde{\rho}(\mathbf{x_{2}})\right\rangle 
\label{eq:fifth}
\end{eqnarray}
For the terms with correlations of third order we need to introduce
the so called two-point third order moments
\begin{equation}
B_{ij,l}(\mathbf{r})=\langle u_{i}(\mathbf{x_{1}})u_{j}(\mathbf{x_{1}})u_{l}(\mathbf{x_{2}})\rangle.
\label{eq:Bijl}
\end{equation}
Which can be generalized to include cross-correlations of density
and velocity as explained for equation (\ref{eq:Bvrho}). Analogously
to the second order cross-correlations, considering an isotropic (four-dimensional)
field, and decomposing it in terms of longitudinal and
lateral components. It can be shown (for more details we refer the
reader to \citealt{MY75}) that
\begin{mathletters}
\begin{eqnarray}
\left\langle \tilde{v}_{z}(\mathbf{x_{1}})\tilde{v}_{z}(\mathbf{x_{2}})\tilde{\rho}(\mathbf{x_{1}})\right\rangle  & = & \left\langle \tilde{v}_{z}(\mathbf{x_{1}})\tilde{v}_{z}(\mathbf{x_{2}})\tilde{\rho}(\mathbf{x_{2}})\right\rangle ,\label{eq:v1v1rho1}\\
\left\langle \tilde{v}_{z}(\mathbf{x_{1}})\tilde{\rho}_{z}(\mathbf{x_{1}})\tilde{\rho}(\mathbf{x_{2}})\right\rangle  & = & -\left\langle \tilde{v}_{z}(\mathbf{x_{1}})\tilde{\rho}_{z}(\mathbf{x_{1}})\tilde{\rho}(\mathbf{x_{2}})\right\rangle .
\label{eq:rho1rho2v1}
\end{eqnarray}
\end{mathletters}
Equation (\ref{eq:v1v1rho1}), combined with $\langle\tilde{\rho}(\mathbf{x_{1}})\tilde{v}_{z}^{2}(\mathbf{x_{1}})\rangle=\langle\tilde{\rho}(\mathbf{x_{2}})\tilde{v}_{z}^{2}(\mathbf{x_{2}})\rangle$
reduces equation (\ref{eq:fifth}) to:
\begin{eqnarray}
2\rho_{0}\left\langle \left[\tilde{v}_{z}(\mathbf{x_{1}})-\tilde{v}_{z}(\mathbf{x_{2}})\right]\left[\tilde{v}_{z}(\mathbf{x_{1}})\tilde{\rho}(\mathbf{x_{1}})-\tilde{v}_{z}(\mathbf{x_{2}})\tilde{\rho}(\mathbf{x_{2}})\right]\right\rangle  & = & 4\rho_{0}\left\langle \tilde{\rho}(\mathbf{x_{1}})\tilde{v}_{z}^{2}(\mathbf{x_{1}})\right\rangle \nonumber \\
 &  & -4\rho_{0}\left\langle \tilde{v}_{z}(\mathbf{x_{1}})\tilde{v}_{z}(\mathbf{x_{2}})\tilde{\rho}(\mathbf{x_{1}})\right\rangle .
\label{eq:fifth_final}
\end{eqnarray}
Similarly the last term in equation (\ref{eq:dexp}) can be written
\begin{eqnarray}
2v_{0}\left\langle \left[\rho(\mathbf{x_{1}})-\rho(\mathbf{x_{2}})\right]\left[\tilde{v}_{z}(\mathbf{x_{1}})\tilde{\rho}(\mathbf{x_{1}})-\tilde{v}_{z}(\mathbf{x_{2}})\tilde{\rho}(\mathbf{x_{2}})\right]\right\rangle  & = & 2v_{0}\left\langle \tilde{v}_{z}(\mathbf{x_{1}})\tilde{\rho}^{2}(\mathbf{x_{1}})\right\rangle \nonumber \\
 &  & +2v_{0}\left\langle \tilde{v}_{z}(\mathbf{x_{2}})\tilde{\rho}^{2}(\mathbf{x_{2}})\right\rangle \nonumber \\
 &  & -2v_{0}\left\langle \tilde{\rho}(\mathbf{x_{1}})\tilde{\rho}(\mathbf{x_{2}})\tilde{v}_{z}(\mathbf{x_{1}})\right\rangle \nonumber \\
 &  & -2v_{0}\left\langle \tilde{\rho}(\mathbf{x_{1}})\tilde{\rho}(\mathbf{x_{2}})\tilde{v}_{z}(\mathbf{x_{2}})\right\rangle .
\label{eq:sixth}
\end{eqnarray}
But in this case ,
$\left\langle \tilde{v}_{z}(\mathbf{x_{1}})\tilde{\rho}^{2}(\mathbf{x_{1}})\right\rangle =\left\langle \tilde{v}_{z}(\mathbf{x_{2}})\tilde{\rho}^{2}(\mathbf{x_{2}})\right\rangle =0$,
combined with equation (\ref{eq:rho1rho2v1}) , yields
\begin{equation}
2v_{0}\left\langle \left[\rho(\mathbf{x_{1}})-\rho(\mathbf{x_{2}})\right]\left[\tilde{v}_{z}(\mathbf{x_{1}})\tilde{\rho}(\mathbf{x_{1}})-\tilde{v}_{z}(\mathbf{x_{2}})\tilde{\rho}(\mathbf{x_{2}})\right]\right\rangle =0.
\label{eq:sixthfinal}
\end{equation}
Finally, combining (\ref{eq:dexp}, \ref{eq:third_final} , \ref{eq:fourth}
, \ref{eq:fifth_final}, \ref{eq:sixthfinal})
\begin{eqnarray}
D(\mathbf{r}) & \approx & v_{0}^{2}\left\langle \left[\rho(\mathbf{x_{1}})-\rho(\mathbf{x_{2}})\right]^{2}\right\rangle +\rho_{0}^{2}\left\langle \left[v_{z}(\mathbf{x_{1}})-v_{z}(\mathbf{x_{2}})\right]^{2}\right\rangle +2\left\langle \tilde{v}_{z}(\mathbf{x_{1}})\tilde{\rho}(\mathbf{x_{2}})\right\rangle ^{2}\nonumber \\
 &  & +\left\langle \tilde{v}_{z}^{2}\right\rangle \left\langle \left[\rho(\mathbf{x_{1}})-\rho(\mathbf{x_{2}})\right]^{2}\right\rangle +\left\langle \tilde{\rho}^{2}\right\rangle \left\langle \left[v_{z}(\mathbf{x_{1}})-v_{z}(\mathbf{x_{2}})\right]^{2}\right\rangle \nonumber \\
 &  & -\frac{1}{2}\left\langle \left[v_{z}(\mathbf{x_{1}})-v_{z}(\mathbf{x_{2}})\right]^{2}\right\rangle \left\langle \left[\rho(\mathbf{x_{1}})-\rho(\mathbf{x_{2}})\right]^{2}\right\rangle +4\rho_{0}\left\langle \tilde{\rho}(\mathbf{x_{1}})\tilde{v}_{z}^{2}(\mathbf{x_{1}})\right\rangle \nonumber \\
 &  & -4\rho_{0}\left\langle \tilde{v}_{z}(\mathbf{x_{1}})\tilde{v}_{z}(\mathbf{x_{2}})\tilde{\rho}(\mathbf{x_{1}})\right\rangle .
\label{eq:D_all}
\end{eqnarray}
Grouping some terms:
\begin{eqnarray}
D(\mathbf{r}) & \approx & \left\langle \tilde{\rho}^{2}+\rho_{0}^{2}\right\rangle \left\langle \left[v_{z}(\mathbf{x_{1}})-v_{z}(\mathbf{x_{2}})\right]^{2}\right\rangle +\left\langle \tilde{v}^{2}+v_{0}^{2}\right\rangle \left\langle \left[\rho(\mathbf{x_{1}})-\rho(\mathbf{x_{2}})\right]^{2}\right\rangle \nonumber \\
 &  & -\frac{1}{2}\left\langle \left[v_{z}(\mathbf{x_{1}})-v_{z}(\mathbf{x_{2}})\right]^{2}\right\rangle \left\langle \left[\rho(\mathbf{x_{1}})-\rho(\mathbf{x_{2}})\right]^{2}\right\rangle +2\left\langle \tilde{v}_{z}(\mathbf{x_{1}})\tilde{\rho}(\mathbf{x_{2}})\right\rangle ^{2}\nonumber \\
 &  & +4\rho_{0}\left\langle \tilde{\rho}(\mathbf{x_{1}})\tilde{v}_{z}^{2}(\mathbf{x_{1}})\right\rangle -4\rho_{0}\left\langle \tilde{v}_{z}(\mathbf{x_{1}})\tilde{v}_{z}(\mathbf{x_{2}})\tilde{\rho}(\mathbf{x_{1}})\right\rangle .
\label{eq:D_group}
\end{eqnarray}
And lastly, with $\langle\rho^{2}\rangle=\langle\tilde{\rho}^2+\rho_{0}^{2}\rangle$,
and $\langle v^{2}\rangle=\langle\tilde{v}^{2}+v_{0}^{2}\rangle$
we arrive to equations (\ref{eq:D2}) and (\ref{eq:c}):
\begin{eqnarray}
D(\mathbf{r}) & \approx & \left\langle \rho^{2}\right\rangle \left\langle \left[v_{z}(\mathbf{x_{1}})-v_{z}(\mathbf{x_{2}})\right]^{2}\right\rangle +\left\langle v^{2}\right\rangle \left\langle \left[\rho(\mathbf{x_{1}})-\rho(\mathbf{x_{2}})\right]^{2}\right\rangle \nonumber \\
 &  & -\frac{1}{2}\left\langle \left[v_{z}(\mathbf{x_{1}})-v_{z}(\mathbf{x_{2}})\right]^{2}\right\rangle \left\langle \left[\rho(\mathbf{x_{1}})-\rho(\mathbf{x_{2}})\right]^{2}\right\rangle +c(\mathbf{r}),
\label{eq:D2A}
\end{eqnarray}
with
\begin{equation}
c^{'}(\mathbf{r})=2\left\langle \tilde{v}_{z}(\mathbf{x_{1}})\tilde{\rho}(\mathbf{x_{2}})\right\rangle ^{2}+4\rho_{0}\left\langle \tilde{\rho}(\mathbf{x_{1}})\tilde{v}_{z}^{2}(\mathbf{x_{1}})\right\rangle -4\rho_{0}\left\langle \tilde{v}_{z}(\mathbf{x_{1}})\tilde{v}_{z}(\mathbf{x_{2}})\tilde{\rho}(\mathbf{x_{1}})\right\rangle .
\label{eq:cA}
\end{equation}
We should note that the second term in equation (\ref{eq:cA}) has no
contribution to the structure function of centroids because when
substituted into equation (\ref{eq:S1S22A}) cancels 
($\langle\tilde{\rho}(\mathbf{x_1})\tilde{v_z}(\mathbf{x_1})^2\rangle = \langle\tilde{\rho}(\mathbf{x_1})\tilde{v_z}(\mathbf{x_1})^2\rangle\vert_{\mathbf{X_1}=\mathbf{X_2}}$). Therefore we omitted this term in the
body of the paper. Moreover, this term does not appear in the
correlation function of unnormalized centroids, not including it in
the definition of $c(\mathbf{r})$ shows better the symmetry between
structure functions with correlation functions and spectra.

\section{Centroids and the ``cross-term'' for power-law statistics
\label{app:cross_term}}

We have already discussed the differences between power-law long-wave
and short-wave dominated fields ({\it steep} and {\it shallow} spectra).
In particular, the fact that although in all cases power spectra are
power-laws, but structure functions are only power-laws for steep
spectra whereas correlation functions are only power-laws (at small
scales) for shallow spectra. In this section we will sketch the
expectations for the cross-term ($I3[\mathbf{R}]$) for the
idealized case of infinite power-law spectra, considering all the
different shallow and steep combinations.

\subsection{Steep density and steep velocity}

In this case the structure function of both fields is well described
by a power law. Consider power-law isotropic underlying statistics
of the form 
$d_{v_{z}}(\mathbf{r})=C_{v}r^{m}$, $d_{\rho}(\mathbf{r})=C_{\rho}r^{n}$,
where $m,\  n>0$, and $C_{v}$, $C_{\rho}$ are constants. Here the
cross-term will be
\begin{equation}
I3(\mathbf{R})=-\frac{1}{2}\alpha^{2}C_{v}C_{\rho}\iint{\rm d}z_{1}{\rm d}z_{2}\left[r^{m+n}-\left.r^{m+n}\right|_{\mathbf{X_{1}=\mathbf{X_{2}}}}\right],
\label{eq:I3stst}
\end{equation}
which is analogous to the projection of a steep field with a spectral
index $\gamma_{3D}=-(m+n)-3$. The resulting cross-term is negative
and steeper than both density and velocity. Thus at sufficiently small
scales its contribution can be safely neglected compared to the integrated
density and velocity.

\subsection{Shallow density and steep velocity}

Here the structure function of velocity is a power-law
$d_{v_{z}}(\mathbf{r})\propto C_{v}r^{m}$,
with $m>0$ and $C_{v}$ constant. For small separations (relative
to a critical scale $r_{c}$ as discussed in the main body of the
paper) the density fluctuations will have a power-law correlation
function $\langle\tilde{\rho}(\mathbf{x_{1}})\tilde{\rho}(\mathbf{x_{2}})\rangle\approx C_{\rho}r^{n}$,
where $n<0$ and $C_{\rho}$ is constant. Therefore the structure
function of density can be presented as
$d_{\rho}(\mathbf{r})\approx2(\langle\tilde{\rho}^{2}\rangle-C_{\rho}r^{n})$.
For this combination of indices the cross-term becomes
\begin{equation}
I3(\mathbf{R})\approx-\left\langle \tilde{\rho}^{2}\right\rangle \left\langle \left[V(\mathbf{X_{1}})-V(\mathbf{X_{2}})\right]^{2}\right\rangle +\alpha^{2}C_{v}C_{\rho}\iint{\rm d}z_{1}{\rm d}z_{2}\left[r^{m+n}-\left.r^{m+n}\right|_{\mathbf{X_{1}=\mathbf{X_{2}}}}\right].
\label{eq:I3shst}
\end{equation}

The first term on the right, although negative has the same slope
as $I2(\mathbf{R})$. The two terms inside the integrals are equivalent
to the projection of a field with an spectral index $\gamma_{3D}=-(m+n)-3$.
Which in this case is going to be shallower than the velocity, and
could be either positive or negative. Positive for $m+n>0$, and negative
for $m+n<0$. If the level of density fluctuations $\langle\tilde{\rho}^{2}\rangle$
is large the first term in the right of eq. (\ref{eq:I3shst}) will
dominate the slope of $I3(\mathbf{R})$. However such term will be
canceled with part of $I2(\mathbf{R})$, changing effectively the
weighting of the integrated velocity from $\langle\rho^{2}\rangle=\rho_{0}^{2}+\langle\tilde{\rho}^{2}\rangle$
to $\rho_{0}^{2}$. The resulting structure function of unnormalized
centroids can be written as\footnote{Equations (\ref{eq:s1s1shst},
\ref{eq:s1s1stsh}, and \ref{eq:s1s1shsh}) coincide with the decomposition we derived in \citet{OELS05}.}

\begin{eqnarray}
\left\langle \left[S(\mathbf{X_{1}})-S(\mathbf{X_{2}})\right]^{2}\right\rangle  & \approx & \left(v_{0}^{2}+\left\langle \tilde{v}_{z}^{2}\right\rangle \right)\left\langle \left[I(\mathbf{X_{1}})-I(\mathbf{X_{2}})\right]^{2}\right\rangle +\rho_{0}^{2}\left\langle \left[V(\mathbf{X_{1}})-V(\mathbf{X_{2}})\right]^{2}\right\rangle \nonumber \\
 &  & +\alpha^{2}C_{v}C_{\rho}\iint{\rm d}z_{1}{\rm d}z_{2}\left[r^{m+n}-\left.r^{m+n}\right|_{\mathbf{X_{1}=\mathbf{X_{2}}}}\right]+I4(\mathbf{R}).
\label{eq:s1s1shst}
\end{eqnarray}
The exact contribution of the terms inside the double integrals to
$I3(\mathbf{R})$, and in general the importance of the cross-term
to the centroid statistics will depend on the constants $C_{v}$ and
$C_{\rho}$. If at large scales the magnitude of the term containing
the structure function of integrated velocity and the cross-terms
are comparable, then at small scales the centroids could fail tracing
velocity.

\subsection{Steep Density and shallow velocity}

This case is somewhat analogous to that described above. Consider
for the density a power-law structure function
$d_{\rho}(\mathbf{r})\propto C_{\rho}r^{n}$,
with $n>0$ and $C_{\rho}$ constant. For the velocity, a power-law
correlation function that translates into a structure function
$d_{v_{z}}(\mathbf{r})\approx2(\langle \tilde{v}_{z}^{2} \rangle-C_{v}r^{m})$
with $m<0$ and $C_{v}$ constant. We have already discussed that
a shallow velocity is not physically motivated. However if we have
a steep velocity field without infinite power-law behavior we could have a
structure function that resembles the shallow case. For
instance if the structure function ``saturates'' at large separations
it will be similar to that of a shallow field, which
grows rapidly at small scales and then flattens at large scales. The
corresponding cross-term is
\begin{equation}
I3(\mathbf{R})\approx-\left\langle \tilde{v}_{z}^{2}\right\rangle \left\langle \left[I(\mathbf{X_{1}})-I(\mathbf{X_{2}})\right]^{2}\right\rangle +\alpha^{2}C_{v}C_{\rho}\iint{\rm d}z_{1}{\rm d}z_{2}\left[r^{m+n}-\left.r^{m+n}\right|_{\mathbf{X_{1}=\mathbf{X_{2}}}}\right].
\label{eq:I3stsh}
\end{equation}
It has a component that scales as the column density (first term on
the right). The remaining component could be positive (if $m+n>0$) or
negative (if $m+n<0$). Unnormalized centroids for these indices
can be expressed as
\begin{eqnarray}
\left\langle \left[S(\mathbf{X_{1}})-S(\mathbf{X_{2}})\right]^{2}\right\rangle  & \approx & v_{0}^{2}\left\langle \left[I(\mathbf{X_{1}})-I(\mathbf{X_{2}})\right]^{2}\right\rangle +\left(\rho_{0}^{2}+\left\langle \tilde{\rho}^{2}\right\rangle \right)\left\langle \left[V(\mathbf{X_{1}})-V(\mathbf{X_{2}})\right]^{2}\right\rangle \nonumber \\
 &  & +\alpha^{2}C_{v}C_{\rho}\iint{\rm d}z_{1}{\rm d}z_{2}\left[r^{m+n}-\left.r^{m+n}\right|_{\mathbf{X_{1}=\mathbf{X_{2}}}}\right]+I4(\mathbf{R}).
\label{eq:s1s1stsh}
\end{eqnarray}

\subsection{Shallow density and shallow velocity}

For this combination of indices we can consider power-law correlation
functions (for small separations). Yielding structure functions of the form
$d_{\rho} (\mathbf{r})\approx2(\langle \tilde{\rho}^{2}\rangle-C_{\rho}r^{n})$,
$d_{v_{z}}(\mathbf{r})\approx2(\langle \tilde{v}_{z}^{2}\rangle-C_{v}r^{m})$,
with $m,\  n<0$, and $C_{\rho},\  C_{v}$ constants. As a result
the cross-term becomes
\begin{eqnarray}
I3(\mathbf{R}) & \approx & -\left\langle \tilde{v}_{z}^{2}\right\rangle \left\langle \left[I(\mathbf{X_{1}})-I(\mathbf{X_{2}})\right]^{2}\right\rangle -\left\langle \tilde{\rho}^{2}\right\rangle \left\langle \left[V(\mathbf{X_{1}})-V(\mathbf{X_{2}})\right]^{2}\right\rangle \nonumber \\
 &  & -2\alpha^{2}C_{v}C_{\rho}\iint{\rm d}z_{1}{\rm d}z_{2}\left[r^{m+n}-\left.r^{m+n}\right|_{\mathbf{X_{1}=\mathbf{X_{2}}}}\right].
\label{eq:I3shsh}
\end{eqnarray}
Now we have a term that scales as the column density, a term that scales
the integrated velocity, and the term inside the integrals (now
negative). Notice also that for large values of the velocity
or density dispersion, the contribution of column density or integrated
velocity respectively is increased within $I3(\mathbf{R})$. But such
contributions will be canceled for the unnormalized centroids that
will have a structure function:
\begin{eqnarray}
\left\langle \left[S(\mathbf{X_{1}})-S(\mathbf{X_{2}})\right]^{2}\right\rangle  & \approx & v_{0}^{2}\left\langle \left[I(\mathbf{X_{1}})-I(\mathbf{X_{2}})\right]^{2}\right\rangle +\rho_{0}^{2}\left\langle \left[V(\mathbf{X_{1}})-V(\mathbf{X_{2}})\right]^{2}\right\rangle \nonumber \\
 &  & -2\alpha^{2}C_{v}C_{\rho}\iint{\rm d}z_{1}{\rm d}z_{2}\left[r^{m+n}-\left.r^{m+n}\right|_{\mathbf{X_{1}=\mathbf{X_{2}}}}\right]+I4(\mathbf{R}).
\label{eq:s1s1shsh}
\end{eqnarray}

\section{Correlation function and power spectra of unnormalized centroids
\label{app:cf_ps_cent}}

Structure functions are considered to be preferable statistics according to
\citet{MY75}. 
They are subjected to less errors related to averaging.
In any case, correlation functions are trivially related to each other by the
formula in equation(\ref{eq:Dij_dec}).
Using the expression for the structure function of unnormalized centroids from
LE03 one gets (see also \citealt{Lev04} for a direct derivation): 
\begin{eqnarray}
\left\langle S(\mathbf{X_{1}})S(\mathbf{X_{2}})\right\rangle  & \approx & \langle\rho^2\rangle \langle v_z^2\rangle(\alpha z_{tot})^{2}+\alpha^{2}v_{0}^{2}\iint\left\langle \tilde{\rho}(\mathbf{x_{1}})\tilde{\rho}(\mathbf{x_{2}})\right\rangle {\rm d}z_{1}{\rm d}z_{2}+\alpha^{2}\rho_{0}^{2}\iint\left\langle \tilde{v}_{z}(\mathbf{x_{1}})\tilde{v}_{z}(\mathbf{x_{2}})\right\rangle {\rm d}z_{1}{\rm d}z_{2}\nonumber \\
 &  & +\alpha^{2}\iint\left\langle \tilde{\rho}(\mathbf{x_{1}})\tilde{\rho}(\mathbf{x_{2}})\right\rangle {\rm \left\langle \tilde{v}_{z}(\mathbf{x_{1}})\tilde{v}_{z}(\mathbf{x_{2}})\right\rangle d}z_{1}{\rm d}z_{2}-\alpha^{2}\iint\left\langle \tilde{\rho}(\mathbf{x_{1}})\tilde{v}_{z}(\mathbf{x_{2}})\right\rangle ^{2}{\rm d}z_{1}{\rm d}z_{2}\nonumber\\
 &  & +2\ \alpha^{2}\rho_{0}\iint\left[\left\langle \tilde{\rho}(\mathbf{x_{1}})\tilde{v}_{z}(\mathbf{x_{1}})\tilde{v}(\mathbf{x_{2}})\right\rangle \right]{\rm d}z_{1}{\rm d}z_{2}.
\label{eq:cf_S1S2exp}
\end{eqnarray}

In analogy with equation (\ref{eq:S1S2Dec}) we can rewrite (\ref{eq:cf_S1S2exp}) as:
\begin{equation}
\left\langle S(\mathbf{X_{1}})S(\mathbf{X_{2}})\right\rangle \approx\langle\rho^2\rangle \langle v_z^2\rangle(\alpha z_{tot})^{2}+B1(\mathbf{R})+B2(\mathbf{R})+B3(\mathbf{R})+B4(\mathbf{R}),\label{eq:cf_S1S2dec}
\end{equation}
where
\begin{mathletters}
\begin{eqnarray}
B1(\mathbf{R}) & = & \alpha^{2}v_{0}^{2}\iint\left\langle \tilde{\rho}(\mathbf{x_{1}})\tilde{\rho}(\mathbf{x_{2}})\right\rangle {\rm d}z_{1}{\rm d}z_{2},\label{eq:B1}\\
B2(\mathbf{R}) & = & \alpha^{2}\rho_{0}^{2}\iint\left\langle \tilde{v}_{z}(\mathbf{x_{1}})\tilde{v}_{z}(\mathbf{x_{2}})\right\rangle {\rm d}z_{1}{\rm d}z_{2},\label{eq:B2}\\
B3(\mathbf{R}) & = & \alpha^{2}\iint\left\langle \tilde{\rho}(\mathbf{x_{1}})\tilde{\rho}(\mathbf{x_{2}})\right\rangle \left\langle \tilde{v}_{z}(\mathbf{x_{1}})\tilde{v}_{z}(\mathbf{x_{2}})\right\rangle{\rm d}z_{1}{\rm d}z_{2},\label{eq:B3}\\
B4(\mathbf{R}) & = & - \alpha^{2}\iint\left\langle \tilde{\rho}(\mathbf{x_{1}})\tilde{v}_{z}(\mathbf{x_{2}})\right\rangle ^{2}{\rm d}z_{1}{\rm d}z_{2}+2\ \alpha^{2}\rho_{0}\iint\left[\left\langle \tilde{\rho}(\mathbf{x_{1}})\tilde{v}_{z}(\mathbf{x_{1}})\tilde{v}(\mathbf{x_{2}})\right\rangle \right]{\rm d}z_{1}{\rm d}z_{2}.
\label{eq:B4}
\end{eqnarray}
\end{mathletters}
Here $B1(\mathbf{R})$ is $v_{0}^{2}$ times the correlation function
of fluctuations of column density, $B2(\mathbf{R})$ is $\alpha^{2}\rho_{0}^{2}$
the correlation function integrated velocity, $B3(\mathbf{R})$ is
different from $2\,I3(\mathbf{R})$ only by a constant, and $B4(\mathbf{R})$
contains the density-velocity cross-correlations found in $I4(\mathbf{R})$.

The Power spectrum of unnormalized centroids is the Fourier Transform
of equation (\ref{eq:cf_S1S2exp}):
\begin{equation}
P_{2D,S}(\mathbf{K})=\langle \rho^2\rangle\langle v_z^2\rangle(\alpha z_{tot})^{2}\delta(\mathbf{K})+v_{0}^{2}P_{2D,I}(\mathbf{K})+\alpha^{2}\rho_{0}^{2}P_{2D,V_{z}}(\mathbf{K})+\mathcal{F}\left\{ B3(\mathbf{R})\right\} +\mathcal{F}\left\{ B4(\mathbf{R})\right\} .
\label{eq:PS_S}
\end{equation}
It is interesting of this is that the weighting factors are no longer
$\langle v_{0}^{2}+\tilde{v}^{2}\rangle$ and
$\langle\rho_{0}^{2}+\tilde{\rho}^{2}\rangle$,
but only $v_{0}^{2}$ and $\rho_{0}^{2}$. And in fact the contribution
of the column density alone can be eliminated by removing the mean
LOS velocity $v_{0}$. However the cross-term, and density-velocity
cross-correlations do affect the scaling properties of the centroids
of velocity. And as the turbulence increases in strength and the ratio
$\langle \tilde{\rho}^2 \rangle / \rho_0^2 $ grows,
the ``contamination'' will also be stronger \citep[see][]{OELS05}.

The correlation function of MVCs can be obtained by subtracting
$\langle v_z^2 \rangle$ times the correlation function of column density
from the correlation function of normalized centroids. Or equivalently:
\begin{eqnarray}
CF_{MVC}(\mathbf{R}) & = & 
\left\langle S(\mathbf{X_{1}})S(\mathbf{X_{2}})\right\rangle-
\left\langle I(\mathbf{X_{1}})I(\mathbf{X_{2}})\right\rangle \nonumber\\
&  \approx & \langle\rho^2\rangle \langle v_z^2\rangle(\alpha z_{tot})^{2}
+B2(\mathbf{R})+B3^{'}(\mathbf{R})+B4(\mathbf{R}),
\label{eq:cf_MVC}
\end{eqnarray}
with
\begin{eqnarray}
B3^{'}(\mathbf{R}) & = &
-\alpha^{2}\iint\left\langle\tilde{\rho}(\mathbf{x_{1}})\tilde{\rho}(\mathbf{x_{2}})\right\rangle
\left[ \left\langle\tilde{v}_{z}^2\right\rangle- \left\langle\tilde{v}_{z}(\mathbf{x_{1}})\tilde{v}_{z}(\mathbf{x_{2}})\right\rangle\right] {\rm d}z_{1}{\rm d}z_{2}\nonumber\\
& = & -\frac{\alpha^{2}}{2}\iint\left\langle\tilde{\rho}(\mathbf{x_{1}})\tilde{\rho}(\mathbf{x_{2}})\right\rangle d_{v_z}(\mathbf{r})\,{\rm d}z_{1}{\rm d}z_{2}.
\label{eq:B3_MVC}
\end{eqnarray}
And, thus the power spectrum of MVCs is:
\begin{equation}
P_{2D,MVC}(\mathbf{K})=\langle \rho^2\rangle\langle v_z^2\rangle(\alpha z_{tot})^{2}\delta(\mathbf{K})+\alpha^{2}\rho_{0}^{2}P_{2D,V_{z}}(\mathbf{K})+\mathcal{F}\left\{ B3^{'}(\mathbf{R})\right\} +\mathcal{F}\left\{ B4(\mathbf{R})\right\} .\label{eq:PS_MVC}
\end{equation}

As the three measures (spectra, correlation, and structure functions) are trivially related, they  represent the same statistics. They can be used interchangeably as dictated by practical convenience.

\section{Unnormalized centroids vs. MVCs \label{app:mvc_good}}

When are MVCs better than regular unnormalized centroids?
To answer this question let us consider MVCs in terms of correlation
functions. 
First, we can minimize the contribution of density fluctuations
to the unnormalized centroids by setting $v_0=0$.
In this case we can fold the difference of unnormalized and modified
centroids in the cross-terms $B3(\mathbf{R})$, and $B3^{'}(\mathbf{R})$
respectively.
Thus, the criterion for MVCs to be an improvement over
unnormalized centroids would be 
$\left|B3(\mathbf{R})\right|>|B3^{'}\mathbf{R})| $. 
In other words:
\begin{equation}
\frac{\left|\iint\left\langle \tilde{\rho}(\mathbf{x_{1}})\tilde{\rho}(\mathbf{x_{2}})\right\rangle \left\langle \tilde{v}_{z}(\mathbf{x_{1}})\tilde{v}_{z}(\mathbf{x_{2}})\right\rangle{\rm d}z_{1}{\rm d}z_{2} \right|}{\frac{1}{2}\left|\iint\left\langle\tilde{\rho}(\mathbf{x_{1}})\tilde{\rho}(\mathbf{x_{2}})\right\rangle d_{v_z}(\mathbf{r})\,{\rm d}z_{1}{\rm d}z_{2} \right|}>1.
\label{eq:MVC_good}
\end{equation}

\begin{figure}
\epsscale{0.35}
\plotone{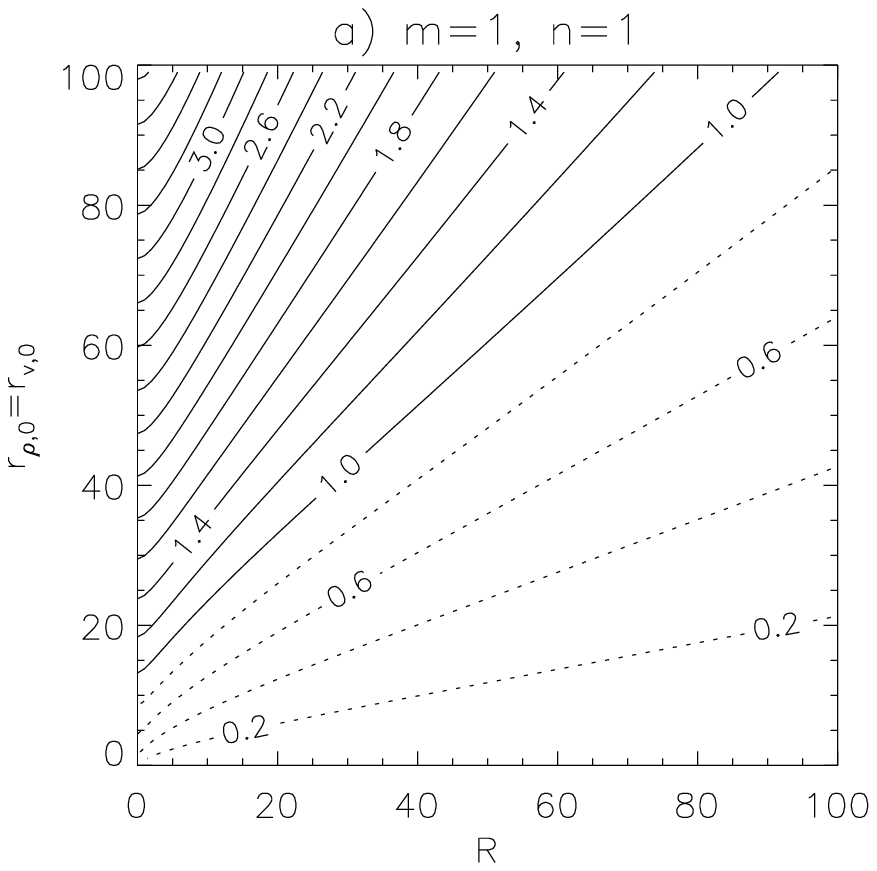}\plotone{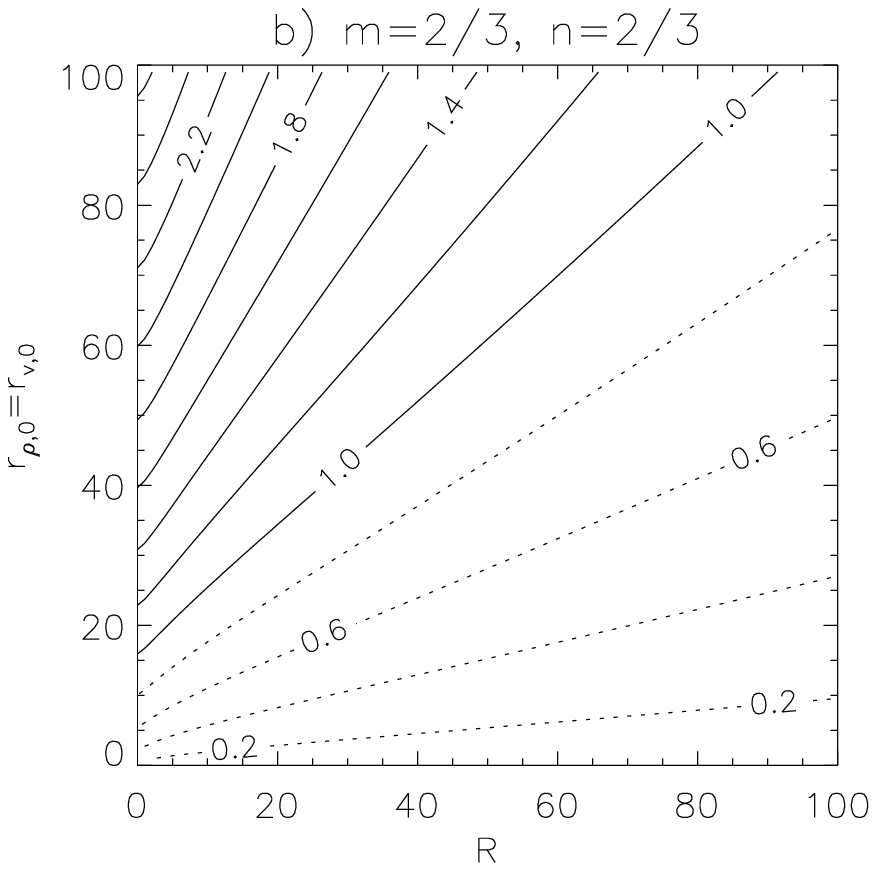}\plotone{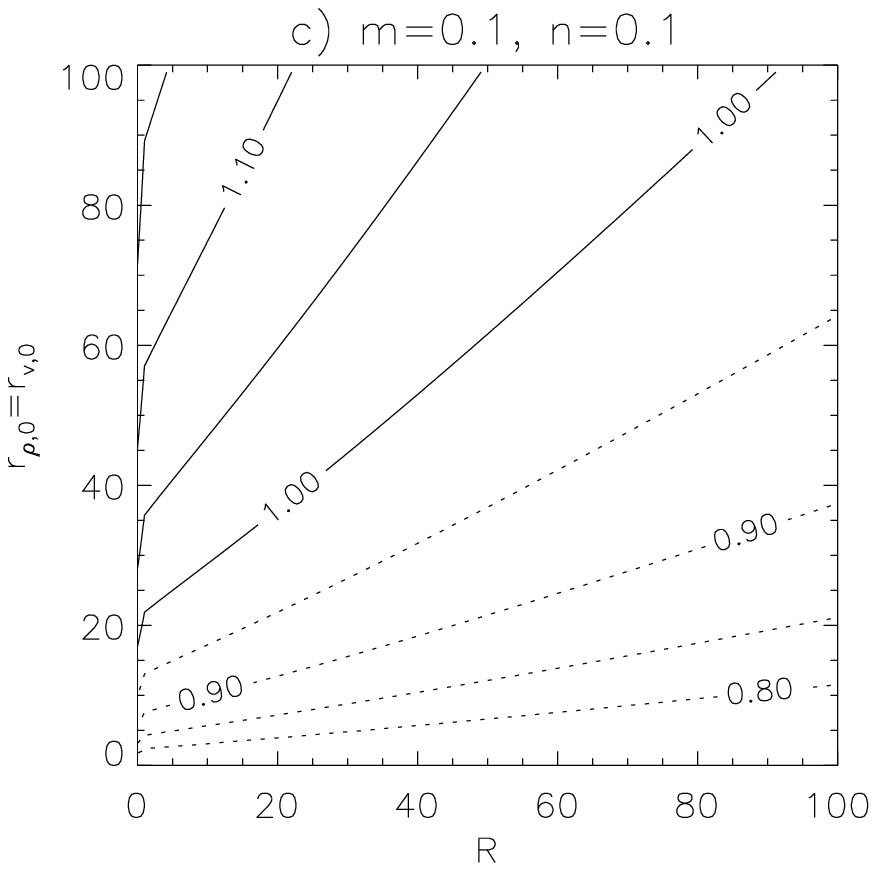}
\plotone{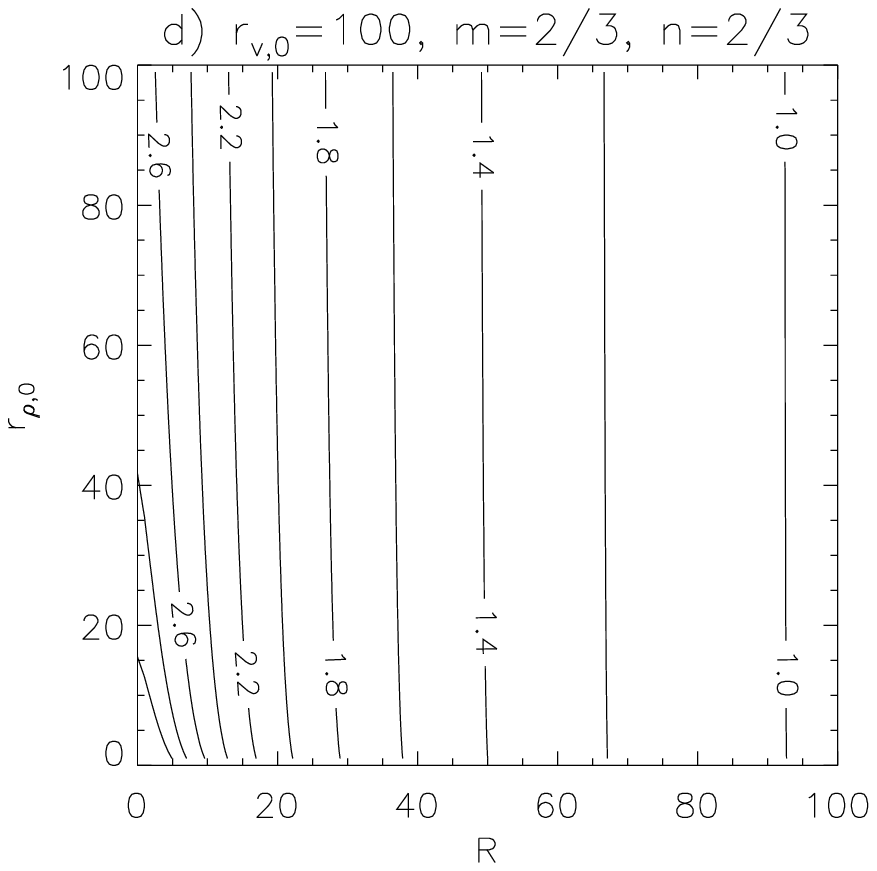}\plotone{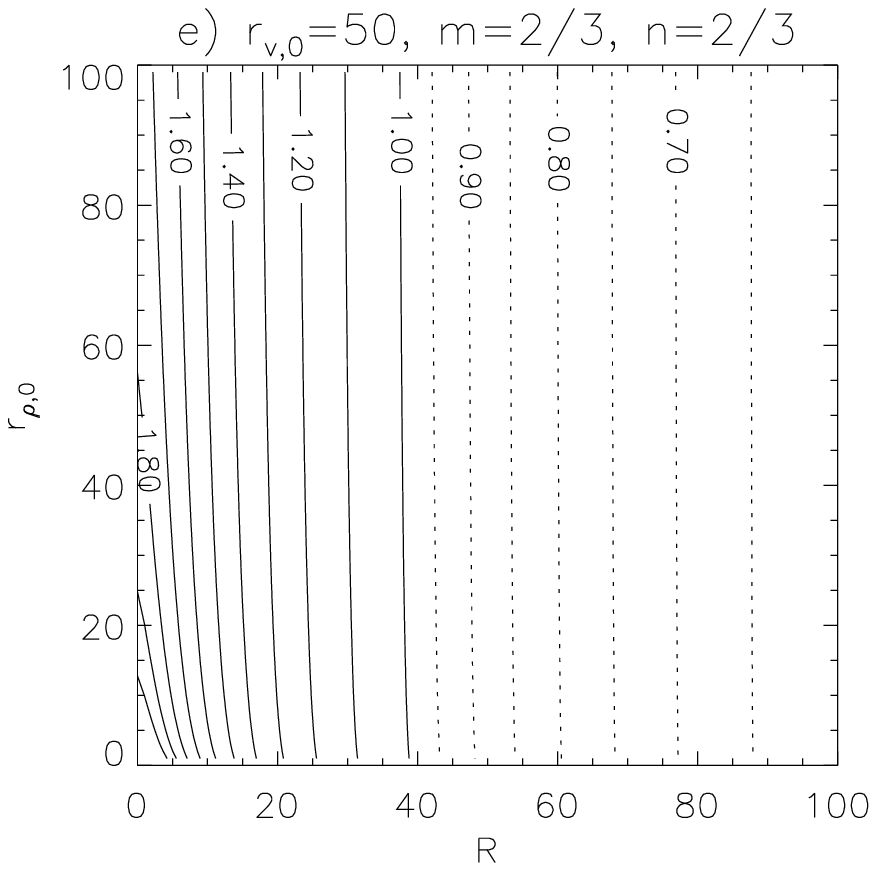}\plotone{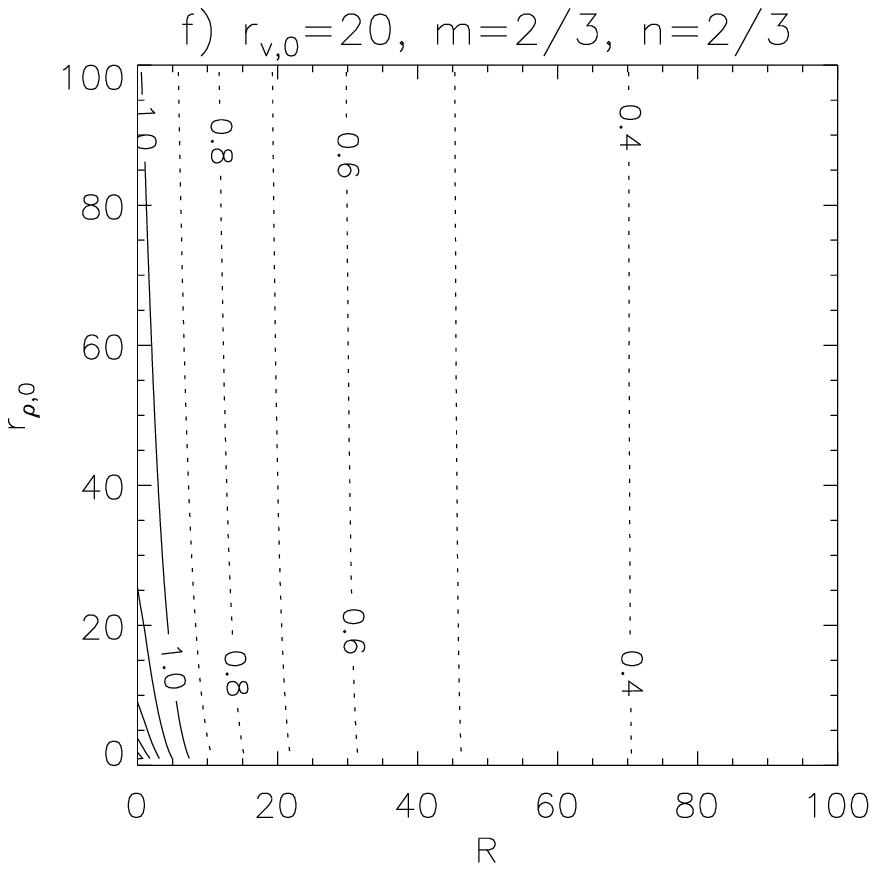}
\figcaption{Criterion for MVCs to trace velocity better than unnormalized centroids for an example of steep density and velocity. (see equation [\ref{eq:MVC_good_st_st}]). The parameters are labeled in the title and axis for each panel, $z_{tot}=100$ in all cases. The criterion is met for parameters that correspond to iso-contours above $1$. Panels ({\it a}), ({\it b}), and  ({\it c}) show the low sensitivity of the position of the $1$ iso-contour with the spectral indices. Panels ({\it d}), ({\it e}), and  ({\it f})  show the strong dependence on $r_{v,0}$ and the lag $R$y, as well as a weak dependence on $r_{\rho,0}$.
\label{fig:mvc_good}}
\end{figure}

Let us consider a more realistic {\it steep} velocity field, with a 
power-law structure function that saturates at some cutoff scale $r_{v,0}$
(see LP00, \citealt{LP04}):
\begin{equation}
d_{v_z}(\mathbf{r})=2\left\langle \tilde{v}_z^2\right\rangle
\frac{r^{m}}{r^{m}+r_{v,0}^{m}}
\label{eq:sf_cut}.
\end{equation}
We use $r_{v,0}$ to avoid
confusion with the critical scale for the short-wave dominated spectra,
and because this cutoff scale can also be related to a correlation length
for the correlation function:
\begin{equation}
\left\langle \tilde{v}_z(\mathbf{x_1})\tilde{v}_z(\mathbf{x_2})\right\rangle
= \left\langle \tilde{v}_z^2\right\rangle
\frac{r_{v,0}^{m}}{r^{m}+r_{v,0}^{m}}.
\label{eq:cf_cut}
\end{equation}
If we combine this steep velocity field, with a steep density field
--i.e. 
$\langle \tilde{\rho}(\mathbf{x_1}\tilde{\rho}(\mathbf{x_1}) \rangle=
\langle\tilde{\rho}^2\rangle\,r_{\rho,0}^{n}/(r^{n}+r_{\rho,0}^{n})$--,
with the same change of variables as before ($z_{-}=z_2-z_1$)
the criterion in equation (\ref{eq:MVC_good}) becomes
\begin{equation}
\frac
{\left|\int_{0}^{z_{tot}}
\left[\frac{r_{\rho,0}^{n}}{(R^2+z_{-}^2)^{n/2}+r_{\rho,0}^{n}}\right]
\left[\frac{r_{v,0}^{m}}{(R^2+z_{-}^2)^{m/2}+r_{v,0}^{m}}\right]
{\rm d}z_{-}\right|}
{\left|\int_{0}^{z_{tot}}
\left[\frac{r_{\rho,0}^{n}}{(R^2+z_{-}^2)^{n/2}+r_{\rho,0}^{n}}\right]
\left[\frac{(R^2+z_{-}^2)^{m/2}}{(R^2+z_{-}^2)^{m/2}+r_{v,0}^{m}}\right]
{\rm d}z_{-}\right|}
>1.
\label{eq:MVC_good_st_st}
\end{equation}
In Figure \ref{fig:mvc_good} we plotted iso-contours of this ratio for various
combinations of parameters. Within the parameter space corresponding to
iso-contour above $1.0$ MVCs are expected to trace the statistics of
velocity better than unnormalized centroids.
In the first three panels ({\it a}, {\it b}, {\it c}) we present, for
$r_{\rho,0}=r_{v,0}$ how the ratio in eq.(\ref{eq:MVC_good_st_st}) varies
with the lag $R$ over a extreme contrast of possible spectral indices
(within the long-wave dominated regime).
In the last three panels  ({\it d}, {\it e}, {\it f}) we choose a
particular set of spectral indices and relax the condition
$r_{\rho,0}=r_{v,0}$. In all cases $z_{tot}=100$.
We find that the criterion for MVCs to be better than unnormalized
centroids is satisfied for lags smaller than the cutoff for the velocity
field, with very little sensitivity to the particular spectral index or
the cutoff scale of density.

We performed the same exercise for the other physically motivated case,
that is, steep velocity and shallow density. A shallow density can be
approximated by the same expression for the correlation function of density,
but with a small $r_{\rho,0}$. We tested for spectral indices from
$0.1$ to $1.0$ for the velocity, and from $-1.0$ to $-0.1$ for the density.
The same behavior as for the steep-steep case was found. That is, MVC are
expected to trace the statistics of velocity better than unnormalized centroids
for lags smaller than the velocity correlation-scale ($r_{v,0}$), and smaller
than the LOS extent of the object ($z_{tot}$).

\end{appendix}

\end{document}